\documentclass[epsfig,usegraphicx,useAMS,usenatbib,caption]{mn2e}

\def\sqdeg{\,deg$^2$}
\def\etal{{et~al.~}}
\def\msolh{\,{\it h}^{-1}\, {\rm M_{\odot}}}
\def\msol{\,{\rm M_{\odot}}}

\def\kpc{\,{\rm kpc}}

\def\mpc{\,{\rm Mpc}}

\def\kms{{\rm \,km \, s^{-1}}}

\def\mpc3h3{\,{\it h}^{-3}\, {\rm Mpc}^{3}}
\def\G3C{G$^3$Cv1}
\def\MMW{$\mathcal{M}_{S,MW}$}

\title[GAMA: The Life and Times of $L^*$ Galaxies]{Galaxy And Mass Assembly (GAMA): The Life and Times of $L^*$ Galaxies}
\author[A.S.G. Robotham~\etal]{A.S.G.~Robotham$^{1,2}$\thanks{E-mail:asgr@st-and.ac.uk},
J.~Liske$^3$,
S.P.~Driver$^{1,2}$,
A.E.~Sansom$^{4}$,
I.K.~Baldry$^5$,
\newauthor
A.E.~Bauer$^6$,
J.~Bland-Hawthorn$^7$,
S.~Brough$^6$,
M.J.I.~Brown$^8$,
M.~Colless$^6$,
\newauthor
L.~Christodoulou$^9$,
M.J.~Drinkwater$^{10}$,
M.W.~Grootes$^{11}$,
A.M.~Hopkins$^6$,
L.S.~Kelvin,$^{1,2,12}$,
\newauthor
P.~Norberg$^{13}$,
J.~Loveday$^9$,
S.~Phillipps$^{14}$,
R.~Sharp$^{15}$,
E.N.~Taylor,$^{7,16}$,
R.J.~Tuffs,$^{11}$\\\\
$^1$SUPA\thanks{Scottish Universities Physics Alliance}, School of Physics \& Astronomy, University of St Andrews, North Haugh, St Andrews, KY16 9SS, UK\\
$^2$ICRAR\thanks{International Centre for Radio Astronomy Research}, The University of Western Australia, 35 Stirling Highway, Crawley, WA 6009, Australia\\
$^3$European Southern Observatory, Karl-Schwarzschild-Str.~2, 85748 Garching, Germany\\
$^4$Jeremiah Horrocks Institute, University of Central Lancashire, Preston PR1 2HE, UK\\
$^5$Astrophysics Research Institute, Liverpool John Moores University, Egerton Wharf, Birkenhead, CH41 1LD, UK\\
$^6$Australian Astronomical Observatory, PO Box 296, Epping, NSW 1710, Australia\\
$^7$Sydney Institute for Astronomy, School of Physics, University of Sydney, NSW 2006, Australia\\
$^8$School of Physics, Monash University, Clayton, Victoria 3800, Australia\\
$^9$Astronomy Centre, University of Sussex, Falmer, Brighton BN1 9QH, UK\\
$^{10}$Department of Physics, The University of Queensland, Brisbane, QLD 4072, Australia\\
$^{11}$Max Planck Institute for Nuclear Physics (MPIK), Saupfercheckweg 1, 69117 Heidelberg, Germany\\
$^{12}$Institut f\"{u}r Astro- und Teilchenphysik, UniversitŠt Innsbruck, Technikerstra§e 25, 6020 Innsbruck, Austria\\
$^{13}$Institute for Computational Cosmology, Department of Physics, Durham University, South Road, Durham DH1 3LE, UK\\
$^{14}$HH Wills Physics Laboratory, University of Bristol, Tyndall Avenue, Bristol, BS8 1TL, UK\\
$^{15}$Research School of Astronomy \& Astrophysics, The Australian National University, Cotter Road, Weston Creek, ACT 2611, Australia\\
$^{16}$School of Physics, the University of Melbourne, Parkville, VIC 3010, Australia\\
\vspace*{-3em}
}

\begin{document}

\date{\vspace*{-4em}\noindent07/01/2013}

\pagerange{\pageref{firstpage}--\pageref{lastpage}} \pubyear{2013}

\maketitle

\label{firstpage}

\vspace*{-1em}

\begin{abstract}

In this work we investigate in detail the effects local environment (groups and pairs) has on galaxies with stellar mass similar to the Milky-Way ($L^*$ galaxies).
A volume limited sample of 6,150 galaxies is visually classified to determine emission features, morphological type and presence of a disk. This large sample allows for the significant characteristics of galaxies to be isolated (e.g.\ stellar mass and group halo mass), and their codependencies determined.

We observe that galaxy-galaxy interactions play the most important role in shaping the evolution within a group halo, the main role of halo mass is in gathering the galaxies together to encourage such interactions. Dominant pair galaxies find their overall star formation enhanced when the pair's mass ratio is close to 1, otherwise we observe the same galaxies as we would in an unpaired system. The minor galaxy in a pair is greatly affected by its companion galaxy, and whilst the star forming fraction is always suppressed relative to equivalent stellar mass unpaired galaxies, it becomes lower still when the mass ratio of a pair system increases.

We find that, in general, the close galaxy-galaxy interaction rate drops as a function of halo mass for a given amount of stellar mass. We find evidence of a local peak of interactions for Milky-Way stellar mass galaxies in Milky-Way halo mass groups. Low mass halos, and in particular Local Group mass halos, are an important environment for understanding the typical evolutionary path of a unit of stellar mass.

We find compelling evidence for galaxy conformity in both groups and pairs, where morphological type conformity is dominant in groups, and emission class conformity is dominant in pairs. This suggests that group scale conformity is the result of many galaxy encounters over an extended period of time, whilst pair conformity is a fairly instantaneous response to a transitory interaction.

\end{abstract}


\newpage

\section{Introduction}

$L^*$ galaxies exist at the knee of the popular \citet{sche76} parameterisation of the galaxy luminosity function. The galaxy stellar mass function has a similar form, and as such the knee of the stellar mass function ($\mathcal{M}^*$) represents broadly the same galaxy population. Due to the shape of the galaxy luminosity function, $L^*$ galaxies dominate the integrated mass and luminosity densities in the Universe at low redshift \citep[e.g.][]{bald12,driv12}.

To understand the integrated properties of all galaxies we can, to the first order, just consider the properties that affect the evolution of $\sim L^*$ galaxies. Simply doing this offers insight into the fate of the median and modal unit of stellar mass in a cosmological volume. Further to this, apparent magnitude limited surveys at moderate redshifts \citep[e.g.\ 2dFGRS, SDSS, MGC, GAMA, see][respectively]{coll01,adel08,lisk03,driv11} are dominated by $L^*$ galaxies in terms of raw number counts, even though they are not the most populous type of galaxy. Thus they are both extremely important in determining integrated properties of mass and light in cosmological volumes of the Universe, and the most easily studied variety of galaxy by virtue of the double power-law shape of the luminosity function { \citep[see][for a discussion on the origin of $L^*$]{coor05}}. By a happy coincidence our own galaxy, the Milky-Way, is also an $L^*$ galaxy. { This observational bias is not surprising, since it just means our sun resides in the most typical type of galaxy to find stellar mass.}

The combination of these factors makes a detailed investigation of $L^*$ galaxies an instructive exercise, informing us both about the integrated properties of galaxies in cosmological volumes, and the typicality of our own galaxy. It is important that we fully understand any statistical biases present in the Milky-Way since, by virtue of proximity, it will always be the galaxy we can investigate in the most detail, and from which we can derive the most information about its formation history.

However, question marks remain over how typical the MW halo is in the context of the Universe and how unusual its galaxy occupation statistics are \citep[e.g.][]{boyl09,boyl10,boyl11,jame11,love12,toll11,weis11,robo12}. We are set to learn a vast amount about the Milky-Way in the coming decades. In the near future GAIA \citep{wilk05} will measure space motions and properties for 2 billion stars in the Local Group (LG) which includes all known member galaxies as part of the new era of Galactic archeology studies. Amongst likely discoveries, we will learn about dynamical equilibrium, or lack of it, for the first time. Building up to these hugely detailed surveys it is important we discover where the Milky-Way fits into the bigger picture. Only then can we apply what we know about the Milky-Way to  cosmological galaxy formation models. Combining near-field cosmology (LG scale) and far-field cosmology (redshift surveys) is key to completing the full picture of galaxy formation \citep{free02}.

$L^*$ galaxies are observed to have extremely wide ranging properties, inhabiting a regime where we see the most variation in star formation, morphology and colour. This indicates that there is no typical evolutionary history for these `typical' galaxies. We can theorise about a large number of factors that might directly alter the evolution of any galaxy, including $L^*$ ones. A non-comprehensive list includes galaxy-galaxy interactions, local scale environment (e.g.\ its dark matter halo), and the large scale environment of the cosmic web.

Many recent studies have investigated the role of environment in driving the evolution of galaxies, with much contradiction in the literature. Studies on SDSS by \citet{wein06} suggest strong evidence of `conformity' between central and satellite galaxies in halos for a given amount of stellar mass, the suggestion from \citet{wang12} being that redder central galaxies of a given stellar mass tend to occupy larger halos. This conclusion suggests that the halo environment has a significant role in determining the properties of galaxies, on top of those driven purely by stellar mass. Counter to this is recent work by \citet{peng10} using z-COSMOS and \citet{thom10}, who both broadly conclude that once `environment' has influenced the stellar mass of a galaxy { (particularly a central galaxy)}, the remaining properties are predominantly driven by the stellar mass. This is in alignment with earlier work by \citet{bald06}, which finds the colour--stellar mass and colour--concentration locus of galaxies is uncorrelated with environment (which in this case is density defined).

This work puts the investigation of $L^*$ galaxies into an observational cosmological context by using data from the Galaxy and Mass Assembly project \citep[GAMA:][]{driv11}. GAMA is a multi-wavelength photometric and spectroscopic survey, and is designed to answer questions about how matter has assembled on a huge variety of scales: galaxies, groups, clusters and filaments (i.e.\ all of the environmental scales outlined above). The first phase of the redshift survey was conducted on the AAT (known as GAMA-I) and these data are used in this work \citep{driv11}.

GAMA aims to describe the physics occurring in dark matter haloes across the largest mass range ever probed within a single survey, spanning at least 3 orders of magnitude in mass: $\sim10^{12} \msolh < M_{\rm halo} < 10^{15} \msolh$. The GAMA galaxy group catalogue (\G3C) was recently created using the first 3 years of GAMA-I redshift data and is described in detail in \citet[][R11 hereafter]{robo11}. This catalogue was constructed using an adaptive Friends-of-Friends (FoF) grouping algorithm that was extensively tested on mock simulations. The \G3C contains LG mass haloes out to $z\sim 0.1$, allowing us to better understand how typical Milky-Way type galaxies are within a cosmologically significant volume of the Universe.

This work has been divided into multiple papers. The first paper investigated the specific occurrence rates of Magellanic Cloud mass galaxies around Milky-Way mass ($L^*$) galaxies--- this is a novel question in its own right since much cosmology depends upon the typicality of the Magellanic Cloud satellites of the Milky-Way \citep{robo12}. This second paper investigates how the defining characteristics of $L^*$ galaxies (type, morphology, star formation rates, halo mass and stellar mass amongst others) vary within the total galaxy population, grouped galaxies and close pair systems. This will answer questions on what physics are driving the characteristics of $L^*$ galaxies like our own MW. The final planned paper will measure the luminosity function of MW mass halos, revealing how typical the dwarf galaxy population of the MW is.

There has been much written in the literature about the different physical processes that have a measurable effect on the properties of a galaxy. Recent key observational discussions regarding the influence of environment on the properties of galaxies include \citet{bald06,wein06,peng10,thom10,kauf12}. The complicating factor has always been in separating out the clear primary role that local environment has in producing a particular distribution of galaxies \citep[e.g.\ luminosity or stellar mass:][]{robo06,wein06} and second order processes that affect properties of these galaxies \citep[e.g.\ colour and morphology:][]{driv06,kelv12}.

The major roles that a group (or halo) has in assembling stellar mass are: feeding gas into galaxies to enhance star formation \citep{kere05,deke06}; heating and stripping gas out of galaxies \citep[e.g.][]{gunn72,abad99,balo00,deke06} or stripping of stars directly \citep{bekk01} and providing a location for rapid galaxy-galaxy interactions \citep{barn92a}, be they harassment, minor or major merger processes \citep[e.g.][respectively]{moor96,cole00}. All of these processes provide mechanisms for stellar mass build-up in galaxies to be either accelerated or decelerated, and will have noisy average net effects as a function of stellar mass, halo mass and epoch. Except for the tidal dwarf variety of galaxies \citep{barn92b} all galaxies are assumed to be located within a dark matter halo of some description, though to date no rotation curve has been measured for a `tidal dwarf' that does not require some `dark' component \citep{bour07}. We can therefore say that galaxy stellar mass is entirely coupled to environment, which is not to say the coupling is trivial. The fact it is not trivial has invited a large variety of feedback models in the literature \citep[e.g.][]{bens03,bowe06,crot06}. Beyond this, it is less clear-cut what additional role the environment has on the observed properties of galaxies.

The second order properties of galaxies (those things that are not stellar mass) can be largely divided into two categories: luminous energy characteristics (any aspect of a galaxy defined by light output as a function of wavelength: star formation rate (SFR) and galaxy colour etc) and orbital energy characteristics (aspects concerning the position and velocity of stars: morphology and colour gradients etc). Processes that can affect the luminous energy characteristics are mostly those that directly enhance or suppress star formation (which is then necessarily coupled to the stellar mass build-up discussed above) and more contentiously any process that affects the initial mass function (IMF) that stars are forming with \citep[see][for recent work on variable IMFs]{guna11}.

Orbital properties within galaxies can evolve naturally through `secular' processes that allow for orbital energy redistribution, which we observe as brief and varying manifestations such as pattern bars and spiral arms \citep[e.g.][respectively]{sell93,bert89}, or permanent transitions such as the formation of thick disks and pseudo bulges \citep[e.g.][respectively]{hayw08,korm04}. Orbital properties are also able to be changed more dramatically and violently during encounters with other galaxies \citep{barn92a}. Minor merger events can produce subtle changes to stellar orbits, providing another mechanism for the creation of a thick disk \citep{read08}. Major mergers can be incoherent enough to transform a galaxy with regular rotational disk-like stellar structure (the canonical late-type galaxy) into a system with random orbits (the canonical early-type galaxy).

Self-evidently `secular' processes are not a direct consequence of environment since they describe processes that should occur regardless of outside influence / triggering, therefore they might be expected to be entirely predictable from the instantaneous properties of a galaxy. However, environment can play a role in disrupting secular processes, e.g.\ a major merger would destroy any bar formation process. Clearly there is scope for environment to play at least a secondary role in almost all aspects of galaxy evolution, if only by virtue of disruption, delay and prevention mechanisms. The purpose of this paper is to simultaneously separate out the stellar mass, galaxy-galaxy interaction, and halo mass terms in order to quantify the codependencies of secularly and environmentally driven stellar mass evolution.

Secondary properties of galaxies clearly display differing sensitivity to modifying external drivers. E.g.\ H$\alpha$ equivalent width (EW) is a relatively instantaneous response to an increase in star formation, accompanied by a change in integrated bandpass colour to bluer colours. In the longer term the in-situ gas and locked-up stellar metallicty will increase, which will create a redder SED. If this burst of star formation occurred due to a minor merger event then we might also expect orbits to become disrupted, leading to a change in the velocity dispersion and light profile of the galaxy. If the change is dramatic enough (e.g.\ a major merger) it might also invoke a change in the `type' classification of the galaxy. Thus we have galaxy properties that are highly sensitive to current star formation invocations (H$\alpha$-EW) and ones that betray more about the integrated history of a galaxy (some component of colour and all aspects of morphology). Depending on the physical processes that caused star formation we might also observe longer term changes to the morphology of the galaxy (in the case of minor or major merger events). In this work we aim to separate out these different secondary properties into those most dependent on different likely primary drivers of galaxy properties: stellar mass, halo mass and galaxy-galaxy interactions.

In Section \ref{sec:Data} we discuss the data used in this work in detail. In Section \ref{sec:selection} we discuss the sample selections applied to the data. In Section \ref{sec:eyeball} we describe the visual classification process that was used to categorise the morphological type, emission class and presence of a disk. In Section \ref{sec:galcat} we discuss the different varieties of galaxy categorisation used for constructing later plots. In Section \ref{sec:galdep} we analyse how properties of MW mass galaxies are affected both by the group-scale and pair-scale environment, where the codependence of galaxy properties is discussed in detail. In Section \ref{sec:discussion} we pull together the main results of Section \ref{sec:galdep} into a coherent narrative and put the findings into context. In Section \ref{sec:conclusions} we briefly summarise the major findings of the work.

In this work we use $\Omega_M=0.25$, $\Omega_\Lambda=0.75$ and $H_{0}=70\,{\rm km}\,{\rm s}^{-1}$. This is consistent with the first paper in this series that dealt specifically with the commonality of Magellanic Cloud galaxies around $L^*$ systems \citep{robo12}.

\section{Data}
\label{sec:Data}

\subsection{GAMA}

The Galaxy and Mass Assembly project (GAMA) is a major new
multi-wavelength photometric and spectroscopic galaxy survey \citep{driv11}. The final
redshift survey will contain $\sim$300,000 redshifts to $r_{\rm AB}=19.8$
over $\sim280$~\sqdeg, with a survey design aimed at providing an
exceptionally uniform spatial completeness
\citep{robo10a,bald10,driv11}. 

Extensive details of the GAMA survey characteristics are given in
\citet{driv11}, with the survey input catalogue described in
\citet{bald10} and the spectroscopic tiling algorithm in
\citet{robo10a}. The first 3 years of data obtained has been frozen
for internal team use and is referred to as GAMA-I.

Briefly, the GAMA-I survey covers three regions  
each $12 \times 4$ degrees centred at 09h, 12h and 14h30m
(respectively G09, G12 and G15 from here). The survey
depths and areas relevant for this study are: $\sim96$~\sqdeg\ to
$r_{\rm AB}=19.4$ mag (G09 and G15) and $\sim47$~\sqdeg\ to $r_{\rm AB}=19.8$
(G12)\footnote{See \citet{bald10} for additional GAMA-I selections.}.
All regions are more than 98\% complete 
\citep[see][for precise completeness details]{driv11}, with special
emphasis on a high close pair completeness, which is greater than 95\%
for all galaxies with up to 5 neighbours within 40$''$ of them 
\citep[see Fig.~19 of][]{driv11}. Stellar masses used for this work are taken from \citet{tayl11}.

\subsection{Spectral Analysis}

The spectra collected during the GAMA observations have been flux calibrated, Balmer decrement corrected, aperture corrected, and the emission line equivalent widths measured. The spectral pipeline process is discussed in detail in Hopkins et~al. 2013 (in prep). The line measurements have also been converted into star formation rates, a process that is described in Gunawardhana et~al. 2013 (in prep). These SFRs are used directly in this work.

These emission line measurements are also used for automatic spectral classifications by comparing the [OIII]/H$\beta$ as a function of [NII]/H$\alpha$ in the popular manner outlined in \citet{bald81}, widely known as BPT classification. This process, and the classifications selections used, are discussed in detail in Hopkins et~al. 2013 (in prep) and Gunawardhana et~al. 2013 (in prep). These automatic classifications are supplemented by extensive visual checks to minimise classification errors (discussed below).

For this work we keep the classifications as simple as possible: H$\alpha$ emission that is dominated by star formation irradiation, H$\alpha$ emission that is dominated by AGN irradiation and no H$\alpha$ emission. This binary definitions of star formation means we are highly insensitive to the details of flux calibration, Balmer decrement correction and aperture correction for the majority of our work. The fidelity of the spectral classifications is investigated below.

\subsection{\G3C}

One of the principal science goals of
GAMA is to make a statistically significant analysis of low mass
groups ($M \leq 10^{13}\msol$), helping to constrain the low
mass regime of the dark matter halo mass function
and galaxy formation efficiency in Local Group like haloes. This low mass regime contains groups
that have halo masses similar to the LG, allowing us to investigate how common
the LG environment is, and putting LG science into a cosmological context.

The GAMA-I spectroscopic data was used to construct the first GAMA galaxy group catalogue \citep[\G3C:][]{robo11}.
This group catalogue was constructed using an adaptive Friends-of-Friends (FoF) algorithm that was optimised to recover
groups in the magnitude limited GAMA-I spectroscopic data.

In total \G3C contains 14,388 groups (with multiplicity $\ge 2$), including 44,186 galaxies out of a possible 110,192 galaxies,
implying $\sim$40\% of all galaxies are assigned to a group. Since GAMA-I has an uneven $r$-band selection function between
GAMA regions we use an $r < 19.4$ mag subset of the full \G3C for this work. This catalogue has 37,576 galaxies in 12,200 groups.

The \G3C contains a wealth of ancillary data regarding the properties of the groups, including dynamically estimated halo masses. These
halo masses are used extensively in this work. For details regarding the reliability and biases the reader should refer to \citet{robo11} and \citet{alpa12}.

\subsection{Pair Catalogue}
\label{sec:paircat}

This paper is investigating the effect that various kinds of environment have on the characteristics of $L^*$ galaxies. Close galaxy interactions have a significant role on the evolution of galaxies, so these systems are selected for close analysis in this work. Part of the process of creating the \G3C catalogue involves the construction of all galaxy pairs \citep[see][for details]{robo11}. This catalogue contains galaxies with potentially quite large radial (velocity) and tangential (spatial distance) separations between galaxies. Using the full pair catalogue would include galaxy pairs with very large dynamical times, and such pairs will have weak evolutionary effects on the component galaxies. Instead we select a narrow window of interaction phase space in order to preferentially extract galaxies that will be most affected by close galaxy-galaxy interactions. The `close pair' sample selected here is based on that presented in \citet{robo12}, where we aimed to recover galaxy pairs that are similar to the MW Magellanic Clouds system. This ensures we can investigate a variety of interaction that our own MW $L^*$ galaxy has experienced, helping to place it in the broader family of $L^*$ galaxies.

The SMC is $\sim60\kpc$ from the MW and travelling radially away at $\sim 17 \kms$ and $301\kms$ tangentially with respect to the MW: a net velocity of $302 \kms$ \citep{nich11}. The LMC is $\sim50\kpc$ from the MW and travelling radially away at $\sim 89 \kms$ radially and $367 \kms$ tangentially with respect to the MW: a net velocity of $378 \kms$ \citep{nich11}. To conservatively recover all systems where the galaxies are in such spatial--velocity proximity, we create a catalogue of pairs for this work where the projected separation is $r_{\rm sep-proj}<70\kpc$ and the radial velocity separation is $v_{\rm sep-rad}<400 \kms$. See \citet{robo12} for detailed discussion on galaxy pair selections.

Applying these selection limits to recover all systems that have similar pairwise dynamical properties to the MW and the Magellanic Clouds creates a catalogue containing 5,964 galaxy pairs and 10,416 unique galaxies. These numbers are for the whole GAMA-I dataset, so include stellar masses that might be highly dissimilar to the real MW Magellanic system.

Some galaxies will be members of more than one pair (the MW is paired with two galaxies--- the SMC and LMC). We create groupings that contain all galaxy-galaxy associations and count this as a single `pair' system, e.g.\ the MW is in a pair system of 3, as are the SMC and LMC. In the same way that simple galaxy pairs are only counted once, higher order pair systems are also only counted once. Table \ref{tab:pairstat} shows the frequency of different pair systems of galaxies. Whilst we recover 5,964 galaxy pairs, once common links are explored we are left with 5,446 systems.

\begin{table}
\begin{center}
\begin{tabular}{rrr}
Galaxies in paired systems  	& 	Frequency	&	Total galaxies\\
\hline
2	&	4,550	&	9,100	\\
3	&	370		&	1,110	\\
4	&	45 		&	180		\\
5	&	4		&	20		\\
6	&	1		&	6		\\
\end{tabular}
\end{center}
\caption{Number of pair systems with different numbers of galaxies.}
\label{tab:pairstat}
\end{table}

\section{Sample Selection}
\label{sec:selection}

A large component of the analysis required for this paper required comprehensive visual classification (described in detail below). The task of visually classifying the data in this manner was extremely labour intensive--- each galaxy could take a minute to analyse and GAMA-I contains 129,765 galaxies with redshifts above 0.01. Doing this for 8 hours a day would take 270 days. Instead we selected a subset of galaxies concentrated on $L^*$ systems (the focus of this paper), and systems that share common properties with the MW and its local environment (allowing us to identifying the evolutionary drivers for the MW).

Since this work is predominantly interested in the properties of MW mass ($L^*$) galaxies we had to comfortably include all GAMA galaxies with stellar mass similar to the MW \citep[$5.0 \times 10^{10} \msol$:][]{flyn06}. A conservative lower stellar mass limit of $\mathcal{M_{S}}=10^{10} \msol$ was chosen. Due to the rapid decrease in the number of objects as we move into the higher stellar mass regime there was no need to implement a higher stellar mass cut-off. In total this stellar mass selected sample contains 3,550 galaxies.

Since we are also investigating the environment that $L^*$ galaxies reside in, we add all galaxies that are associated with the $L^*$ selection to the sample. There are two varieties of `environment' investigated in this work: galaxy pairs and galaxy groups. These are just two of an infinite number of possible definitions of environment, but they probe the smallest physical scale environments possible, and as such should possess the cleanest possible relationship between environment and galaxy properties due to the recent dynamical events. To link this work closely with our own MW environment (important since so much effort is being invested in surveys of the MW and local galaxies), we are careful to ensure that our definitions of `pair' fully capture interactions like the MW is undergoing with the LMC (its dominant pairwise interaction). We also select all groups that are similar mass to the Milky-Way halo, allowing us to investigate how common the MW system is, given its halo mass as a prior.

First we add our pair selection criteria. We include all pairs within a redshift limit of $z=0.089$ that have at least one member selected in the mass sample described above. This allows us to observe LMC equivalent galaxies throughout our sample \citep[see][ for details of the selection limits]{robo12}. Together with the lower redshift limit of $z=0.01$ this gives a sample of 3,503 galaxies. In this work we use stellar mass rather than $r$-band magnitude to describe the galaxies, and the redshift limit $z=0.089$ is specifically the implied $r$-band magnitude limit to observe LMC type galaxies. However, since the LMC is blue, late-type and star forming this will act as a very conservative LMC stellar mass limit. Indeed the $\mathcal{M}_{s}>\mathcal{M}_{S,LMC}$ number counts drop off at a $z \sim 0.1$. For consistency with \citet{robo12} we use the more conservative limit of $z=0.089$ to select down to LMC mass galaxies.

In order to determine what role different environments may have on the evolution of $L^*$ systems, we then select all galaxies that are in any \G3C group or galaxy-galaxy pair containing any of the 3,503 galaxies in the above subset. This gives us an expanded sample of 5,054 galaxies. We add to this all galaxies in all groups that have a \G3C halo mass between $7\times10^{11} \msol < M_{FoF} < 10^{13} \msol$ where $M_{FoF}$ is defined using the functional form of the halo mass used in \citet{robo11}. This halo mass range comfortably covers both sides of the MW halo mass used in \citet[][$2.5 \times 10^{12} \msol$]{robo12} and the total LG mass \citep[MW+M31 haloes = $\sim 5 \times 10^{12} \msol$:][]{li08}. Including this halo mass range guarantees that all systems with properties similar to the MW by virtue of stellar mass, and halo mass or close pair interactions, are certain to be included in our sample. This gives us a final sample of 6,150 galaxies to visually classify in detail. As long as the stellar mass and halo mass ranges are kept within the limits described above, this sample is volume limited.

The result is our sample has two major components:

\begin{itemize}
\item All GAMA galaxies with $\mathcal{M_{S}} \ge 10^{10} \msol$ (mass sample, with 3,550 galaxies).
\item Within $0.01<z<0.089$, all pairs and groups are included that have at least one member in the above mass selection. These two varieties of environment are therefore complete down to LMC mass galaxies. In addition all \G3C groups with a halo mass in the range $7\times10^{11} \msol < M_{FoF} < 10^{13} \msol$ and $0.01<z<0.089$ (volume sample, with 6,150 galaxies).
\end{itemize}

Plots that only require a mass selection are defined by the mass sample. Any plot the makes use of pair or group information is defined by the second volume limited selection. By using this combination of stellar mass, group and pair selections we are able to investigate all of the key characteristics that define the MW and the LG environment.

\section{Visual Inspection of the Data}
\label{sec:eyeball}

In this work we are primarily interested in how the presence of a group environment, or another nearby galaxy, can affect the characteristics of a given stellar mass galaxy; particularly in the regime of MW mass galaxies (i.e.\ $L^*$ galaxies).

To undertake such analysis it is necessary to know various characteristics about $L^*$ mass galaxies (i.e.\ the knee of the stellar mass function, $\mathcal{M}^*$). Some of these characteristics can be determined directly from the data: colour, mass. Others are more difficult to extract from the data due to the subjective nature of the variable or the imperfect quality of the data.

Galaxy type, by which we mean morphologically early-type or late-type, is difficult to determine from the catalogue data directly. When analysing a fiducial sample of galaxies taken from \citet{kelv12} the relationship between S\'{e}rsic index and morphological type is highly imperfect: it is not true to say that high S\'{e}rsic index ($n>3$) galaxies are early-type and low S\'{e}rsic index ($n<3$) galaxies are late-type. Inclination in particular can distort catalogue values, and the reliability of colour as a galaxy `type' discriminant is heavily compromised by the presence of dust.

Star formation rate (or specific star formation rate) appears to be a characteristic that is well suited to catalogued data. However, analysis of a fiducial sample revealed that the current emission line based SFRs (Hopkins et~al. in prep) fail to detect SFRs for $\sim$20\% of galaxies where the emission line is very weak and the spectrum has low $S/N$. The lack of a measurement is not an error, rather a pragmatic decision to improve the reliability of quoted SFRs by ignoring marginal lines. A particular problem is that these undetected weakly star forming galaxies are typically fainter lower mass systems, creating a stellar mass bias in our analysis of star formation. Also, at present GAMA incorporates spectra from non-AAOmega sources (e.g.\ SDSS, 2dFGRS) and emission line analysis has not yet been run on the entire sample. This has a larger effect at lower redshifts, which is the regime of interest in this work, since a larger fraction of spectra are non-AAOmega for brighter magnitudes. Ultimately, independently identifying spectral emission features visually was deemed to be an appropriate route forward for the science investigated in this paper, where we are mostly interested in the binary divide of star formation: on or off. This route of investigation is informed by recent studies \citep[e.g.][]{wije12} showing the star forming fraction, not the rate of star formation, is the property sensitive to `environment'. { This requires us to push the usable $S/N$ limits of the catalogues, which is discussed in more detail below.}

The presence, or otherwise, of a disk is also a key characteristic that is difficult to extract from catalogued data. Using the $data-model$ residual images produced by \citet{kelv12} it was clear that a large number of early-type galaxies, even with high S\'{e}rsic indices, possess weak but distinct disks. Whilst late-type galaxies, by their definition, always possess a disk like structure, there exists a large fraction of early-type galaxies where this might be a key distinguishing variable (we include S0 and lenticulars in this class).

To determine classes for these three categories a large subset of the GAMA-I survey was studied in detail by ASGR. Using the single object viewer (SOV) tool developed by JL it was possible to view the following information simultaneously for each galaxy:

\begin{itemize}

\item colour image;
\item basic catalogue information;
\item preferred spectrum;
\item 2D modelling information, including difference image;
\item 1D surface brightness profile.

\end{itemize}

\begin{figure}
\centerline{\mbox{\includegraphics[width=3.7in]{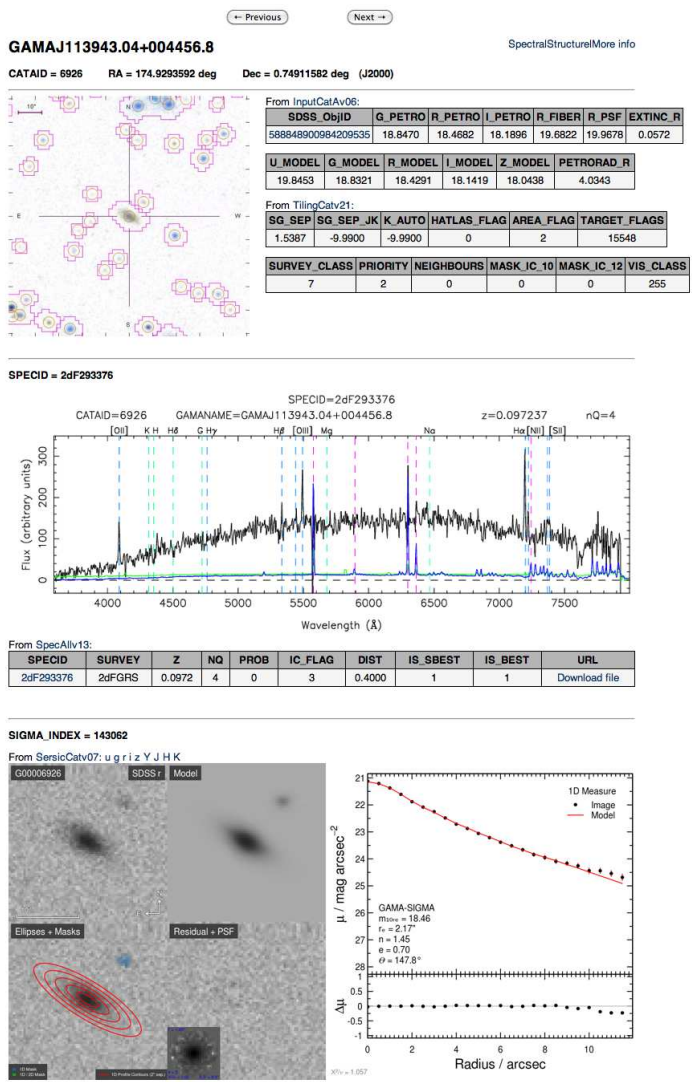}}}
\caption{\small Screen grab of a typical display using the SOV tool available for the GAMA database. Top-left is an image taken directly from the SDSS image server. Top-right are some key catalogue values for the current galaxy. In the middle is the preferred GAMA spectrum with positions of emission and absorption lines over-plotted. Directly below the spectrum is a list of other available spectra in the GAMA database (useful when there are data quality issues in key regions of the preferred spectrum). In this example the only available spectrum is from 2dFGRS. Bottom-left is the image output of \citet{kelv12}. Within this panel the top-left is the $u$ to $K$-band image ($r$-band is the default, but others can be selected by clicking on the links above), top-right is the final Galfit model, bottom-left shows the ellipses used to construct the 1D surface brightness profile, bottom-right is the $data-model$ residual image with the local PSF overplotted. The bottom-right of the whole figure is the 1D surface brightness profile estimated using IRAF Ellipse, with key values regarding the fit printed inside the plot window.}
\label{fig:SOV}
\end{figure}

Figure \ref{fig:SOV} is a screen grab of the SOV tool for GAMA galaxy 6926. Using the information available in the SOV ASGR assessed the following 3 galaxy characteristics:

\begin{itemize}

\item whether the galaxy is late-type (visibly disk dominated) or early-type (visibly bulge dominated);
\item whether the galaxy has any evidence for a disk (binary true or false);
\item whether the galaxy exhibits star forming H$\alpha$ emission, no H$\alpha$ emission or is an obvious AGN (broad emission features).

\end{itemize}

\begin{figure}
  \center{\mbox{\includegraphics[width=3.5in]{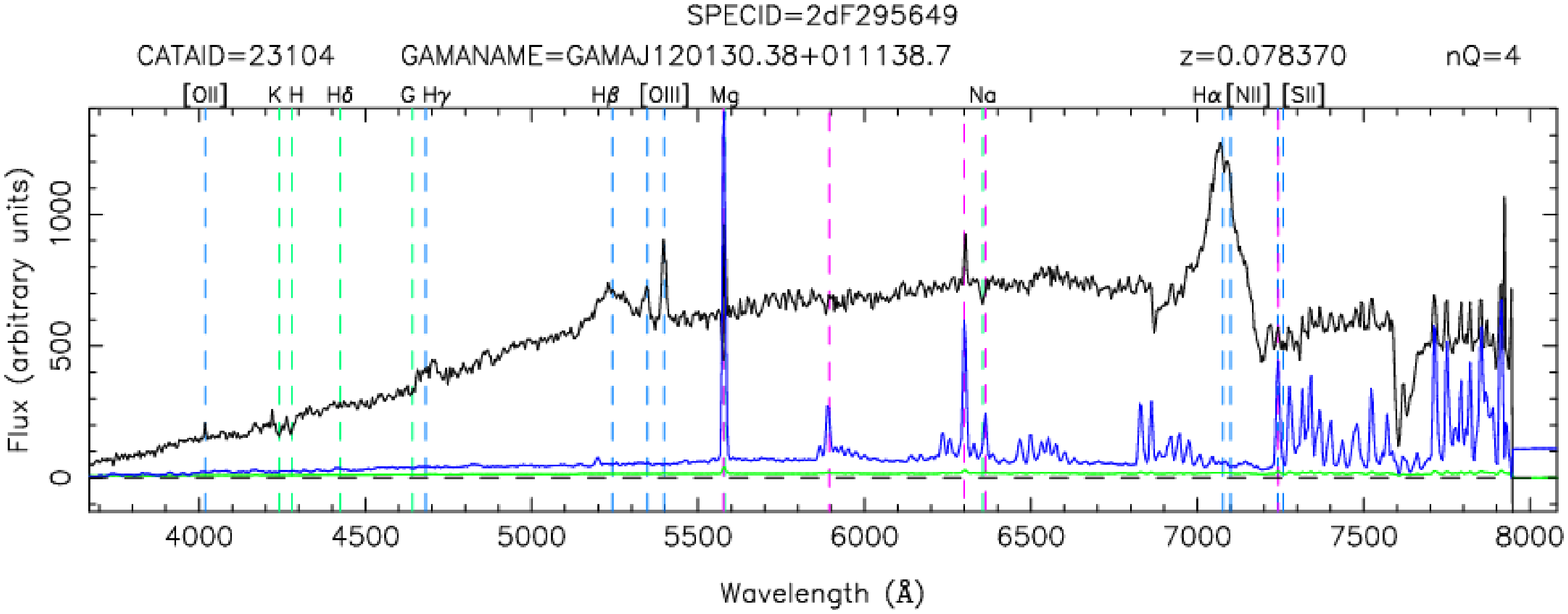}}}
  \center{\mbox{\includegraphics[width=3.5in]{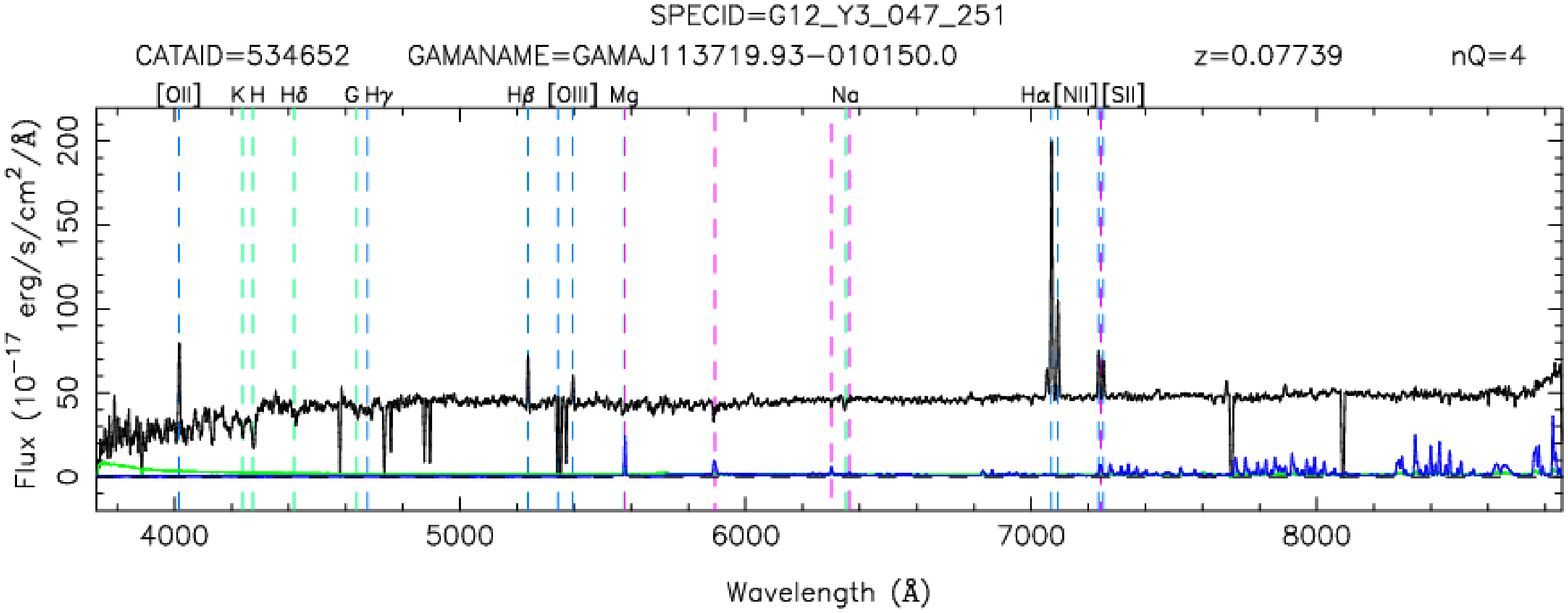}}}
  \center{\mbox{\includegraphics[width=3.5in]{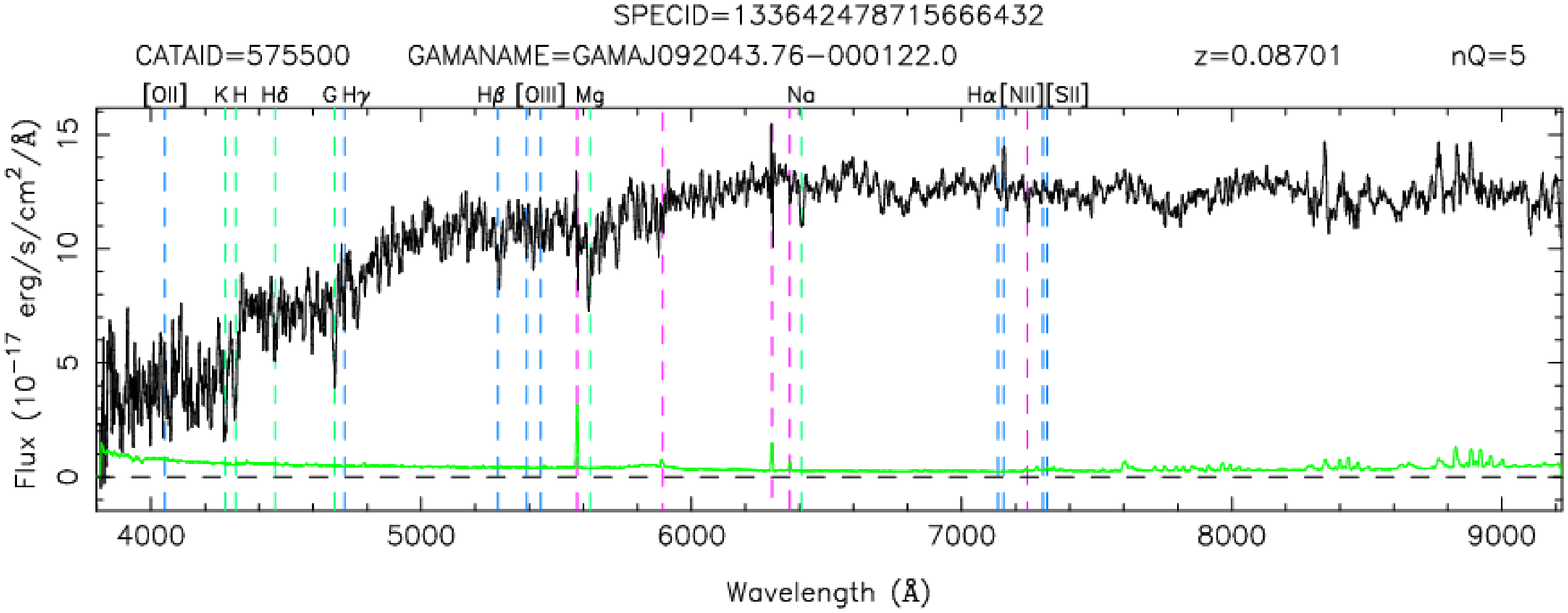}}}
   \caption{ \small Examples of the three emission classes used for visual classification. Top panel shows broad line features due to high velocity emission from an AGN (class A). Middle panel shows high equivalent width H$\alpha$ emission relative to all other emission lines, so a star forming spectrum (class S). Bottom panel shows no signs of any H$\alpha$ emission on top of the continuum (X).}
\label{fig:emtype}
\end{figure}

The three classifications used to quantify emission class are shown in Figure \ref{fig:emtype}. As well as assessing galaxies for these different classes, data quality was simultaneously checked. Galaxies were assessed as to whether they were sub-structure of larger galaxies or misclassified stars that have a redshift virtue of background galaxy flux. 16 galaxies were re-classified as substructure and 6 galaxies were re-classified as stars, i.e.\ 99.64\% of galaxies in our original sample were correctly identified as being galaxies \citep[see][for details of the original GAMA star galaxy separation]{bald10}.

Finer resolution grading of properties was not possible over the redshift range probed, i.e.\ it was not possible to reliably identify the Hubble class or similar lower order classifications. The properties assessed were subjectively considered to be the most information that could be robustly extracted given the quality of data. To guide later discussions, lenticular galaxies and early-type spiral galaxies would both be classified as early-type galaxies possessing disks.

\section{Galaxy Categorisation}

\label{sec:galcat}

For multi-dependent data mosaic plots are a very powerful and information rich means of visualisation. They differ from principal component analysis (PCA) in that they readily handle discrete data classes, which PCA does not. For readers unfamiliar with how mosaic plots are constructed, and best interpreted, Appendix A describes them in detail. It is strongly recommended that novices to the mosaic plotting technique start there. To ease the comprehension of the large number of plots presented, the major observations drawn from each figure are included in the relevant captions. This should be particularly helpful in explaining the mosaic plots to those unfamiliar with them.

In order to present information in the form of mosaic plots all information must be discretised. Some categories are naturally discrete: morphological type and presence of disk. Others have to be coerced into discrete categories: SFR (as outlined above). Here we describe all of the galaxy categorisations that are used to construct later mosaic plots, along with the abbreviation in parentheses used for labelling.

\subsection{Naturally Discrete Galaxy Categories}

Below are listed all of the naturally discrete classes of galaxy properties investigated. The determination of these divisions require less subjectivity than for continuous data. However, not all categories are fully objective--- opinion is still required to determine whether a galaxy is early-type or late-type.
 
\subsubsection{Galaxy Type}
\label{sec:galtype}

Early-Type (e) / Late-Type (l)
\newline\newline
As part of the visual classification process outlined in Section \ref{sec:eyeball}, galaxies are categorised by simple Hubble type. The divide used was whether the galaxy appeared to be bulge dominated ($R_e$ inside the bulge regime, { which is determined by a strong visual break in the 1D light profile}) or disk dominated (either no bulge, or $R_e$ outside the bulge regime). This estimation is informed by both multi-band and $r$-band SDSS images, and the 1D profile of 2D single S\'{e}rsic fits generated by LSK \citep[see][for details on the single S\'{e}rsic fitting process]{kelv12}. Further resolution beyond early-type and late-type is not attempted due to the quality of the data. Lenticulars (S0) galaxies would be classified as early-type in this scheme, as would early-type spiral galaxies.

\subsubsection{Galaxy Disk}
\label{sec:galdisk}

False (F) / True (T)
\newline\newline
As part of the visual classification process outlined in Section \ref{sec:eyeball}, galaxies are assessed to either possess some form of disk, or not. This subjective observation is based on the $r$-band S\'{e}rsic profile subtracted residual images created by LSK \citep[see][for details on the single S\'{e}rsic fitting process]{kelv12}. Further detail beyond this binary status is not attempted due to the quality of the data. Lenticulars (S0) galaxies would be classified as possessing a disk in this scheme.

\subsubsection{Galaxy in Group}
\label{sec:galgrouped}

False (F) / True (T)
\newline\newline
Logic class defining whether a galaxy is found within a \G3C group that makes the selection criteria outlined in Sec \ref{sec:selection}.

\subsubsection{Galaxy is Central or Satellite}
\label{sec:galbgg}

Central Galaxy (C) / Satellite Galaxy (S)
\newline\newline
Class defining whether a galaxy is the largest (in terms of stellar mass) in its group, otherwise referred to in this work as the biggest group galaxy (BGG). If it is then it is defined as the `central' galaxy (C), otherwise it is a `satellite' (S). All ungrouped galaxies are defined as central galaxies in their group of multiplicity 1 (but these 1 member groups do not count as part of the \G3C).

\subsubsection{Galaxy in Pair System}
\label{sec:galpaired}

False (F) / True (T)
\newline\newline
Logic class defining whether a galaxy found within a \G3C pair system that makes the selection criteria outlined in Sec \ref{sec:selection}.

\subsubsection{Galaxy is Largest in Pair System}
\label{sec:galbpg}

False (F) / True (T)
\newline\newline
Logic class defining whether a galaxy is the largest (in terms of stellar mass) in its pair system.

\subsection{Naturally Continuous Galaxy Categories}

Below are listed all of the naturally continuous classes of galaxy properties investigated. The determination of these divisions contain some subjectivity on our part. Making continuous data discrete obviously reduces the information content available to us, but in practice it is often the only way to visualise behaviour of noisy relationships: it is common practice to plot the running median in scatter plots, which is a discrete separation of the data. Even binning data when plotting a luminosity function is a discretising process. Obviously any further numerical analysis should make use of the underlying continuous data.

\subsubsection{Emission Strength}
\label{sec:galem}

AGN (A) / Star Forming (S) / No Emission (X)
\newline\newline
As part of the visual classification process outlined in Section \ref{sec:eyeball} galaxies are visually assessed into these three classes of emission. AGN classes (A in plots) are those with clear H$\alpha$ emission, but where the spectral signature is dominated by AGN irradiation. Star forming classes  (S in plots) are those with clear H$\alpha$ emission, but where the spectral signature is dominated by star formation irradiation. No Emission classes (X in plots) are those without any H$\alpha$ emission, though other lines may still be present (typically weak [NII]).

Figure \ref{fig:emtype} gives an example of each. Galaxies were both spectrally classified through a classic BPT separation \citep{bald81} and by eye. The GAMA application of the BPT separation is described in detail in Gunawardhana et~al. 2013 (in prep) and \citet{hopk13}, and includes all GAMA and SDSS origin Spectra (which covers 92.2\% of the sample investigated here). In the automatic classification scheme a galaxy is classed as `X' when it cannot be placed on a BPT, i.e.\ it is missing at least one of the H$\alpha$, H$\beta$, [OIII] or [NII] lines. This differs to the definition desired in this work, that `X' refers only to systems with no H$\alpha$ emission. By using this definition the separation between classes is much clearer.

Separately to this purely automatic process, all spectra were classified by eye. It is easy to identify AGN spectra when there are broad emission line features, but these are a minority of cases. In other cases the decision was made by dividing the height of the [NII] and H$\alpha$ features, and classifying the galaxy as AGN when this ratio was clearly above 0.5. This uses part of the information available in a formal BPT analysis, and will be approximate at best. As such, a clear weakness of visually classifying spectra is separating marginal BPT classifications. All disagreements were followed up at least once and a discernment made between cases where the automatic process was distinguishing between subtle emission ratios that cannot be accurately determined by eye, and those where an fitting error has occurred, creating a poor automated fit.

In cases where the BPT was able to select marginal AGN systems and H$\alpha$, H$\beta$, [NII] and [OIII] were all in emission, the BPT was preferred. However, the majority of BPT selected AGN classifications were tenuous low powered LINER-like systems with no significant H$\alpha$, H$\beta$ or [OIII] flux. In these situations only a weak [NII] line was present (as in the bottom panel of Figure \ref{fig:emtype}), and the classification was changed to `X', since there is no strong H$\alpha$ flux. This is the main source of disagreement revealed during visual classification of spectra. It is important to change these classifications since a significant number of AGN are, by the definitions we wish to use, erroneous.

\begin{figure}
\centerline{\mbox{\includegraphics[width=3.7in]{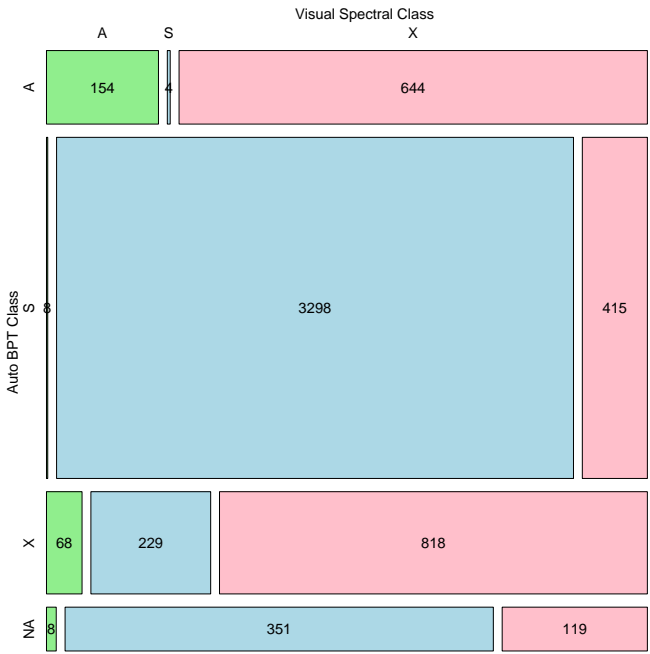}}}
\caption{\small Mosaic plot comparing the automated BPT classifications (horizontal bands) versus the visual classifications (vertical coloured bands). For both classifications A=AGN, S= star forming and X= non star forming. For the visual classifications, green= AGN, blue= star forming and light red= non star forming. The most disagreement comes for AGN classes (A) where very low powered LINER-like systems with no visible H$\alpha$ are being classed as AGN. Since we are most interested in H$\alpha$ emission, and only wish to exclude cases where a large component of the H$\alpha$ flux is due to AGN activity, these objects are mostly moved to the no emission class (X). Spectra taken from surveys other than GAMA and SDSS do not have an auto classification (shown as NA). For a detailed description of how to interpret mosaic plots the reader should refer to the Appendix.
\newline
{\bf Observations:}
\newline
- Strong agreement for `S' class emissions.
\newline
- Large fraction of auto AGN classes have no H$\alpha$ emission.
}
\label{fig:BPTvis}
\end{figure}

Figure \ref{fig:BPTvis} summarises the BPT versus visual classification. Overall 75.4\% of automated BPT classifications agree with the visual classifications. For AGNs the figure drops to 19\% for the reasons described above (only exceptionally obvious AGN can be reliably picked out by eye). For star forming systems the agreement is high, at 88\%. Finally, the non emission classes agree at the 73\% level. In addition, a non-negligible (7.8\%) fraction of sources do not have SDSS or GAMA spectra, therefore they do not have a BPT classification in the GAMA database (see bottom panels of Figure \ref{fig:BPTvis}). These systems had to be classed entirely by eye, but broadly reflect the overall classification distributions seen for the full sample. The main survey source for these galaxies without an automated BPT classification is 2dFGRS.

\begin{figure}
\centerline{\mbox{\includegraphics[width=3.7in]{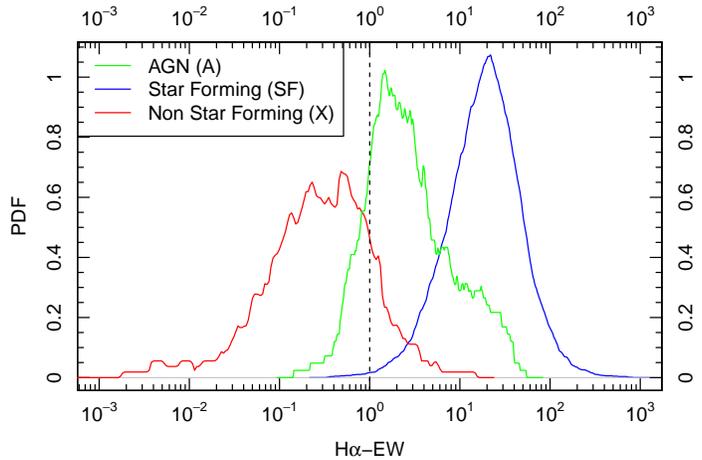}}}
\caption{\small Comparison of the H$\alpha$-EW  PDFs for the AGN (A), star forming (S) and non star forming (X) classifications used in this work, where the H$\alpha$ emission has been measured during the GAMA line measuring process \citep[see][for details]{hopk13}. As expected, where H$\alpha$ has been measured the EW is significantly higher for star forming compared to non star forming galaxies, with AGN classed galaxies sitting broadly between the two. To aid comparison with other work we suggest dividing the H$\alpha$-EW at 1 (vertical dashed black line) and classing galaxies above this threshold as AGN/ star forming and classing galaxies below this threshold as non star forming.}
\label{fig:HAEW}
\end{figure}

{ Figure \ref{fig:HAEW} demonstrates the H$\alpha$-EW distributions for the final classifications used in this work. Star forming galaxies have a typical H$\alpha$-EW at least a factor 10 and on average a factor 100 larger than the visually identified non star forming galaxies. The majority of non star forming galaxies do not have any measured H$\alpha$ flux (83.8\%). This plot shows the PDFs for the small fraction that have some flux detected during the automated line measuring conducted for GAMA \citep[see][for details]{hopk13}. An approximate version of our classifications can be made using H$\alpha$-EW alone: 97.6\% of galaxies that have an H$\alpha$-EW above 1 are either AGN or star forming, and 93.3\% of galaxies with H$\alpha$-EW below 1 (including non measurable) are non star forming. Using H$\alpha$-EW alone does not allow for a clean separation of AGN and star forming galaxies (hence the popularity of BPT diagrams and similar schemes), but these divisions will allow some of the more dramatic results to be compared to other work.}

The AGN classifications contain within them complex subclasses that we do not attempt to resolve in this work: i.e.\ QSO, Seyfert type 1 and 2, and strong LINER galaxies. Since the physical origin of these different classes are quite different and some of the variation is due to observational bias (e.g.\ distinction between Seyfert classes is largely observational orientation), we will focus this work on the star formation (S) and no emission (X) classes. However, the AGN classes will be listed in all mosaics in order to accurately represent the full population fractions. In our final categorisations, 3.9\% of galaxies are AGN with H$\alpha$ emission, 63.1\% have star forming H$\alpha$ emission, and 32.5\% have no H$\alpha$ emission. The final 0.5\% of objects are not real galaxies, either stars or substructure of extended galaxies.

\subsubsection{Colour}
\label{sec:galcol}

Blue (b) / Green (g) / Red (r)
\newline\newline
Colour is an important characteristic of a galaxy. Here we treat colour as a purely relative quantity, { and we use $g-r$ because we have detections in both bands for all galaxies. It is known that the bimodal nature of galaxy colour is enhanced by using a bluer band than $g$ since this only partly samples blue-wards of the 4000 \AA\ break \citep{bald04}, but since colour does not form the main focus of future analysis, completeness was preferred. Also, since we divide colour into relative quantiles, the lack of $g-r$ bimodality should not dramatically alter later plots since bluer galaxies do tend to be bluer in both $X-r$ and $g-r$, where $X$ is any band bluer than $g$.} The Galactic dust corrected and k-corrected \citep[using k-correct and GAMA's $ugriz$YJHK photometry,][respectively]{blan07,hill11} $g-r$ colour is separated into three equal sized quantiles of data for galaxies within 0.15 dex of \MMW. The limits of the quantiles are used to define galaxies as blue, green or red in a relative sense:

\begin{eqnarray*}
g-r<0.77		&=&		{\rm Blue (b)}	\\
0.77<g-r<0.86	&=&		{\rm Green (g)}	\\
g-r>0.86		&=&		{\rm Red (r)}	\\
\end{eqnarray*}

\begin{figure}
  \center{\mbox{\includegraphics[width=3.5in]{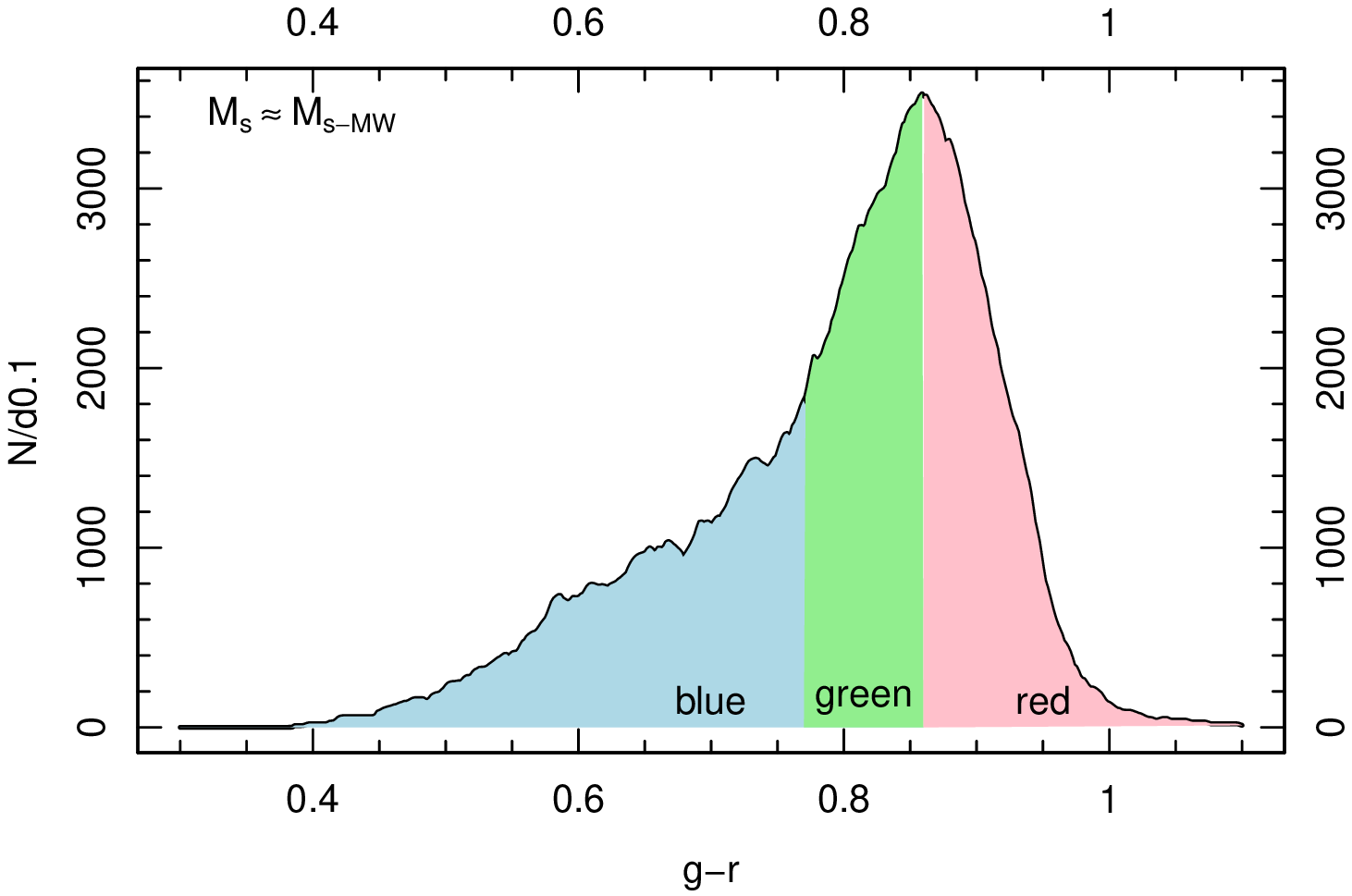}}}
   \caption{ \small The distribution of $g-r$ colours for galaxies with stellar masses within 0.15 dex of \MMW. The blue/ green/ red divisions used are shown by the different coloured shading. Selection shown is the mass sample (Sec \ref{sec:selection}).
\newline
{\bf Observations:}
\newline
- There is no clear bimodality in the $g-r$ colour for the MW stellar mass selection used to define the colour quantiles.
\newline
- The blue population selects the majority of the outlying wing in the $g-r$ distribution, whilst the green and red population sample different sides of the main population peaks. This suggests that the green and red selections are likely to be selecting galaxies driven by similar physical processes.
}
\label{fig:gminr}
\end{figure}

Figure \ref{fig:gminr} shows the distribution of $g-r$ colours for the mass selection used, and highlights the divisions used. There is no clear bimodality for this mass selection and colour combination, although the blue galaxies exist in a distinct wing. The green and red galaxies are largely selecting different sides of the dominant colour distribution peak.

\subsubsection{Stellar Mass}
\label{sec:galstelmass}

\begin{equation}
X \times \mathcal{M}_{S,MW}
\end{equation}

\noindent We split GAMA galaxy stellar masses \citep[taken from][]{tayl11} into bins separated by factors of two, and each bin therefore contains a mass range of $\pm 0.15$ dex. $\mathcal{M}_{S,MW}$ is the reference MW stellar mass. Since we are complete down to $\mathcal{M_{S}}=10^{10} \msol$ we define the bin centre of the $X=1$ bin containing the MW as $5.67\times10^{10} \msol$ \citep[so it includes stellar mass of the MW, which we take to be $\mathcal{M}_{S,MW}=5.0 \times 10^{10} \msol$;][]{flyn06}, meaning we have a complete sample with 2.5 bin widths on the low mass side exactly. The stellar mass range selected is 100\% complete for galaxies where $X \ge \frac{1}{4}$.

For reference the MW is in the $X=1$ bin \citep[$\mathcal{M}_{S,MW}=5.0 \times 10^{10} \msol$;][]{flyn06}; the LMC is in the $X=\frac{1}{32}$ bin \citep[$\mathcal{M}_{S,LMC}=2.3 \times 10^{9} \msol$;][]{jame11}; the SMC is in the $X=\frac{1}{128}$ bin \citep[$\mathcal{M}_{S,MW}=5.3 \times 10^{8} \msol$;][]{jame11}; M31 is in the $X=2$ bin \citep[$\mathcal{M}_{S,M31}=1.0 \times 10^{11} \msol$;][]{hamm07} and M33 is in the $X=\frac{1}{8}$ \citep[$\mathcal{M}_{S,M33}=5.7 \times 10^{9} \msol$;][]{verl09}.
\subsubsection{Halo Mass}
\label{sec:galhalomass}

\begin{equation}
Y \times M_{H,MW}
\end{equation}

\noindent The discrete halo mass bins are defined in a similar vein to the stellar mass bins: adjacent bins vary in mass by a factor two, and each bin therefore contains a mass range of $\pm 0.15$ dex. Halo masses are taken from \citet{robo11}, and are the variably scaled (with redshift and group multiplicity) dynamical masses that are discussed in that work. $M_{H,MW}$ uses the halo mass of the MW calculated in \citet{li08}: $M_{H,MW}=2.5 \times 10^{12} \msol$. The $Y=1$ bin corresponds to haloes with MW and M31 masses. The $Y=2$ bin corresponds to groups that have the same halo mass as the whole LG system ($5.0 \times 10^{12} \msol$). Halo masses are 100\% complete for groups where $\frac{1}{2} \le Y \le 2$.

\section{Galaxy Property Dependencies}
\label{sec:galdep}

\subsection{The Imperfect Relationship Between Colour, Emission, Type \& Disk}

Having established a set of contingencies that describe galaxies in a discrete manner, we can now construct mosaic plots that reveal the most significant dependencies. Combining information in this manner will give us an insight into what drives galaxy formation in galaxies with a stellar mass content similar to the MW. We can also determine which galaxy parameters are degenerate.

Galaxy populations are known to be bimodal in a number of properties \citep[see][and Taylor et~al. in prep]{blan03,bald04,bald06,driv06}. The main ones that we can investigate here are colour, disk presence, morphological type and line emission. A simplistic description of extra-galactic astronomy might bracket galaxies into two main classes (or dependencies): blue, disky, star forming, late-type galaxies and red, diskless, passive, early-type galaxies.

\begin{figure}
\centerline{\mbox{\includegraphics[width=3.7in]{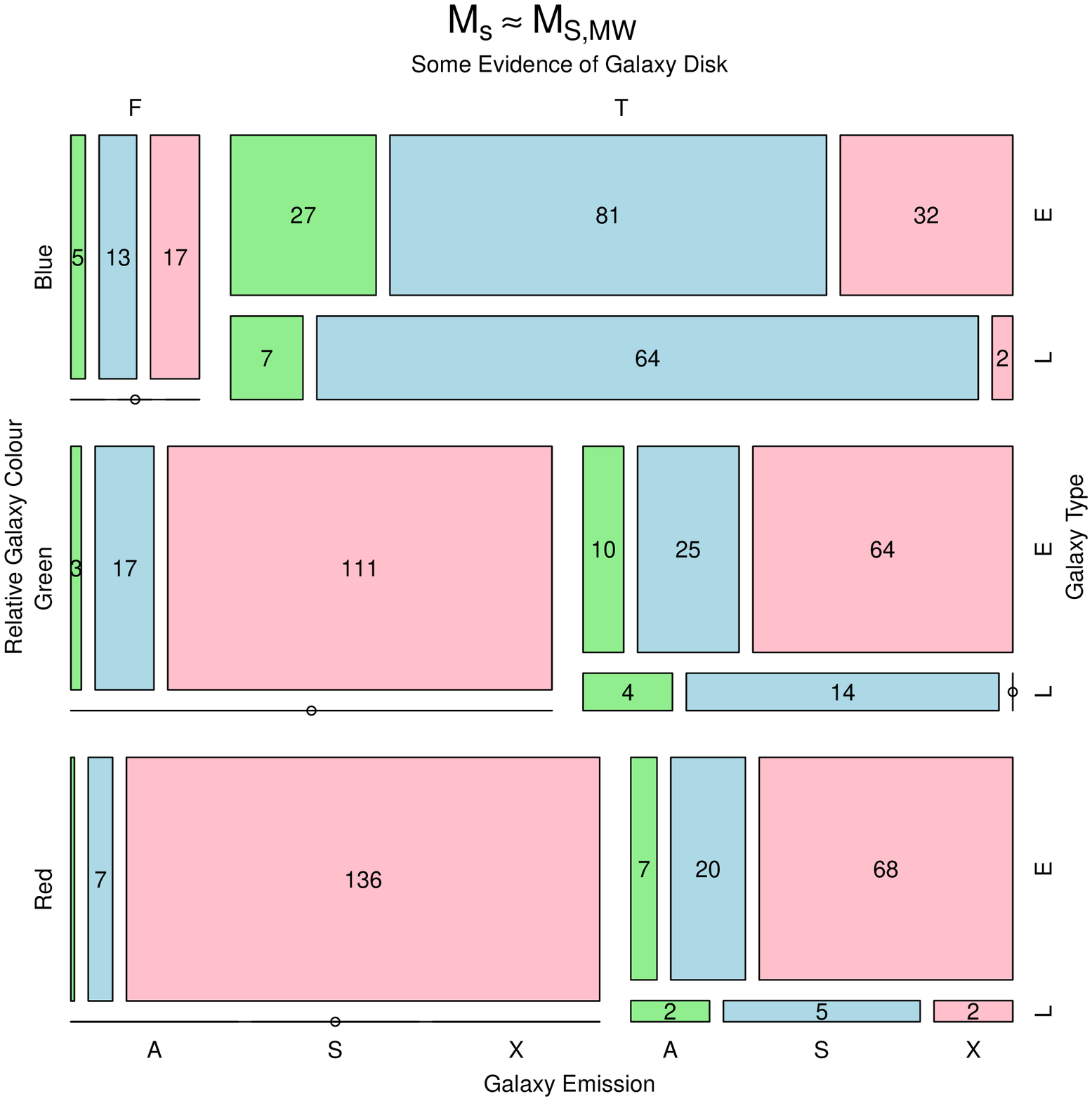}}}
\caption{\small Mosaic plot of Colour (Sec \ref{sec:galcol})/ Disk Presence (Sec \ref{sec:galdisk})/ Type (Sec \ref{sec:galtype})/ Emission (Sec \ref{sec:galem}). Subset used for this figure is $\mathcal{M}_{s} = \mathcal{M}_{S,MW}$ (Sec \ref{sec:galstelmass}). Selection shown is the mass sample (Sec \ref{sec:selection}).
\newline
{\bf Observations:}
\newline
- Star formation predicts galaxy is blue and has a disk.
\newline
- No H$\alpha$ emission predicts the galaxy is early-type.
}
\label{fig:CoDiTyEmLs0}
\end{figure}

Figure \ref{fig:CoDiTyEmLs0} shows the mosaic for Colour/ Disk Presence/ Type/ Emission Class for galaxies where $\mathcal{M}_{s} \sim \mathcal{M}_{S,MW}$ (within $\pm$0.15 dex of the MW stellar mass). Whilst this mosaic plot contains a narrow range of galaxy stellar mass, it is immediately clear that there are no contingencies that unambiguously predict each other. For example, in this mass range almost as large a fraction of green galaxies are non-star forming as red galaxies. This method of presenting bimodality is in contrast to more standard 2D scatter relations, with a $3^{rd}$ dimension simulated through successive cuts in the data \citep[e.g.][]{blan03}. With good clarity we are able to describe codependencies in 4 dimensions, whilst maintaining the relative counts of objects throughout in a graphically intuitive way by representing the number density with the size of the box. This latter point is important because this information is often lost in scatter plot contours. A final consideration is that we are able to display discrete and continuous characteristics on the same set of axes in a consistent manner.

Despite the complex dependencies we see in Figure \ref{fig:CoDiTyEmLs0}, there are two strong predictors in the data. Star forming emission predicts the galaxy being blue and having a disk (but not its morphological type). Also, no-emission predicts the galaxy being early-type (but not the presence of a disk or colour, beyond being not blue). The reverse of these statements are much weaker predictors: blue galaxies with a disk are most likely to have star forming emission, and early-type galaxies display a large mixture of emission classes, where 58\% have no emission. Colour is a strong predictor of disk presence (blue galaxies have a strong bias towards disks) but a much weaker predictor of morphological type. This is mostly because early-type galaxies dominate for all colours. In general it is not accurate to assume any galaxy property inevitably correlates with another.

AGN do not dominate any contingency, but we find they are most common for blue, early-type galaxies with evidence of a disk. Colour is a good AGN discriminator--- they are much more likely ($\sim4:1$) in blue galaxies than red. Also, AGNs are 5 times more common in early-type galaxies compared to late-type galaxies. The strongest predictor though is disk presence. AGN are 6 times more likely to reside in an $L^*$ galaxy with a disk than one without. They are effectively non-existent in red, diskless, early-type galaxies, so truly `dead' systems will also be devoid of AGN H$\alpha$ activity in the stellar mass range considered.

In the following sections we will have to be careful to extract the most significant of these 4 parameters when creating 4-way mosaics that contain other variables. The most interesting (predictive or predicted) of these 4 contingencies vary case-by-case, and obviously depend on the physics being assessed and the timescales being considered.

\newpage

\subsection{The group environment of $L^*$ Galaxies}
\label{sec:groupmosaic}

\begin{figure*}
\centerline{\mbox{\includegraphics[width=3.7in]{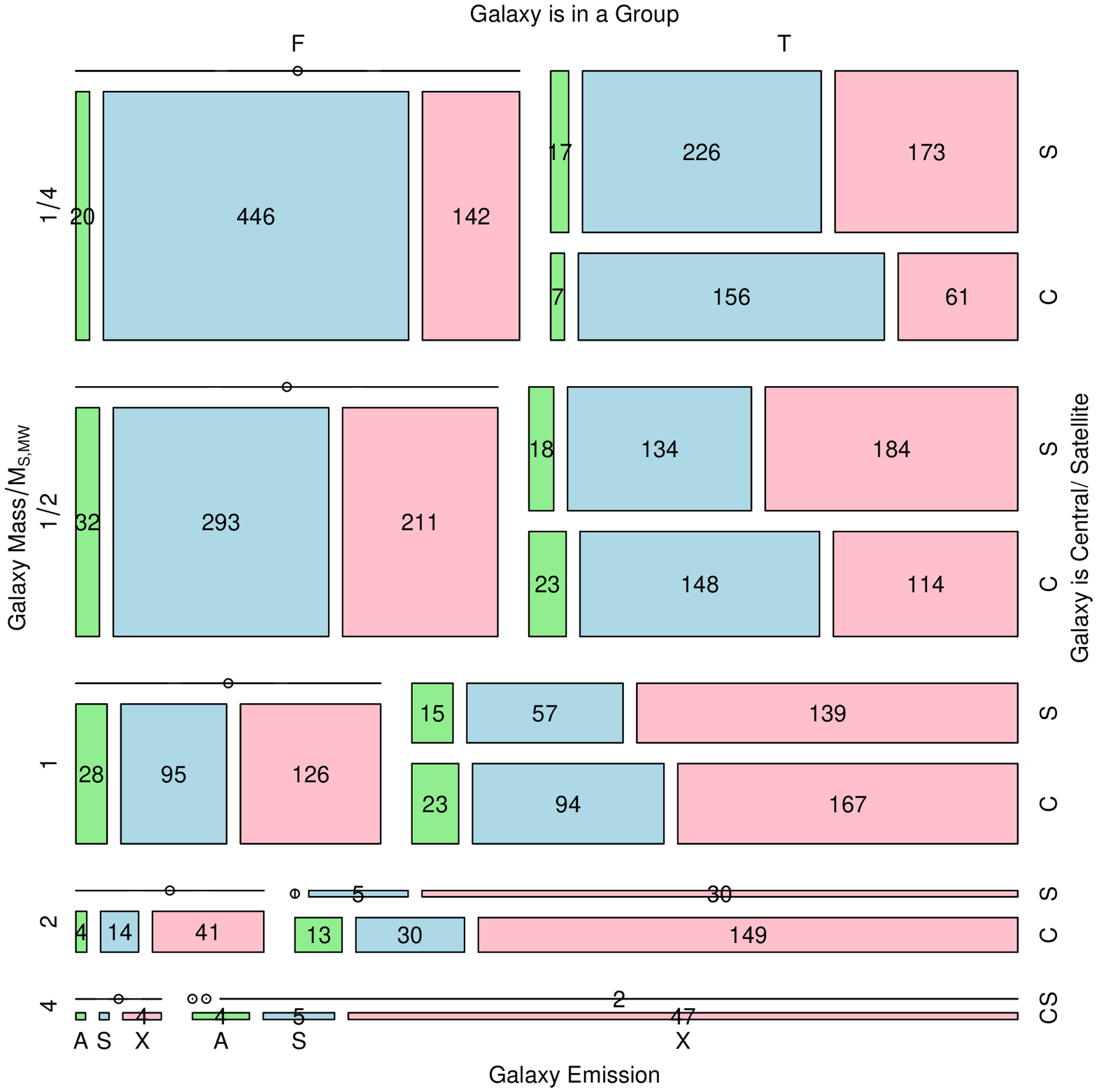}}\mbox{\includegraphics[width=3.7in]{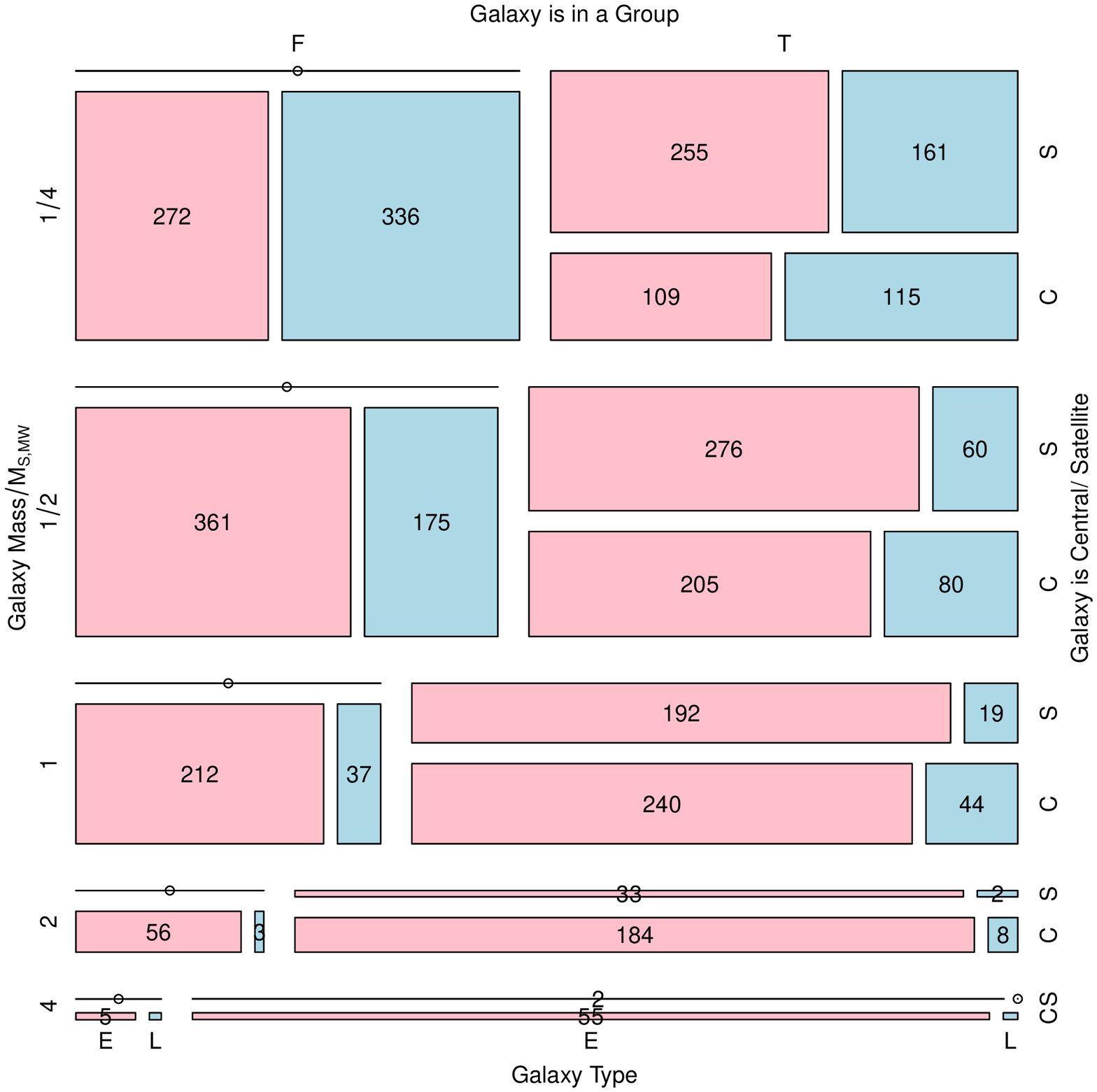}}}
\caption{\small Left Panel is a mosaic plot of Stellar mass (Sec \ref{sec:galstelmass})/ Galaxy in group (Sec \ref{sec:galgrouped})/ Galaxy is Central or Satellite (Sec \ref{sec:galbgg})/ Emission (Sec \ref{sec:galem}). Right Panel is a mosaic plot of Stellar mass (Sec \ref{sec:galstelmass})/ Galaxy in group (Sec \ref{sec:galgrouped})/ Galaxy is Central or Satellite (Sec \ref{sec:galbgg})/ Type (Sec \ref{sec:galtype}). Ungrouped galaxies are referred to us `Central' since they are the largest galaxy in their multiplicity one group. Selection shown is the volume sample (see Sec \ref{sec:selection}).
\newline
{\bf Observations (left):}
\newline
- Larger stellar mass galaxies are more likely to be in a group.
\newline
- The fraction of star forming satellite galaxies is less than observed for central galaxies of the same stellar mass.
\newline
- Central galaxies have similar emission fractions to ungrouped galaxies, which are in reality extremely stellar mass dominant central galaxies in their halos.
\newline
{\bf Observations (right):}
\newline
- The fraction of late-type satellite galaxies is less than observed for central galaxies of the same stellar mass.
\newline
- The fraction of late-type galaxies decreases more rapidly with stellar mass than the fraction of star forming galaxies.
}
\label{fig:LsingroupsLineTy}
\end{figure*}

The group environment plays a key role in the evolution of galaxies. In this section we investigate the effect that being in a group has on galaxies of a given stellar mass.

Figure \ref{fig:LsingroupsLineTy} shows two mosaic plots that only vary on the final contingency. The first three are stellar mass (down to our stellar mass limit of $\mathcal{M}_{s} = \mathcal{M}_{S,MW}/4$), whether the galaxy is grouped and whether the galaxy is a central or satellite. The last contingency shown is emission class for the left panel and morphological type for the right panel. In these plots we are able to observe down to LMC luminosity galaxies, so the term `ungrouped' implies there are no near-by galaxies which have an $r$-band flux brighter than the LMC.

Some simple conclusions can be drawn immediately. Larger stellar mass galaxies are more likely to be grouped, and also more likely to be a central galaxy if they are in a group. Also, a galaxy is more likely to possess no star formation if its stellar mass is larger regardless of whether it is grouped.

More interesting is the clear role the group environment has on the emission class and morphological type of galaxies. Within grouped galaxies, and within the same stellar mass interval, there is a clear indication that satellite galaxies have a larger no-emission fraction than central galaxies. It should be emphasised that here we are comparing galaxies with the {\it same} intrinsic stellar mass, so we are not affected by biases regarding what varieties of galaxies tend to be grouped etc.

Central galaxies in groups have very similar emission class fractions to similar mass ungrouped galaxies, and satellite galaxies have significantly lower star forming fractions. The morphological type fractions remain very similar for all masses between ungrouped galaxies and central galaxies, however satellite galaxies have a consistently larger early-type fraction. Galaxy morphological type shows much more dramatic stellar mass dependencies than the star formation fraction. For all contingencies, a much smaller fraction of galaxies are late-type than star forming for the mass range investigated, and the drop with stellar mass is much more dramatic: 60\% to 9\% star forming between $\frac{1}{4}\mathcal{M}_{S,MW}$ and $4\mathcal{M}_{S,MW}$, and 45\% to 2\% late-type over the same mass range.

\begin{figure*}
\centerline{\mbox{\includegraphics[width=7.4in]{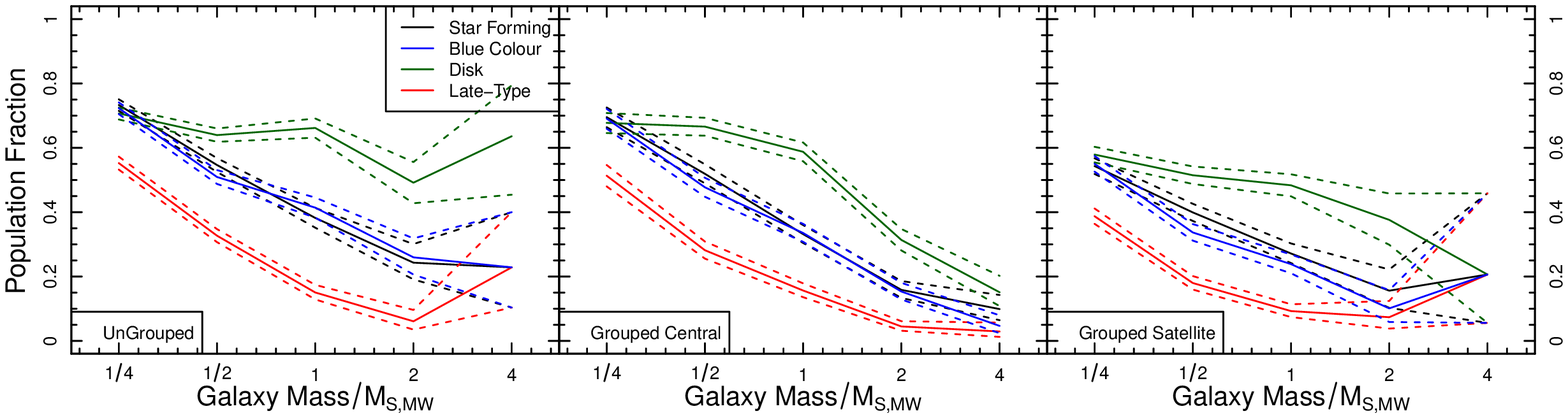}}}
\caption{\small Population fractions for different properties that are commonly associated with late-type populations: being morphologically late-type (by definition), star forming, having a disk and being blue in integrated colour. These are shown for ungrouped (left panel), central (middle) and satellite (right) galaxies. Dotted lines show the formal bayesian error for a bimodal distribution, in this case the chance of galaxies being classed as stated in the legend, or not. Selection shown is the volume sample (see Sec \ref{sec:selection}).
\newline
{\bf Observations:}
\newline
- The disk fraction for grouped galaxies drops rapidly beyond \MMW.
\newline
- Galaxy disks are more likely to be present in ungrouped galaxies for galaxies more massive than \MMW, suggesting they can be easily disrupted by dynamical interactions.
}
\label{fig:PopFracs}
\end{figure*}

Figure \ref{fig:PopFracs}, which shows the raw population fractions as a function of stellar mass. Most fraction gradients trace each other quite closely. The notable exception is the disk fraction. This has a notably flatter dependency with stellar mass, and we see a pronounced break in the population fraction for grouped central galaxies. This population has a disk fraction that stays quite level at $\sim$60\% until the stellar mass is equal to \MMW, after which it drops quite steeply. { A possible interpretation} is that the most massive central galaxies are less able to accrete material (through either minor mergers or disk-like star formation triggered by gas accretion), causing the optical disks to become less common.

In general satellite galaxies experience suppression of all features plotted in Figure \ref{fig:PopFracs}. The ungrouped galaxies look much more like grouped central galaxies. This is consistent with them being central galaxies of groups with undetectably large magnitude gaps, i.e.\ the 2$^{nd}$ rank member is less massive than the LMC (the stellar mass limit of the sample investigated). The main deviation is a significant disk fraction for more massive galaxies compared to the grouped BGGs, a feature consistent with ongoing minor merging. The grouped central galaxies will be more likely to be undergoing major interactions with more massive satellite galaxies, { and such events can suppress evidence of any disk-like structure.}

\begin{figure}
\centerline{\mbox{\includegraphics[width=3.7in]{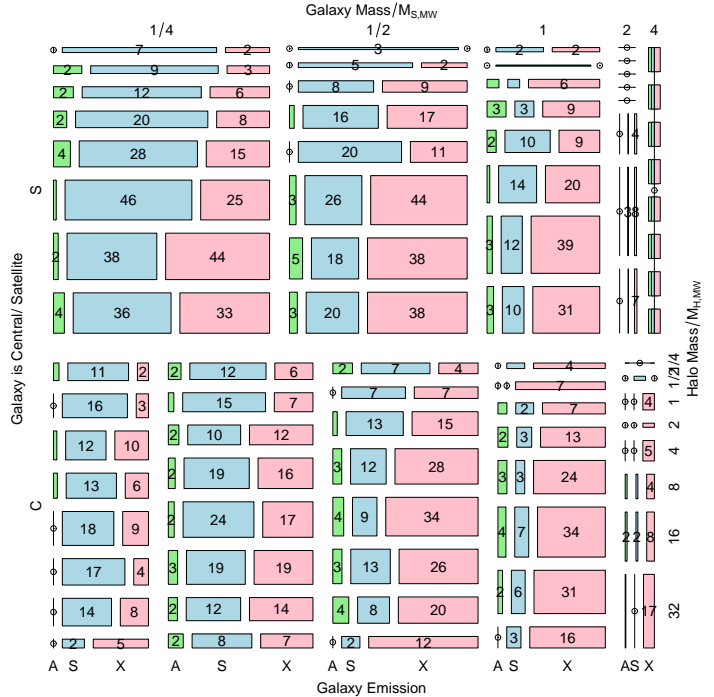}}}
\caption{\small Mosaic plot of Galaxy is Central or Satellite (Sec \ref{sec:galbgg})/ Stellar mass (Sec \ref{sec:galstelmass})/ Halo mass (Sec \ref{sec:galhalomass})/ Emission (Sec \ref{sec:galem}). The subset used for this figure is all grouped galaxies from the selection described in Section \ref{sec:selection}. There are no satellite galaxies with $\mathcal{M}_{s} = 4\mathcal{M}_{S,MW}$, so a line is drawn through these panels. Selection shown is the volume sample (see Sec \ref{sec:selection}).
\newline
{\bf Observations:}
\newline
- Satellites comprise a larger fraction of group stellar mass for more massive halos.
\newline
- The least massive satellites ($\mathcal{M}_{S,MW}/4$) experience a reduction in the star forming fraction as a function of halo mass.
}
\label{fig:GaHmLsLi}
\end{figure}

Figure \ref{fig:GaHmLsLi} allows us to extend the investigation into more detail, adding halo mass relative to the mass of the MW halo as a contingency and removing the grouped logic, i.e.\ only grouped galaxies are shown in the mosaic. We find that halo mass has a very weak effect on the star formation class of the central galaxy, but there is a tendency for more star forming galaxies in less massive haloes. The satellites have a strong variation that depends on halo mass: for the same stellar mass galaxy the star forming fraction is suppressed by a factor $\sim2$ between $M=M_{H,MW}/4$ and $M=32 M_{H,MW}$. The strength of this suppression is strongest for the least massive galaxies in the sample. This is in line with the findings of \citet{groo12}, who found that only the SFR of lower mass (likely to be satellite) galaxies showed a systematic decrease with halo mass. The morphological type is driven in a very similar manner (but the mosaic is not shown here), where the late-type fraction increases in line with the `S' class emission shown in Fig. \ref{fig:GaHmLsLi}. AGN activity shows no statistically significant relationship with any group related property.

Combining this information with the earlier mosaics allows us to build a picture of galaxy evolution drivers for galaxies with similar stellar mass content to the MW. Merely being the central galaxy in a group of any mass provides a clear boost to the star forming fraction of galaxies, but beyond this the halo mass plays no significant role, as shown by the lack of clear systematic effects with halo mass in Fig \ref{fig:GaHmLsLi}.

The smallest satellite galaxies ($\mathcal{M}_{S,MW}/4$) do show a systematic effect, exhibiting suppression of the star forming fraction as a function of halo mass. Since we have isolated out the effects of stellar mass and halo mass in the mosaics, { this indicates that a galaxy's relative stellar mass compared to other galaxies in its shared halo (which is the discriminator between central and satellite galaxies in this work) has an important role in its evolutionary path}. Otherwise we would expect the emission fractions for similar stellar mass and similar halo mass galaxies (BGGs and satellites) to be closer, where the only missing descriptor is how dominant the mass is compared to other galaxies in the group environment.

There are a number of possible explanations for such effects. One possibility is that satellite galaxies are modified during merger events, e.g.\ their gas is stripped upon entering a larger halo and their probability of forming stars is reduced accordingly. Another explanation, and one we investigate in detail later in this work, is that simple galaxy-galaxy interactions may be responsible for a large amount of this effect, i.e.\ the star forming fraction of satellite galaxies is adjusted based on the probability it experiences different stellar mass galaxies. This latter effect will be a function of the stellar mass of the satellite and of the distribution of stellar masses present in the group. By our definition of central galaxy (most massive) there will be {\it at least} one galaxy more massive than any satellite galaxy in any group, whereas a central galaxy will be the most massive in all interactions. The effects of mass ratios on galaxy properties are investigated later in this paper.

By definition the MW is found in a $M_{H,MW}$ mass halo, and has a stellar mass of $\mathcal{M}_{S,MW}$. Since the subset explored is limited to the regime where we would expect to observe (and group) both the MW and the LMC, we can be confident the MW halo would be in the `grouped' subset of data. The emission class of the MW is likely to be `S', i.e.\ it has a small but observable amount of star formation occurring. Using these classifications, we can use information in Figs \ref{fig:LsingroupsLineTy} and \ref{fig:GaHmLsLi} to draw conclusions about the typicality of the MW compared to similar stellar mass galaxies. A similar stellar mass galaxy is more likely to be in a group than not (like the MW) and more likely to be the central galaxy than not (like the MW). Overall such a galaxy is more likely to possess no star formation than some (unlike the MW), and tends to be found in groups with halo masses slightly larger than the MW halo (the MW is in a slightly under-massive halo given its stellar mass).

\subsubsection{Co-dependence of Galaxy Properties in Groups}

{ 
Galaxy conformity is potentially an important phenomenon. The effect was first highlighted in \citet{wein06}, and reproduced in a larger selection of SDSS DR7 data using a different analysis in \citet{kauf12}. The premise is that the properties of central and satellite galaxies are correlated within a halo. This is something we can investigate in using our GAMA data.

\begin{figure*}
\centerline{\mbox{\includegraphics[width=3.7in]{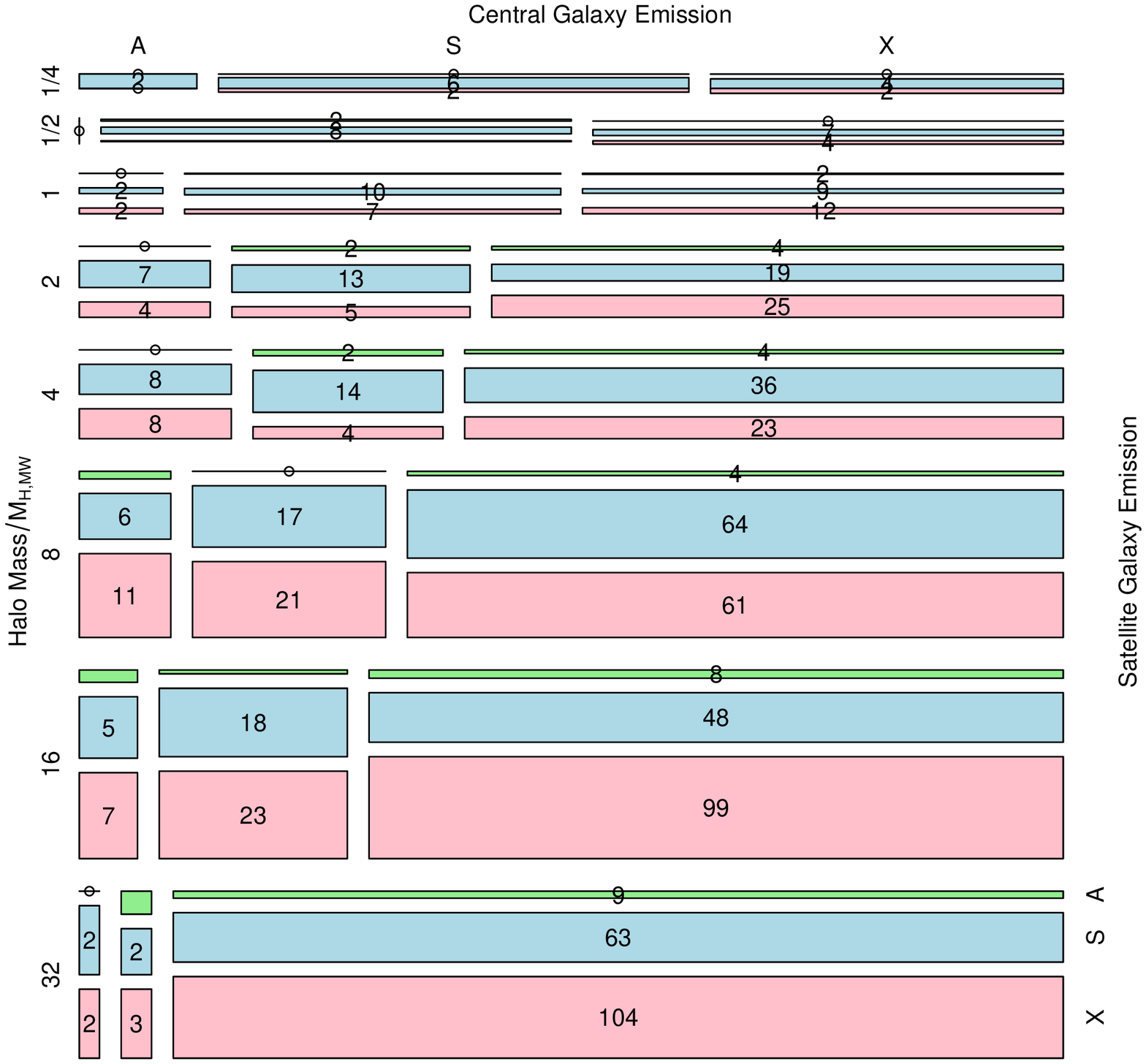}}\mbox{\includegraphics[width=3.7in]{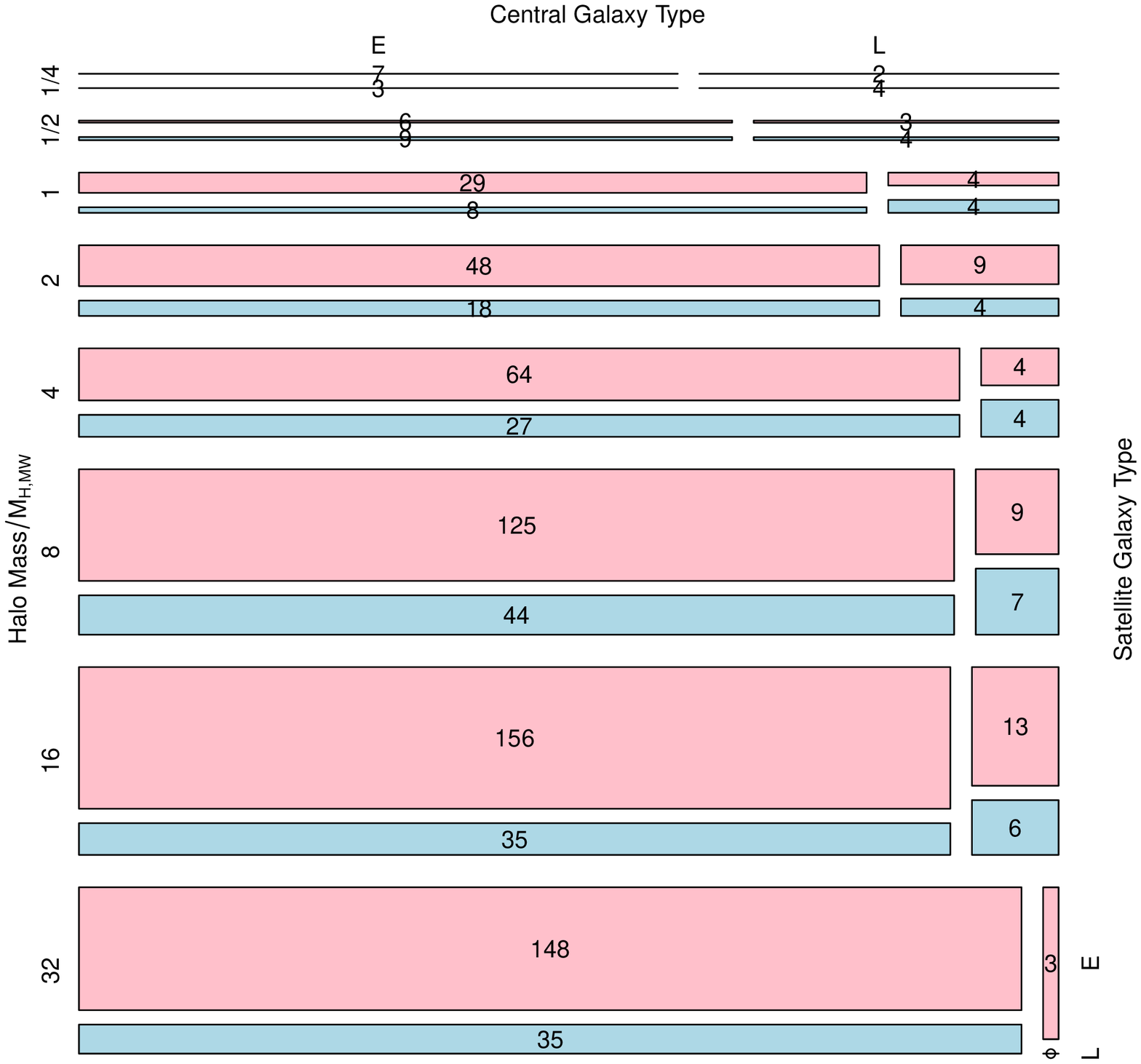}}}
\caption{\small Left panel if a mosaic plot of Halo mass of satellite galaxy (Sec \ref{sec:galhalomass} and \ref{sec:galbgg})/ Emission of central galaxy (Sec \ref{sec:galem} and \ref{sec:galbgg})/ Emission of satellite galaxy (Sec \ref{sec:galem} and \ref{sec:galbgg}). Left panel if a mosaic plot of Halo mass of satellite galaxy (Sec \ref{sec:galhalomass} and \ref{sec:galbgg})/ Type of central galaxy (Sec \ref{sec:galtype} and \ref{sec:galbgg})/ Type of satellite galaxy (Sec \ref{sec:galtype} and \ref{sec:galbgg}) Selection shown is the volume sample (see Sec \ref{sec:selection}).
\newline
{\bf Observations (left):}
\newline
- There is a weak tendency for star formation in a central galaxy to be accompanied by star formation in a satellite galaxy for all halo masses.
\newline
- AGN activity in the central galaxy does not predict any particular response in the satellite galaxy.
\newline
{\bf Observations (left):}
\newline
- There is a stronger tendency for morphological type in a central galaxy to be correlated with morphological type in the accompanying satellite galaxy for all but the largest halo masses.
\newline
- This result is in line with the suggestion of galaxy conformity in \citet{wein06} and \citet{kauf12}.
}
\label{fig:SatdepCen}
\end{figure*}

Figure \ref{fig:SatdepCen} demonstrates the degree to which we observe both emission class (LHS) and morphological type (RHS) conformity. What is clear is that in general we do recover compelling evidence for morphological type conformity (RHS), which is very much in line with the variety discussed in \citet{wein06}, although in that case the `type' was defined using a mixture of galaxy and sSFR (so it was not a purely visual classification scheme, like ours). The emission class demonstrates a weak amount of conformity (LHS), and is certainly not as compelling as the morphological conformity.

Later we investigate the conformity between galaxies in closely interacting pairs. The argument we form is that the conformity we observe for groups reflects the net result of many close interactions, hence the presence of strong morphological type conformity (it takes a long time for the orbits of stars to change radically enough to alter the morphological type) and the weak presence of emission class conformity (this is a fairly instantaneous response to interaction stimulation). This interpretation of the different varieties of conformity that we observe is discussed in detail below.

}

\subsubsection{SFRs in Groups}

\begin{figure}
\centerline{\mbox{\includegraphics[width=3.7in]{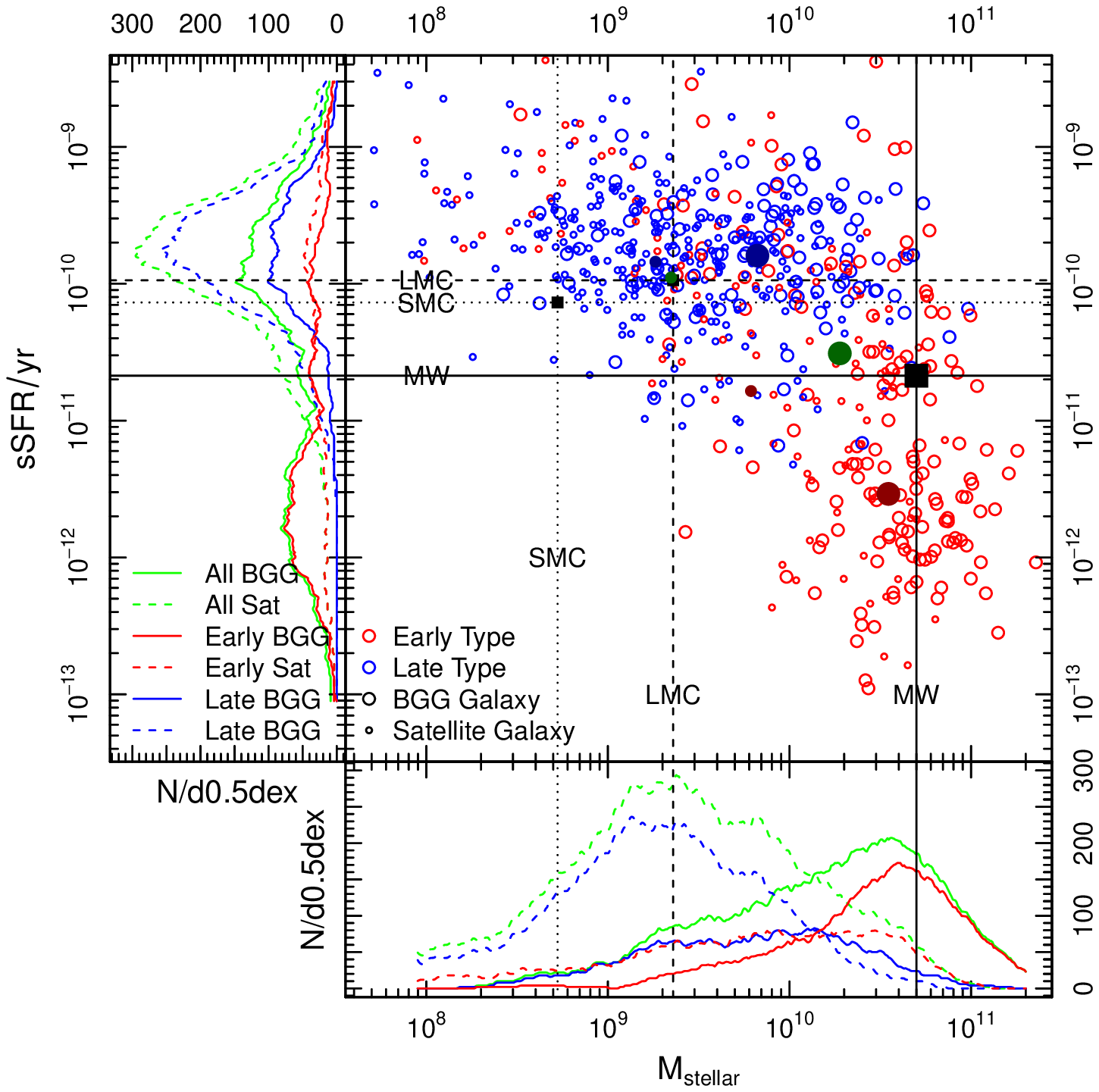}}}
\caption{\small Stellar mass versus specific star formation rate (sSFR) of all pairs where the group is within 0.3 dex of the halo mass of the MW (here assumed to be $2.5 \times 10^{12} \msol$). The colouring refers to morphological type (blue= late-type, red- early-type), and size specifies whether the galaxy is the BGG (large circle) or satellite (small circle). The location of the MW/LMC/SMC are over-plotted for reference. The population median medians are over-plotted, where blue filled points are for late-type, red filled points are for early-type, and green filled points are the full population.
For the MW the stellar mass is from \citet{flyn06} and the SFR is from \citet{robi10}. SMC and LMC stellar masses and SFRs are both from \citet{jame11}.
The density of counts is projected onto the x and y axes, and is separated into BGG / satellite and early / late / combined populations. This plot allows us to see where the MW system sits relative to similar halo mass groups. Selection shown is the volume sample (see Sec \ref{sec:selection}).
\newline
{\bf Observations:}
\newline
- The MW is close to the median sSFR and stellar mass for all central galaxies in MW halo mass groups, but is marginally more massive than typical and has a smaller sSFR (position relative to green point).
\newline
- Adding in the prior knowledge that the MW is late-type, it has an extremely low sSFR and is much more massive than typical for central galaxies in MW halo mass groups.
\newline
- The LMC is almost exactly at the median position expected for late-type satellite galaxies in MW halo mass groups. The caveat here is that to qualify as a group biases the group sample to include groups with satellites as bright as the LMC (since this is the depth limit).
}
\label{fig:GroupScatter}
\end{figure}

\begin{figure}
\centerline{\mbox{\includegraphics[width=3.7in]{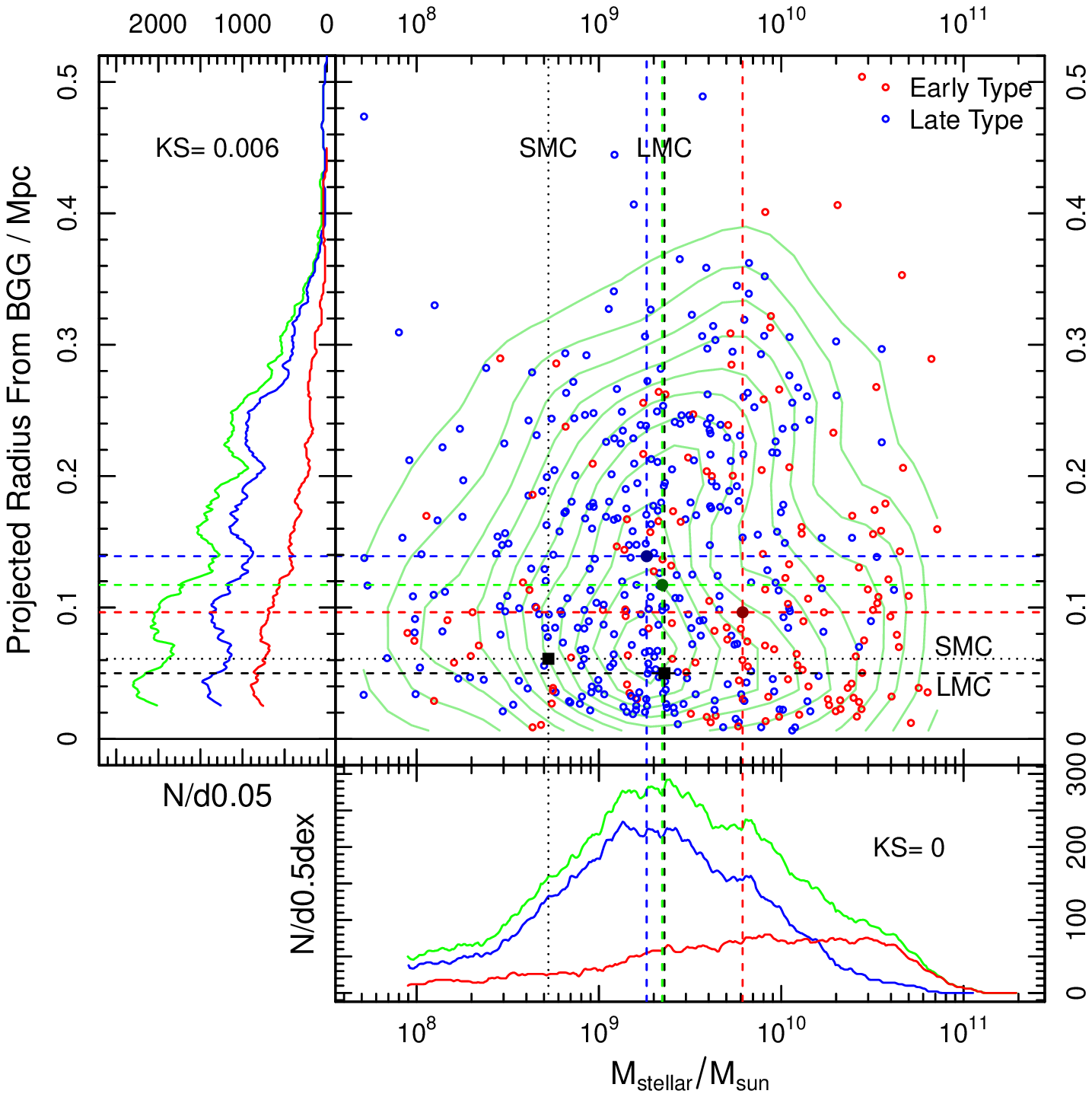}}}
\caption{\small Stellar mass versus projected separation from the BGG for all satellites. The LMC and SMC are over-plotted for comparison. The solid circles show the medians for different subsets, where green, blue and red refers to all, late-type and early-type, and large and small refers to BGG and satellite. The light-green contours show the local density of the full sample.
SMC and LMC stellar masses are both from \citet{jame11}. The separation between the MW and LMC/SMC is taken from \citet{nich11}.
The density of counts is projected onto the x and y axes, and is separated into early / late-type populations. The density plots include a KS test p-value comparison of the early and late-type distributions. Selection shown is the volume sample (see Sec \ref{sec:selection}).
\newline
{\bf Observations:}
\newline
- The LMC is much closer its central group galaxy (the MW) than is typical for satellites in MW mass groups.
\newline
- The KS test comparison of early-type and late-type stellar masses / projected separations for satellites shows it is extremely unlikely the two distributions are sampled from the same parent sample.
\newline
- Early-type satellites tend to be $\sim2/3$ the projected separation to the BGG compared to late-type galaxies.
}
\label{fig:SatScatter}
\end{figure}

Figure \ref{fig:GroupScatter} shows the stellar mass versus specific star formation rate (sSFR) for all galaxies that are found in groups within 0.3 dex of the mass of the MW halo. Figure \ref{fig:SatScatter} the stellar mass versus projected separation to the BGG for all satellites for the same subset of data (right panel). For comparative purposes the median mass and sSFR are shown for different subsets, allowing us to see where the MW-Magellanic system lies.

Considering Figure \ref{fig:GroupScatter} first, it is clear that the MW is in an unusual part of parameter space considering it is a late-type galaxy: all nearby BGGs in stellar mass sSFR space are early-type galaxies. Globally the sSFR is entirely typical, but considering its morphological type (late-type) it is extremely low even allowing for the $\pm 0.15$ dex in SFR error calculated in \citet{robi10}. The MW is $\sim$2.5 times more massive than typical for the global population given its halo mass, and it is $\sim$7 times more massive than the typical late-type BGG in this subset. The LMC is very close to typical in terms of both mass and sSFR compared to the global and the late-type population. The SMC mass galaxies are not complete in the redshift range plotted, so conclusions cannot be drawn. Overall the MW-LMC system appears to be reasonably typical compared to detected \G3C groups if we ignore the fact that the BGG is late-type. Given its other properties, the fact that the MW is late-type makes it appear quite atypical. This finding is in line with the conclusions of \citet{robo12}.

This analysis has to be taken with the caveat that {\it any} group catalogue will have a detection bias towards brighter galaxies being grouped. This is easy to imagine: if only the BGG was bright enough to be assigned a redshift and the 2$^{nd}$ rank galaxy was just below the $r$-band survey limit than the BGG would not be assigned to a group. If the 2$^{nd}$ rank galaxy was brighter (bringing it into our selection limits) then we might be able to assign both galaxies to the same group. The strongest conclusion we can make in this work is that there is at least {\it some} Milky-Way halo mass parameter space that is occupied by $\mathcal{M}_{S,MW}$ mass BGGs and $\mathcal{M}_{S,LMC}$ mass satellites. { This has important implications for galaxy occupation statistic work \citep[e.g.][]{boyl09,boyl10,boyl11,jame11,love12,toll11,weis11,robo12}, since in principle we are biased towards selecting groups which look like our own MW halo, i.e.\ groups where the 2$^{nd}$ rank galaxy (or brightest satellite galaxy) is fairly bright. The typicality of bright companion galaxies around MW stellar mass galaxies was investigated in detail in \citet{robo12}, with the conclusion that $\sim11.9$\% of MW stellar mass galaxies have an LMC stellar mass close pair. This was in broad agreement with the simulated predictions of \citet{boyl10}. }

In Figure \ref{fig:SatScatter} we can see where the LMC and SMC sit in relation to satellites in similar mass groups when comparing stellar mass and separation to the BGG. Concentrating on the LMC, we can see that it lies closer to its BGG (the MW) than the median satellite, both compared to the global and the late-type populations. Early-type galaxies lie closer to the BGG than late-type galaxies, indicating that in an average sense the proximity of a satellite to the BGG affects its evolution. Whilst the LMC might be far closer than the median, it is almost precisely at the {\it mode} of the mass--separation parameter space (shown by light-green contours in Figure \ref{fig:SatScatter}).

\newpage

\subsection{The pair environment of $L^*$ Galaxies}

\begin{figure*}
\centerline{\mbox{\includegraphics[width=3.7in]{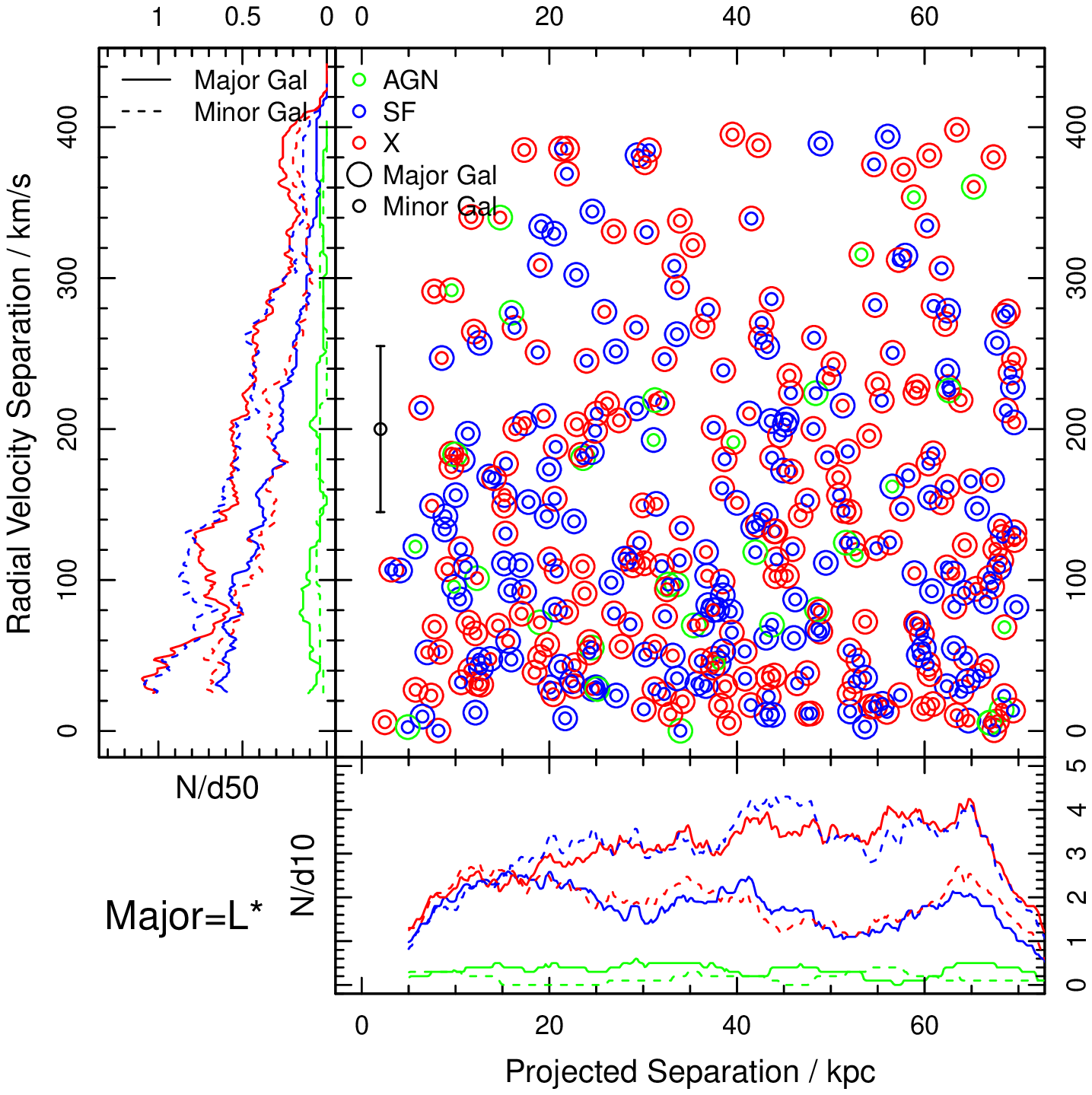}}\mbox{\includegraphics[width=3.7in]{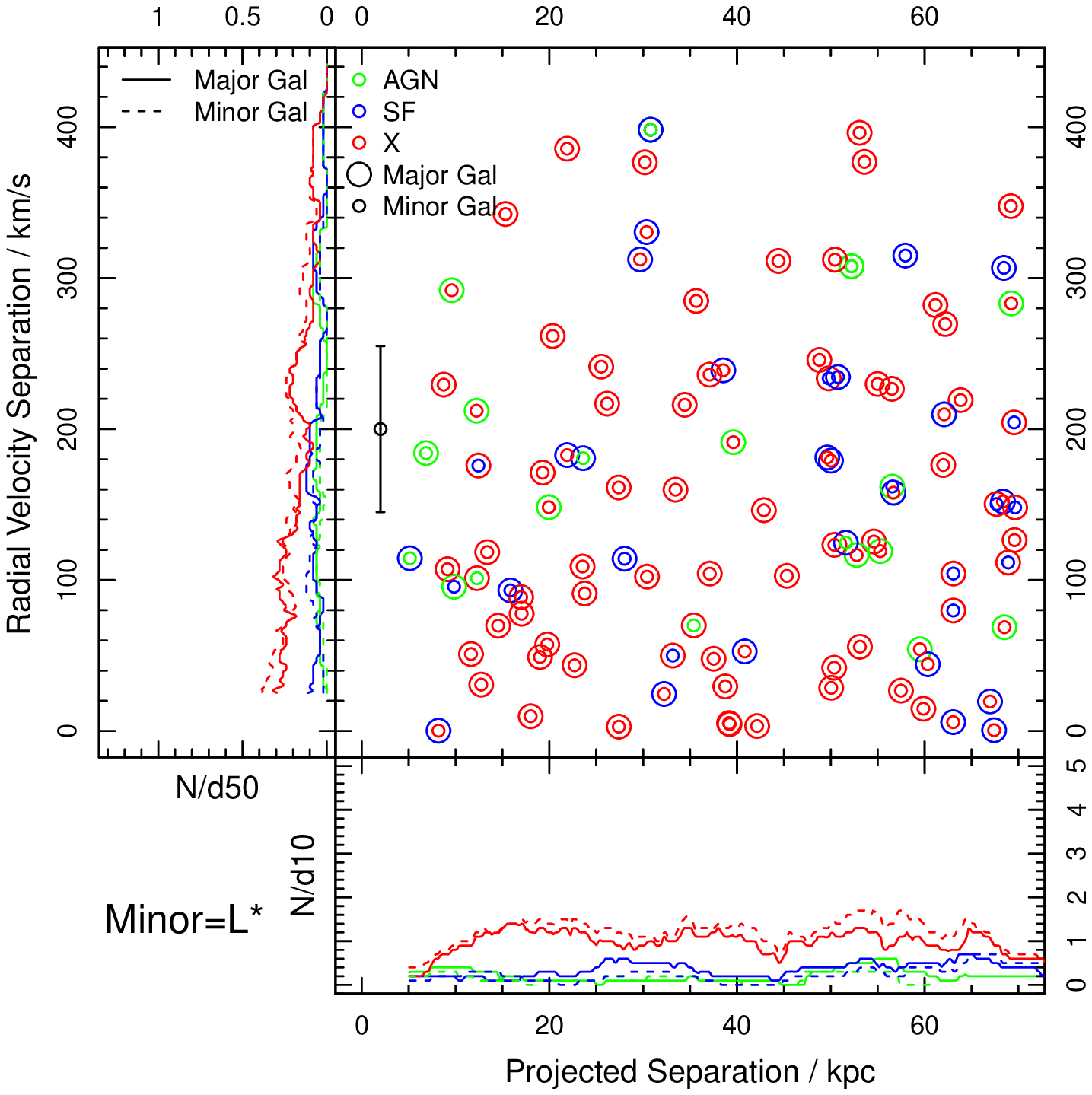}}}
\caption{\small Projected separation versus velocity separation between galaxy pairs, where the major/minor galaxy is within 0.3 dex of stellar mass of the MW (left/right). Galaxy morphological type is indicated by colour, and pair designation by point type. typical velocity errors (55 km/s) are shown on the LHS of the plot windows. Selection shown is the volume sample (see Sec \ref{sec:selection}).
\newline
{\bf Observations (left):}
\newline
- A much larger fraction of $\mathcal{M}_{S,MW}$ mass galaxies are the major galaxy in a pair.
\newline
- When $\mathcal{M}_{S,MW}$ mass galaxies are the major pair galaxy, the star forming--passive (blue--red lines) fraction is close to 1--1 at small projected separations.
\newline
{\bf Observations (right):}
\newline
- When $\mathcal{M}_{S,MW}$ mass galaxies are the minor pair galaxy, they have a much reduced star forming/passive ratio (ratio of blue/red dashed lines) at small projected separations compared to when they are the major pair galaxy.
}
\label{fig:PairPhaseSF}
\end{figure*}

Figure \ref{fig:PairPhaseSF} shows the distribution of emission class in dynamical phase space for galaxies with a stellar mass within 0.3 dex of \MMW. This plot was also assessed as a function of morphological type, disk presence and colour. In all cases there are no strong population gradients within the dynamical phase space explored in this work as either a function of the projected separation or the radial velocity separation. There are clear offsets between the fractions of different classes and due to a galaxy being the major or minor partner in the galaxy pair system (this is discussed in detail later).

There is also an increased likelihood that MW stellar mass minor pair galaxies will be passive at small projected separations, compared to if they are the major pair galaxy. However, in general the constancy of the star forming fraction in these distributions indicate we will not be strongly biased by our somewhat arbitrary definition of a galaxy pair in our analysis, and we will limit the discussion to the full dynamical selection criteria for pairs (rather than refining discussion to sub-bins of Figure \ref{fig:PairPhaseSF}).

\subsubsection{Pair Fractions by Mass}

Having considered the role of the group environment on a given stellar mass galaxy, we can now investigate the effects galaxy pairs have on the evolution of the pair galaxies. This will probe the MW-LMC pair system directly, and allow us to determine how unusual this system is. The pair catalogue used for this work is discussed in Section \ref{sec:paircat}. Due to the redshift, spatial and velocity separation selections we expect all MW-LMC like systems to be observed and complete (i.e.\ for $z<0.089$, $v_{\rm sep-rad}<400 \kms$ and $r_{\rm sep-proj}<70\kpc$).

\begin{figure*}
\centerline{\mbox{\includegraphics[width=3.7in]{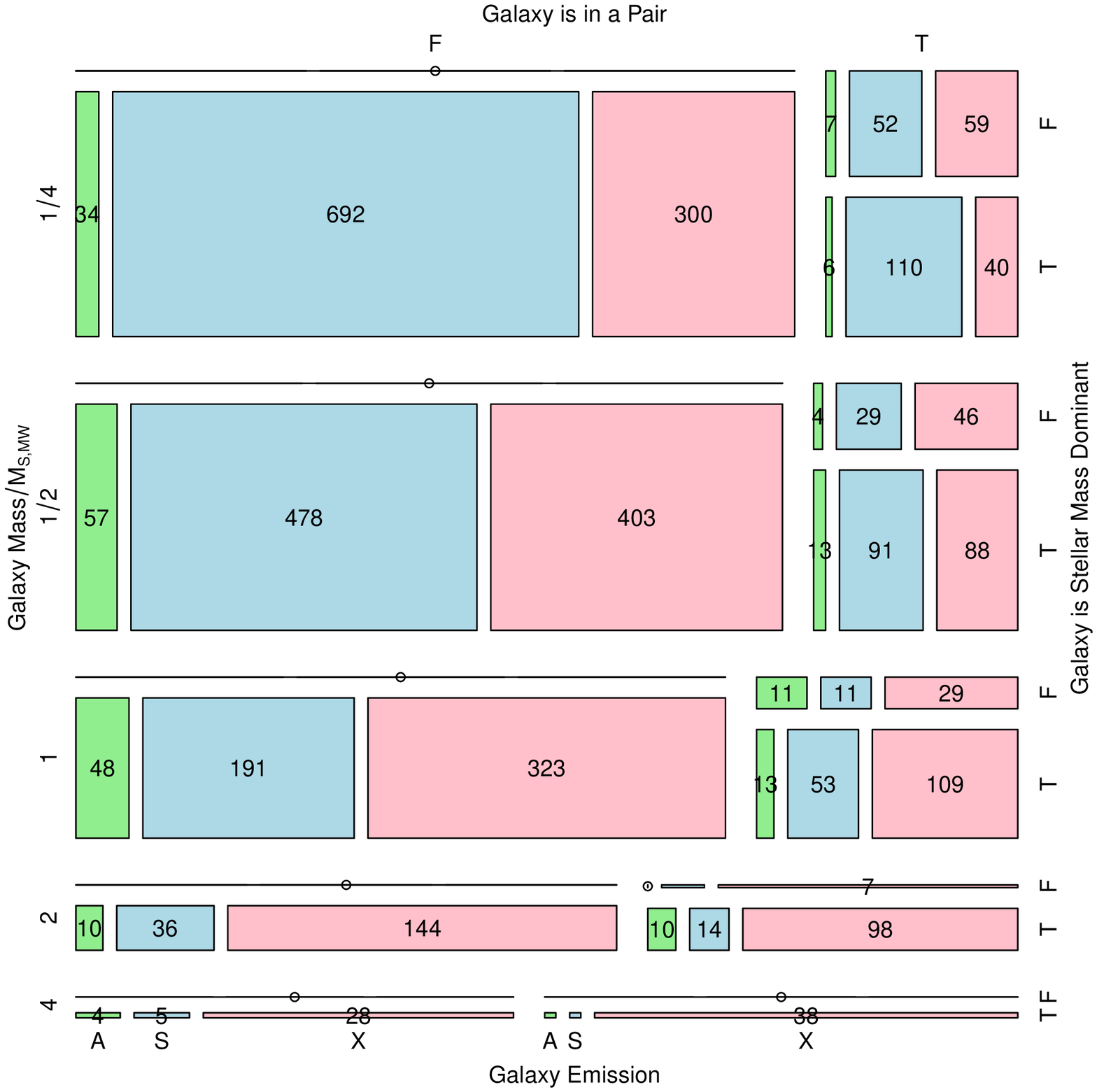}}\mbox{\includegraphics[width=3.7in]{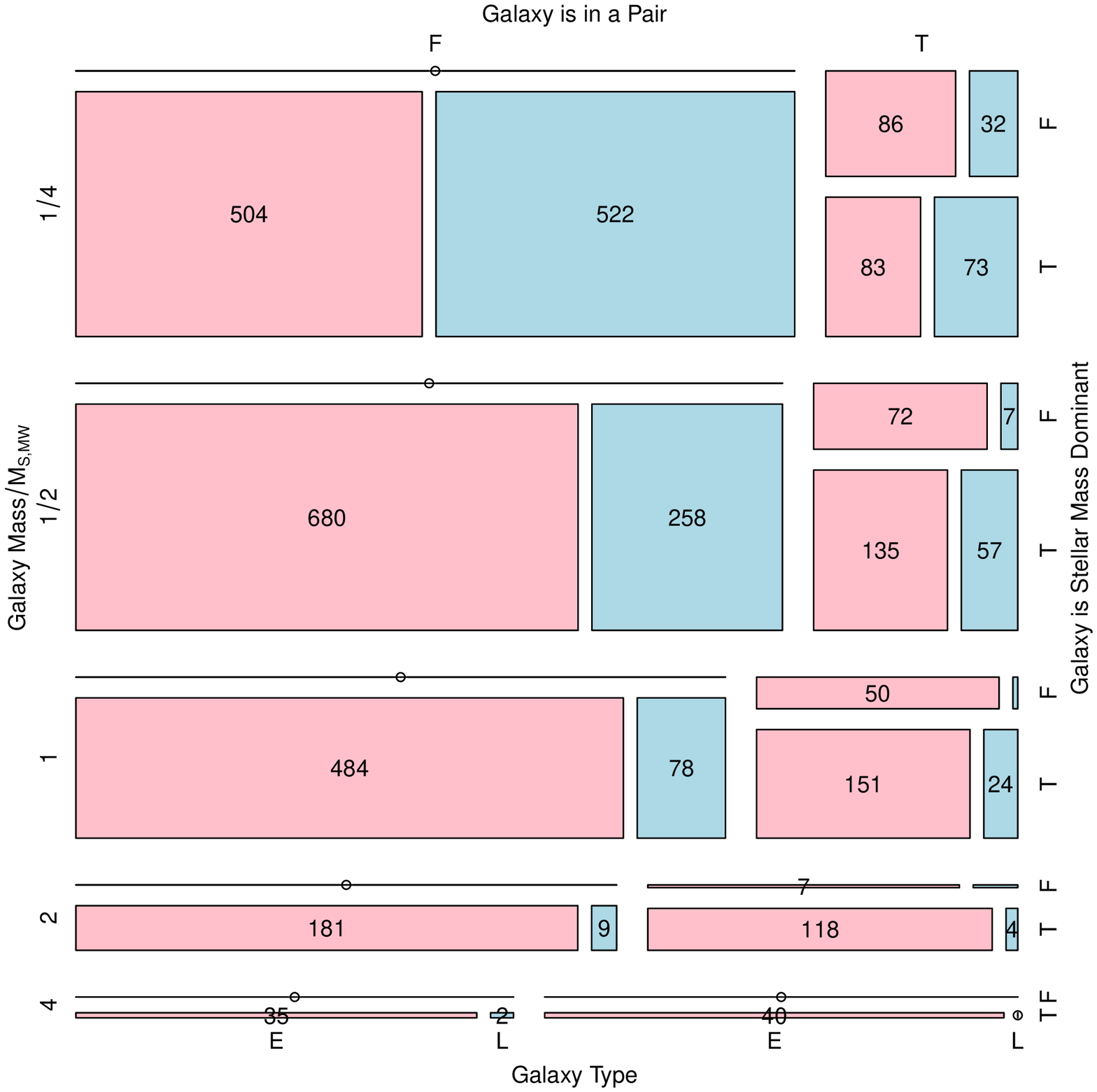}}}
\caption{\small Left panel is a mosaic plot of Stellar mass (Sec \ref{sec:galstelmass})/ Galaxy in pair (Sec \ref{sec:galpaired})/ Largest galaxy in pair (Sec \ref{sec:galbpg})/ Emission (Sec \ref{sec:galem}). Right panel is a mosaic plot of Stellar mass (Sec \ref{sec:galstelmass})/ Galaxy in pair (Sec \ref{sec:galpaired})/ Largest galaxy in pair (Sec \ref{sec:galbpg})/ Type (Sec \ref{sec:galtype}). Selection shown is the volume sample (see Sec \ref{sec:selection}).
\newline
{\bf Observations (left):}
\newline
- More massive galaxies are more likely to be in a pair.
\newline
- The star forming fraction for major pair galaxies is similar to that observed for unpaired galaxies.
\newline
- Minor pair galaxies have a smaller star forming fraction than major pair galaxies for the same stellar mass. Together with the above remark, this indicates that pairwise interactions have a net role of suppressing of star formation, rather than initiating it (though both effects are likely to occur).
\newline
{\bf Observations (right):}
\newline
- Minor pair galaxies have a smaller late-type fraction than major pair galaxies for the same stellar mass, and this dependency is more dramatic than that seen for star formation (above).
\newline
- The late-type fraction drops more rapidly with stellar mass than the star forming fraction, particularly for minor pair galaxies.
\newline
- The MW would be more typical if they were early-type, given its stellar mass. This agrees with the distributions shown in Fig \ref{fig:GroupScatter}.
}
\label{fig:LsinpairLine}
\end{figure*}

The left panel of Fig. \ref{fig:LsinpairLine} shows the mosaic for Galaxy stellar mass/ whether the galaxy is in a pair/ whether the galaxy is the largest in the pair/ emission class. The right panel exchanges the emission class contingency for morphological type.

Immediate observations regarding the left panel of Fig. \ref{fig:LsinpairLine} are that more massive galaxies are more likely to be found in a pair meeting our selection criteria, but in all cases the majority of galaxies are not in pairs. The non-paired emission class distributions are most similar to paired galaxies when the galaxy is more massive. This makes sense because we are only confident the galaxy is not paired down to a minor mass limit of the LMC, so it might be the dominant member of a pair with an even less massive minor companion. There is a weak tendency for the minor galaxy in a pair to have a smaller star forming fraction than it would if it was the largest galaxy in the pair, or if it was unpaired. This is a similar result to the effects discussed for satellite galaxies in Sec \ref{sec:groupmosaic}.

The trends seen in the right panel of Fig. \ref{fig:LsinpairLine} are similar, where the late-type populations follow the same population trends as the `S' class emission galaxies discussed above. In general we see that morphological type is mostly driven by galaxy stellar mass. One stronger dependency noted is that late-type galaxies of a given stellar mass are suppressed if it is the minor component of a pair system relative to when it is the dominant component. The relative amount of suppression is more dramatic than that seen for star formation. When the stellar mass is $\frac{1}{4}$\MMW, $\sim$2 times as many galaxies are dominant pair star forming galaxies compared to minor pair star forming galaxies, and the late-type ratios are similar. However, when the stellar mass is \MMW the ratio is $\sim$5 for the star forming fractions, but 24 for the late-type fractions. The difference between these numbers is independent of stellar mass effects, and shows a real difference in response to close pairwise interactions between star formation and morphology. Perhaps surprisingly, it is morphology that is the best tracer of such events.

Since being in a galaxy pair is potentially a very brief state given our pair selection requirements, this suggests that morphology is actually sensitive to dynamic events on short timescales, but displays the effects for longer periods of time compared to the star forming fraction. I.e.\ there is a higher chance of observing modified morphology due to pairwise interaction than modified star formation.

\begin{figure*}
\centerline{\mbox{\includegraphics[width=7.4in]{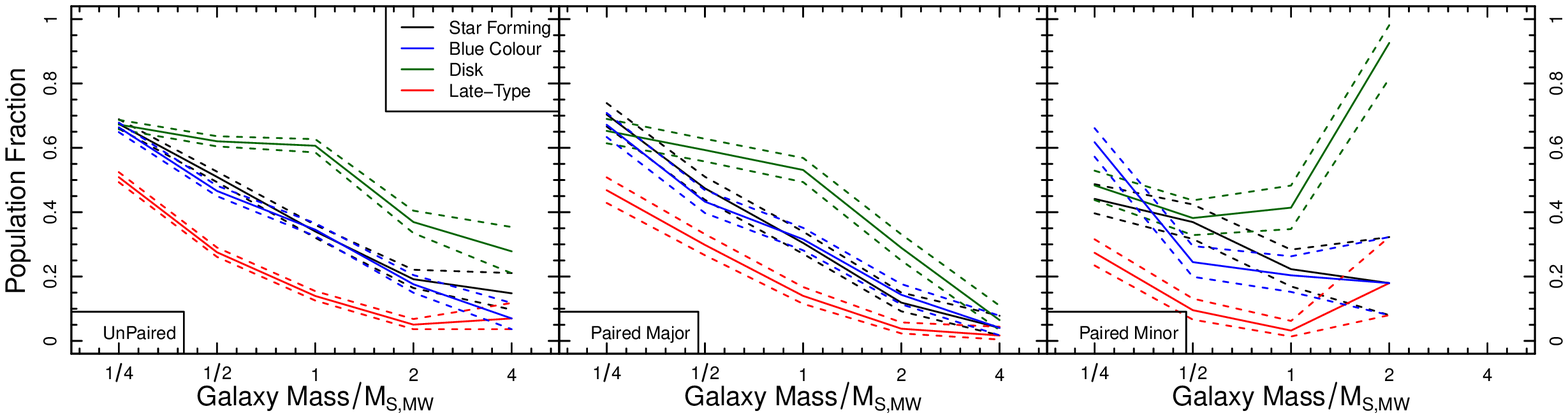}}}
\caption{\small Population fractions for different properties that are commonly associated with late-type populations: being morphologically late-type (by definition), star forming, having a disk and being blue in integrated colour. These are shown for unpaired (left panel), dominant galaxy in pair (middle) and minor galaxy in pair (right) galaxies. Dotted lines show the formal bayesian error for a bimodal distribution, in this case the chance of galaxies being classed as stated in the legend, or not. Selection shown is the volume sample (see Sec \ref{sec:selection}).
\newline
{\bf Observations:}
\newline
- The most massive minor pair galaxies show a statistically significant increase in the disk fraction relative to the unpaired case.
}
\label{fig:PairPopFracs}
\end{figure*}

Figure \ref{fig:PairPopFracs} is similar to Figure \ref{fig:PopFracs}, but is now plotted for galaxies in pairs (rather than groups) and major or minor pair component (rather than group central or satellite galaxy). Galaxy colour displays the most consistency across the three selection classes plotted, whilst disk presence, star formation fraction and late-type fraction show a strong variation between the unpaired and minor pair galaxy classes for a given amount of galaxy stellar mass. A dramatic feature is that the most massive minor pair galaxies show a statistically significant increase in the disk fraction relative to the unpaired case. This sample was reanalysed, and 14/16 of minor pair galaxies with $\mathcal{M}_{s} \ge$\MMW show extremely dramatic and asymmetrical spiral features { associated with the pairwise interaction}. This is in contrast to similar stellar mass systems with disks in the rest of the sample, these tending to be smooth symmetrical disks. Since in these extreme cases the dominant galaxy would have to be extremely large, the disk observed { appears to be} one generated by near-neighbour tidal disruption processes rather than gentle accretion. Such structures are transient (on the order of $\sim$Gyr) since it is the brief close passage between galaxies that generates them.

The grouped central galaxies portray much the same correlations as the dominant paired galaxies. This can be partly understood by the large fraction of population overlap: 63\% of dominant (or major) galaxy pairs are also the central galaxy of a galaxy group in the sample investigated. Equally, the ungrouped galaxies share similar population fraction trends to the unpaired galaxies. The notable exception is the galaxy disk fraction--- there is a continuous downwards trend with stellar mass for the unpaired galaxies, whilst for the ungrouped galaxies we see evidence for a flattened correlation. Considering dynamic timescales, it is likely an ungrouped galaxy has experienced a longer delay since the last major interaction than an unpaired galaxy, since not being grouped is a stronger isolation criterion than not being paired (a larger fraction of galaxies are unpaired compared to ungrouped). This fits with the hypothesis that that presence of a disk typically suggests a longer period of time since a major galaxy-galaxy interaction has ended, since these should disrupt the visible accretion-type disk \citep{barn92a}. The exception is complicated extended disk-like structures observed {\it during} tidal interactions between massive galaxies. These are formed by brief and violent close passages \citep[see early discussion of such systems in][]{korm77}.

{ In Section \ref{sec:groupmosaic} we found that the lowest star forming fractions were found in grouped satellite galaxies. An equivalent statement cannot be made regarding pairs: the lowest star forming fractions are found for paired galaxies where the galaxy is the lowest mass member in the pair. In the next section we investigate how star formation can be suppressed and enhanced in pairs in more detail.}

\subsubsection{Effects of Galaxy Stellar Mass Ratios in Pairs}

A galaxy that is in a pair can be in a number of different mass ratio combinations, i.e.\ a given stellar mass galaxy can either be the larger (dominant or major) or smaller (minor) within a pair. The complication of this is that lower stellar mass galaxies near our survey limit will be incomplete since there will not be any lower mass galaxies within our sample selection for them to be paired with. These galaxies will therefore only appear as the minor partner in a galaxy pair.

\begin{figure}
\centerline{\mbox{\includegraphics[width=3.7in]{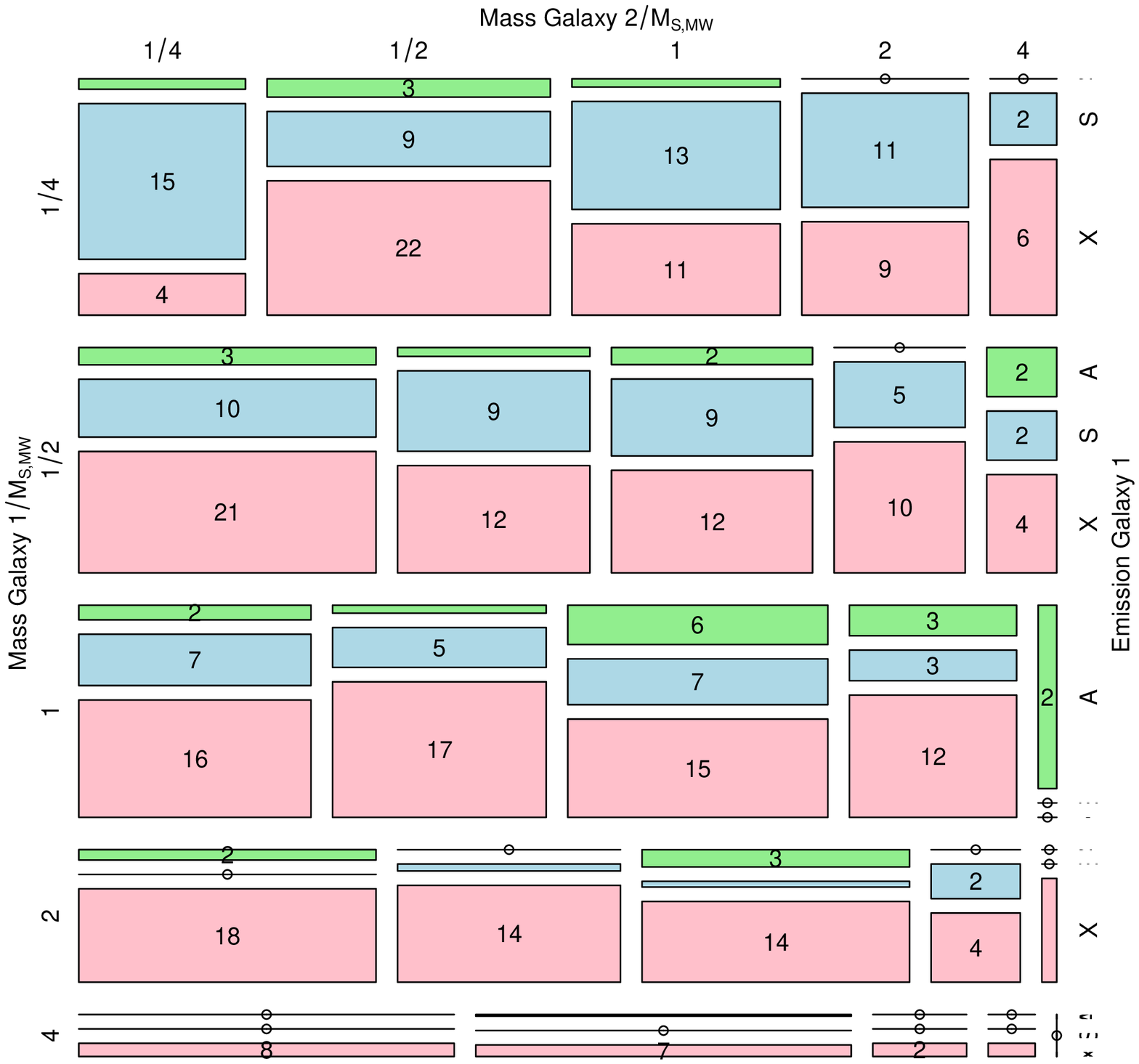}}}
\caption{\small Mosaic plot of Stellar mass of galaxy 1 (Sec \ref{sec:galstelmass})/ Stellar mass of galaxy 2 (Sec \ref{sec:galstelmass})/ Emission of galaxy 1 (Sec \ref{sec:galem}). Selection shown is the volume sample (see Sec \ref{sec:selection}).
\newline
{\bf Observations:}
\newline
- Galaxy-galaxy pairs with similar stellar masses have the largest star forming fraction, though this relationship is noisy.
\newline
- The star forming fraction is driven most strongly by stellar mass, pairwise interactions have a weaker effect.
}
\label{fig:G1depG2vcomp}
\end{figure}

With this effect in mind, Figure \ref{fig:G1depG2vcomp} shows the relationship for galaxy pairs in a narrow mass ratio regime where both the lower mass and higher mass galaxies are within our selection criteria: $\mathcal{M}_{S,MW}/4 < \mathcal{M}_{s} < 4 \mathcal{M}_{S,MW}$.

The grids in Figure \ref{fig:G1depG2vcomp} are aligned to reveal how the emission class of a galaxy varies as a function of the other galaxy pair mass. The number of galaxies within diagonally mirrored emission class cells will add up to the same value, but they will be distributed differently since the emission class is always shown for galaxy 1 (i.e.\ they correspond to the masses plotted on the left). There is a preference for galaxy pairs with similar masses to have the most evidence for star formation. When the secondary galaxy is much more or less massive the star forming fraction drops appreciably. { This implies that major merger (i.e.\ equal mass) interactions have the most significant star formation enhancing effect in galaxy pairs, where unpaired galaxies have as high (or higher) star forming fractions.}

To investigate how these trends continue beyond the stellar mass range considered above, we can explore a larger sample where we force galaxy 1 to always be the larger galaxy in the pair and within the volume and stellar mass limits ($\mathcal{M}_{S,MW}/4 < \mathcal{M}_{s} < 4 \mathcal{M}_{S,MW}$). Galaxy 2 has to be more massive than the LMC and less massive than the least massive galaxies in the galaxy 2 sample in Figure \ref{fig:G1depG2vcomp} to ensure we are only considering a volume limited part of parameter space (i.e.\ $\mathcal{M}_{S,MW}/32 (\sim \mathcal{M}_{S,LMC}) < \mathcal{M}_{s} < \mathcal{M}_{S,MW}/4$).

\begin{figure*}
\centerline{\mbox{\includegraphics[width=3.7in]{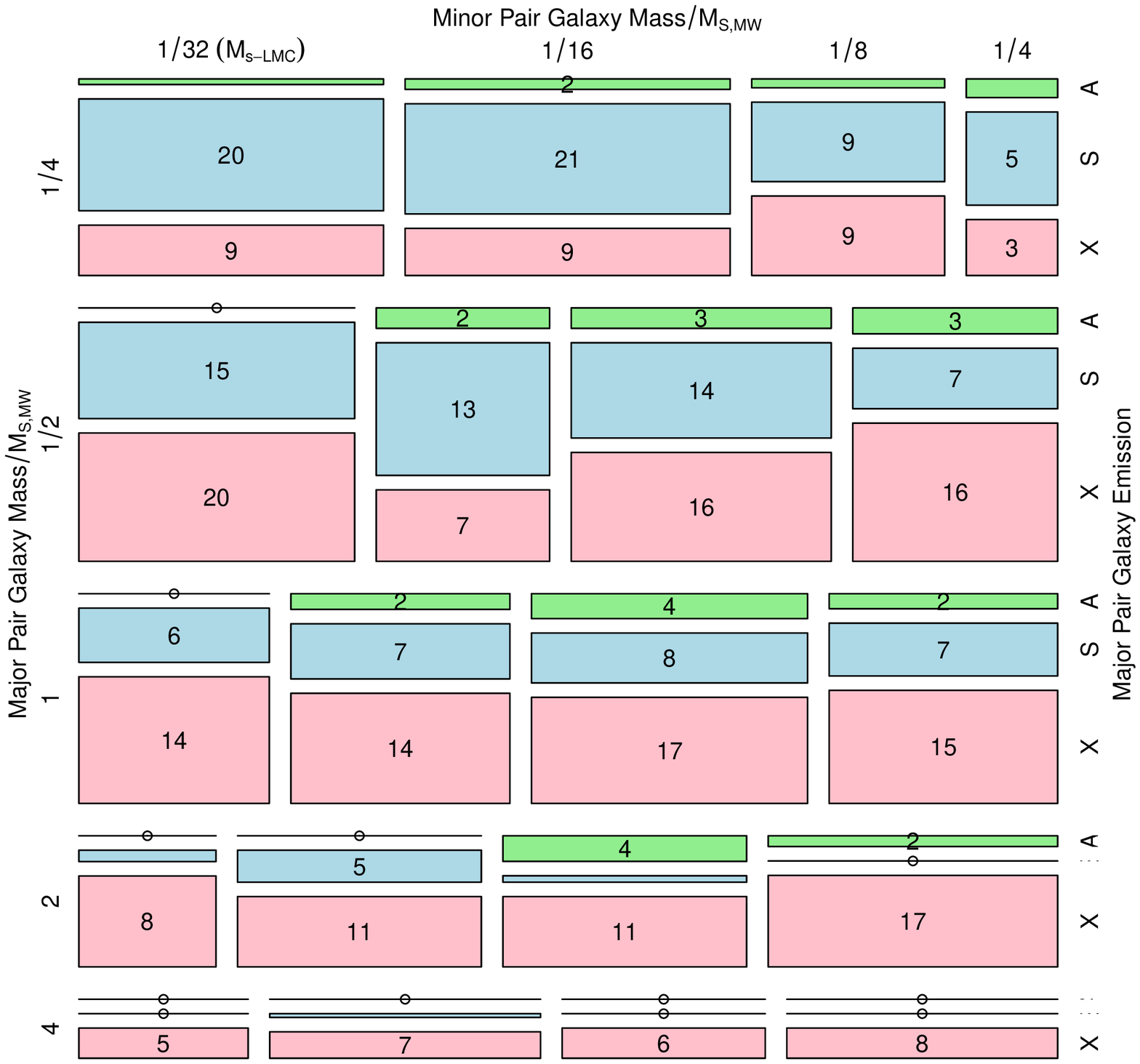}}\mbox{\includegraphics[width=3.7in]{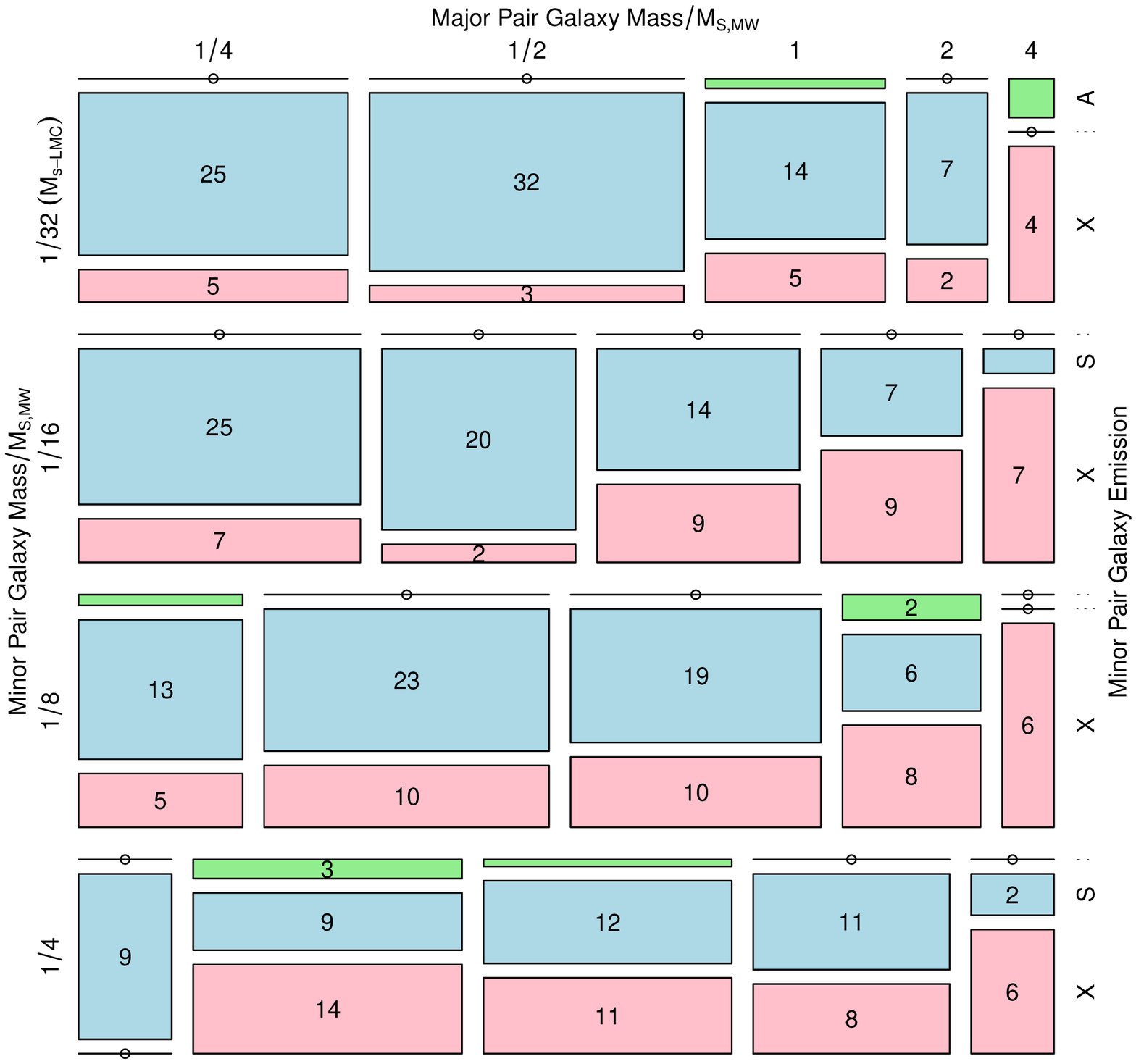}}}
\caption{\small Left panel is a mosaic plot of Stellar mass of the major pair galaxy (Sec \ref{sec:galstelmass} and \ref{sec:galbpg})/ Stellar mass of the minor pair galaxy (Sec \ref{sec:galstelmass} and \ref{sec:galbpg})/ Emission of the major pair galaxy (Sec \ref{sec:galem} and \ref{sec:galbpg}). Right panel is a mosaic plot of Stellar mass of the minor pair galaxy (Sec \ref{sec:galstelmass} and \ref{sec:galbpg})/ Stellar mass of the major pair galaxy (Sec \ref{sec:galstelmass} and \ref{sec:galbpg})/ Emission of the minor pair galaxy (Sec \ref{sec:galem} and \ref{sec:galbpg}). Selection shown is the volume sample (see Sec \ref{sec:selection}).
\newline
{\bf Observations (left):}
\newline
- The star forming fraction of major galaxies is largely unaffected by having pair companions that are much less massive than itself. Star formation in the major pair galaxy of MW-LMC analogues is largely unaffected by the LMC mass minor companion.
\newline
{\bf Observations (right):}
\newline
- Star forming galaxies for minor pair galaxies show a decline as the major pair galaxy mass increase.
\newline
- Where these is population overlap with the left panel, the minor galaxies possess a smaller star forming fraction.
}
\label{fig:G1vG2}
\end{figure*}

The result of this selection is shown in Figure \ref{fig:G1vG2}, where the first two contingencies are the two galaxy masses in the pair and the third contingency is the emission class of the more massive galaxy (left panel) and the emission class of the less massive galaxy (right panel).

Considering the left panel of Figure \ref{fig:G1vG2} first, we can continue the trends observed in Figure \ref{fig:G1depG2vcomp}. There is weak evidence that whilst the star forming fraction peaks near 1--1 mass ratio pairs, it then plateaus as the lower mass companion becomes less massive. The average star forming fractions of the dominant galaxy are similar to unpaired galaxies (see left panel of Figure \ref{fig:LsinpairLine}), so overall they appear enhanced when mass ratio is 1--1, and suppressed otherwise. { This finding requires more data to be robustly confirmed since the number statistics are small, but this work suggests galaxy-galaxy pair stellar mass ratios nearly always suppress star formation, with the net effect at a minimum when the galaxy masses are most similar. This is an interesting point, since gravitational braking is at its most efficient when the mass ratio is close to 1--1, i.e.\ these will be the galaxies most likely to be undergoing full mergers and not simply transitory interactions.}

The right panel of Figure \ref{fig:G1vG2} flips round the first two contingencies of the left panel so that the emission classes always correspond to the left side divisions. Therefore, it should not be immediately surprising that we see much larger star forming fractions in the right panel--- it shows the grids for typically lower mass galaxies. A couple of observations can be drawn though.

The first is that the one mass range that exists for both minor and major pair galaxies is $\mathcal{M}_{s} <=\mathcal{M}_{S,MW}/4$ (the top strip of the left panel and the bottom strip of the right panel) and the star forming fraction is clearly higher when it is the larger partner in the pair. This tallies with the observations regarding Figure \ref{fig:LsinpairLine}, where we saw that dominant pair galaxies of a given stellar mass have more `S' class emission.

The second is that minor galaxy emission is more sensitive to the mass of the major pair galaxy than vice-versa. This is clear from the star forming fractions that vary strongly as a function of major galaxy mass: the over-riding trend being the more dominant the mass of the major galaxy (the larger the mass ratio) the smaller the fraction of strongly emitting galaxies.

\begin{figure}
\centerline{\mbox{\includegraphics[width=3.7in]{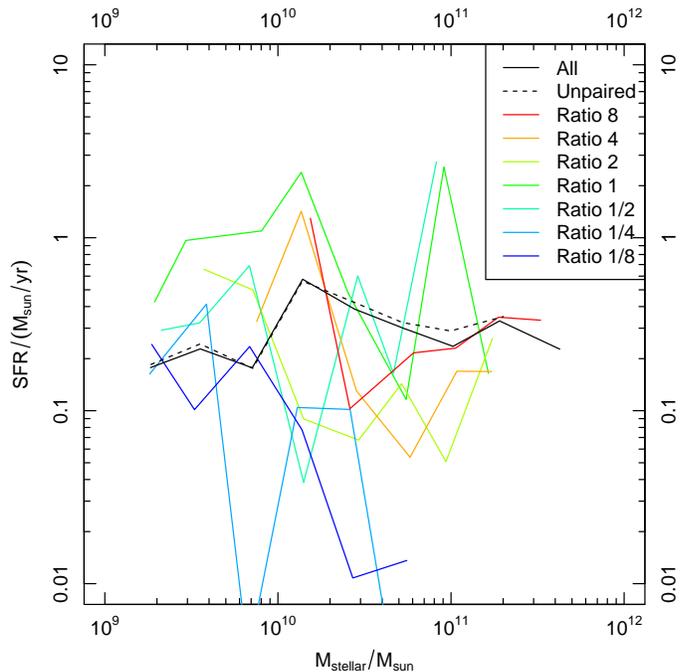}}}
\caption{\small Stellar mass versus SFR for volume limited samples of galaxy pairs with varying pair mass ratios. All samples are volume limited, with the stellar mass limits adjusted as a function of the pair ratio. Black solid line shows the running median for all galaxies in our volume limited sample. Black dashed line shows the running median for all unpaired galaxies (given our observational depth constraints). Different colours refer to different mass ratios, with 1--1 mass ratio pairs shown as a green line. Selection shown is the volume sample (see Sec \ref{sec:selection}).
\newline
{\bf Observations:}
\newline
- Separate to the star forming fraction (shown in the mosaic plots), here we see that the actual amount of star formation is modified by the pair environment: mass ratios close to 1--1 (green line) typically display the highest median star formation rate for a given amount of stellar mass.
\newline
- Star formation is dramatically suppressed when the mass ratio is large and the galaxy under consideration is the minor pair galaxy (blue line).
}
\label{fig:PairSFR}
\end{figure}

Going further, Fig. \ref{fig:PairSFR} shows the runnings median SFR as a function of stellar mass for different volume limited samples of galaxy-galaxy pair mass ratios. The data is of low $S/N$ quality, but a clear feature is that the 1--1 mass ratio population (green solid line) is typically the population with the highest SFR given the stellar mass. So whilst the {\it fraction} of galaxies forming stars is suppressed in all galaxy pairs, the amount of star formation occurring in those galaxies that are still forming stars actually increases when the interaction mass ratio is close to 1--1. As a counter to this, the minor pair galaxies with the smallest mass ratios (bluer lines) are those with the least SFR given the stellar mass. The y-axis separation between the 8--1 mass ratio SFR (red line) and the 1/8--1 mass ratio SFR (blue line) is roughly 1 dex.

The combination of information from the figures presented in this section produce a compelling picture when it is all pieced together.  The emission state of dominant pair galaxies is only significantly altered when the pair ratio is close to 1, otherwise we observe the same galaxies as we would in an unpaired system. The minor galaxy in a pair is greatly affected by its companion galaxy, and whilst all star formation is suppressed relative to equivalent stellar mass unpaired galaxies, it becomes lower still when the mass ratio increases (i.e.\ the major galaxy becomes more massive and dominant within the pair). { The baryonic physics that fully explains these effects is beyond the scope of this work, but there are likely to be a number of physical processes competing for dominance during a galaxy-galaxy interaction, some of these will trigger star formation and some will suppress it. The {\it net} effect of the interactions analysed in this work is to suppress star formation, with the caveat that this is least evident for similar stellar mass galaxy pairs. Whilst the fraction of galaxies forming stars is generally lower in a galaxy pair, the absolute peak of star formation seen is for a 1--1 mass ratio interaction.}

\subsubsection{Co-dependence of Galaxy Properties in Pairs}

Galaxies in pairs can have a number of emission state permutations, hence it is interesting to investigate whether the emission strength in one member of a galaxy in a pair predicts the other: i.e.\ if the larger galaxy in a pair is experiencing star formation, does it make it any more likely that the minor partner is also star forming?

\begin{figure*}
\centerline{\mbox{\includegraphics[width=3.7in]{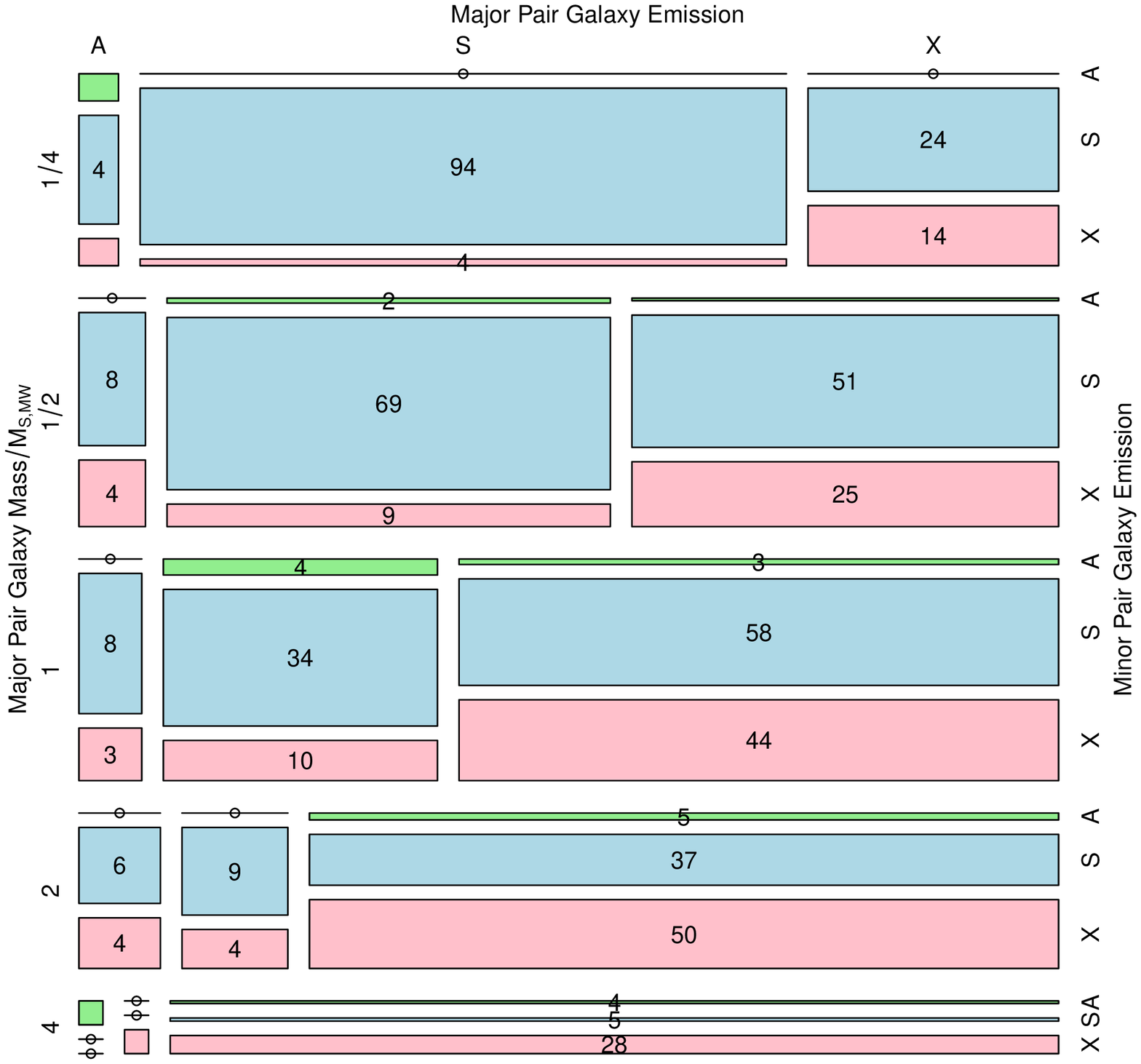}}\mbox{\includegraphics[width=3.7in]{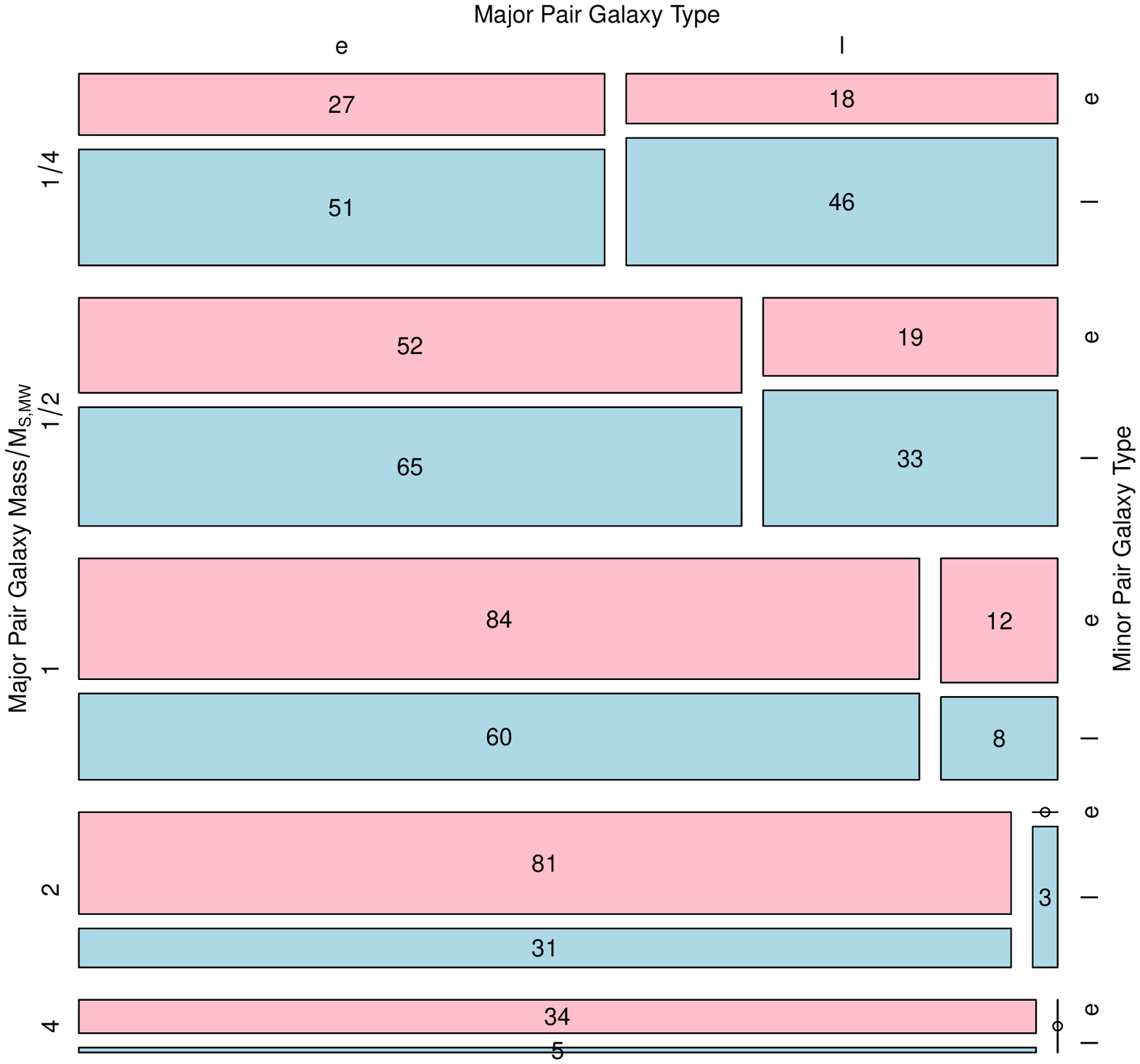}}}
\caption{\small Left panel is a mosaic plot of Stellar mass of the major pair galaxy (Sec \ref{sec:galstelmass} and \ref{sec:galbpg})/ Emission of the major pair galaxy (Sec \ref{sec:galstelmass} and \ref{sec:galbpg})/ Emission of the minor pair galaxy (Sec \ref{sec:galem} and \ref{sec:galbpg}). Right panel is a mosaic plot of Stellar mass of the major pair galaxy (Sec \ref{sec:galstelmass} and \ref{sec:galbpg})/ Type of the major pair galaxy (Sec \ref{sec:galtype} and \ref{sec:galbpg})/ Type of the minor pair galaxy (Sec \ref{sec:galtype} and \ref{sec:galbpg}). Selection shown is the volume sample (see Sec \ref{sec:selection}).
\newline
{\bf Observations (left):}
\newline
- There is a clear tendency for star formation in a major pair galaxy to be accompanied by star formation in a minor pair galaxy for all major pair galaxy stellar masses.
\newline
- If the major pair galaxy does not show evidence for star formation it is less likely that the minor pair galaxy will be forming stars.
\newline
- When the major pair galaxy shows evidence for AGN activity, the likelihood of star formation in the minor pair galaxy is between the above two extremes.
\newline
{\bf Observations (right):}
\newline
- There is weaker tendency for morphological type of a major pair galaxy to be correlated with the morphological type of the accompanying minor pair galaxy.
\newline
- This is the pairwise equivalent of galaxy conformity, described for groups and isolated galaxy systems in \citet{wein06} and \citet{kauf12} respectively.
\newline
- The relative strengths of the galaxy conformity effect seen here is reversed compared to groups (above). There the morphological type showed stronger conformity, whereas when galaxy pairs are considered the conformity is greatest for emission class. 
}
\label{fig:G1depG2em}
\end{figure*}

Figure \ref{fig:G1depG2em} shows a mosaic plot of how the emission class of the dominant pair galaxy predicts the emission class of the minor pair galaxy for a given stellar mass of the dominant galaxy (LHS) or a similar plot showing the co-dependence of morphological type. The clear indication from this plot is that the emission classes are highly co-dependent. Star forming emission in either the dominant or minor galaxy predicts it in the other, but this is particularly true for dominant predicting minor. Taking \MMW stellar mass galaxies, if star formation is present in the dominant galaxy it is 71\% likely that the minor galaxy is also star forming. If the same stellar mass dominant galaxy instead showed no emission (`X' class) then it is only 55\% likely that the minor galaxy has star forming emission. { The strength of this conformity effect is remarkably regular for all stellar masses, although the fraction of star forming dominant galaxies with stellar masses above $\mathcal{M}_{S,MW}$ is very small.}

The `X' class prediction is also very strong, and it becomes more dominant for higher mass dominant galaxies. For all masses, minor galaxy star formation is most likely when the dominant is star forming and least likely when it shows no signs of emission. If the dominant galaxy hosts an AGN then the chance of the minor galaxy showing star forming emission is broadly between these two extremes. { If we hypothesise that AGNs form after a period of star formation, this would suggest that whatever mechanism has stimulated an AGN in the dominant pair galaxy (and shut down star formation) is correlated with a small suppressive effect on the star formation properties of the nearby minor pair galaxies. This correlation does not suggest causation, it could just indicate that there is a natural temporal ordering associated with these observations.

The co-dependence of morphological type is less clear cut, with only a weak tendency for the morphological type of the major pair galaxy to predict the morphological type of the minor pair galaxy. This is the reverse of what we saw for grouped galaxies, where the morphological type conformity between central and satellite galaxies was very strong \citep[in line with the observations of][]{wein06}.

}

\subsubsection{SFRs in MW--LMC Pairs}

\begin{figure}
\centerline{\mbox{\includegraphics[width=3.7in]{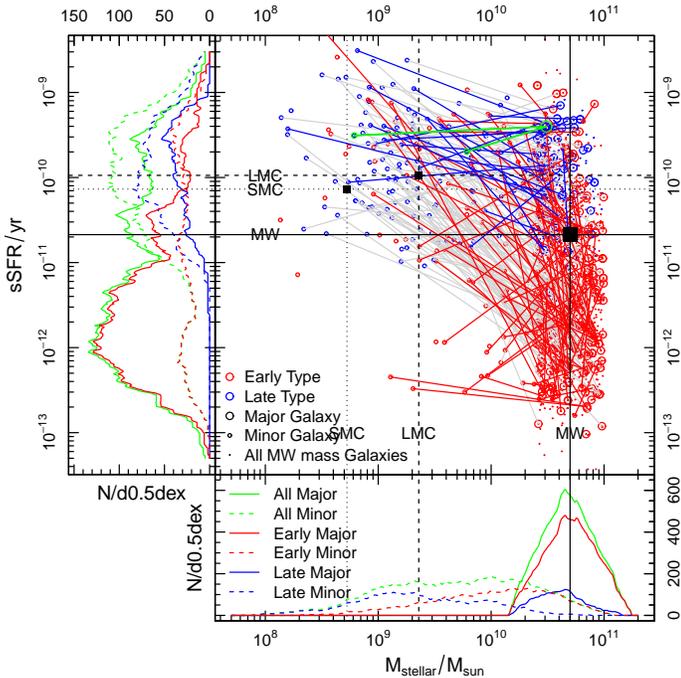}}}
\caption{\small Stellar mass versus sSFR of all pairs where the dominant galaxy is within 0.3 dex of the stellar mass of the MW. Also plotted in the background as points are all galaxies that make the mass selection, whether or not they are in pairs. The colouring refers to morphological type (blue= late-type, red- early-type), and size specifies whether the galaxy is the major (large circle) of minor (small circle) galaxy in the pair, or galaxy that makes the mass selection (point). The coloured lines show connections between galaxy pairs, where blue lines indicate both the major and minor galaxies are late-type, red lines indicate both are early-type, and grey lines show that the differ. Selection shown is the volume sample (see Sec \ref{sec:selection}).
\newline
{\bf Observations:}
\newline
- The MW and the LMC are in a well populated part of stellar mass--sSFR parameter space for all galaxies in the sample.
\newline
- Given it is a late-type dominant pair galaxy, the MW has an unusually low sSFR.
}
\label{fig:PairScatter}
\end{figure}

Figure \ref{fig:PairScatter} shows the stellar mass versus sSFR of all pairs where the dominant galaxy is within 0.3 dex of the stellar mass of the MW. The links between all visible pairs are shown, where blue lines indicate the galaxies are both late-type, red lines connect early-types and grey lines are pairs where the type differs. In the redshift range plotted we only expect to be complete down to the LMC, so we will concentrate the discussion here. 

Qualitatively we can see that the MW and the LMC are in a well populated part of stellar mass--sSFR parameter space if we just compare to the global populations (i.e.\ ignore the morphological type of the galaxies). In particular, the LMC is very typical for pair systems in our sample. There are even plenty of examples where the minor galaxy in the pair is more massive than the LMC, yet both the major and minor galaxies are classified as late-type. There is no parameter space where late-type--late-type pair combinations dominate, however early-type--early-type pairs clearly dominate for low sSFRs.

It is notable that the MW is at the very low extreme for sSFR given it is late-type compared to similar mass dominant pair galaxies. This tallies with our observation in Figure \ref{fig:GroupScatter} that the MW has an exceptionally low sSFR given the mass of its halo. Again, the argument that can be made is that the LMC looks broadly typical as a late-type minor pair galaxy (when LMC stellar mass minor pair galaxies exist), but the MW appear to be anomalous as a late-type galaxy given the combination of its stellar mass, halo mass and sSFR. { This finding is in agreement with the recent work presented in \citet{robo12}, which looked in detail at the occurrence rates of massive satellites around MW stellar mass galaxies. This work also highlighted that the morphological type of the MW and LMC is very unusual--- given their stellar masses and the fact they are in a close pair, it is more likely to observe such stellar systems with early-type galaxies.}

\subsubsection{Effect of Halo Mass on Close Pair Rates}

Above we have discussed a number of effects that are stellar mass independent. The dominant effects are caused by pair interactions, with some effects due to the group scale environment also detectable. However, pairs and groups are clearly not mutually independent parts of galaxy parameters space. Indeed it is almost certain that objects detected as pairs will form some part of a galaxy group as found in \G3C because the compact phase-space selection is usually encompassed by the FoF grouping links used to generate the \G3C.

\begin{figure}
\centerline{\mbox{\includegraphics[width=3.7in]{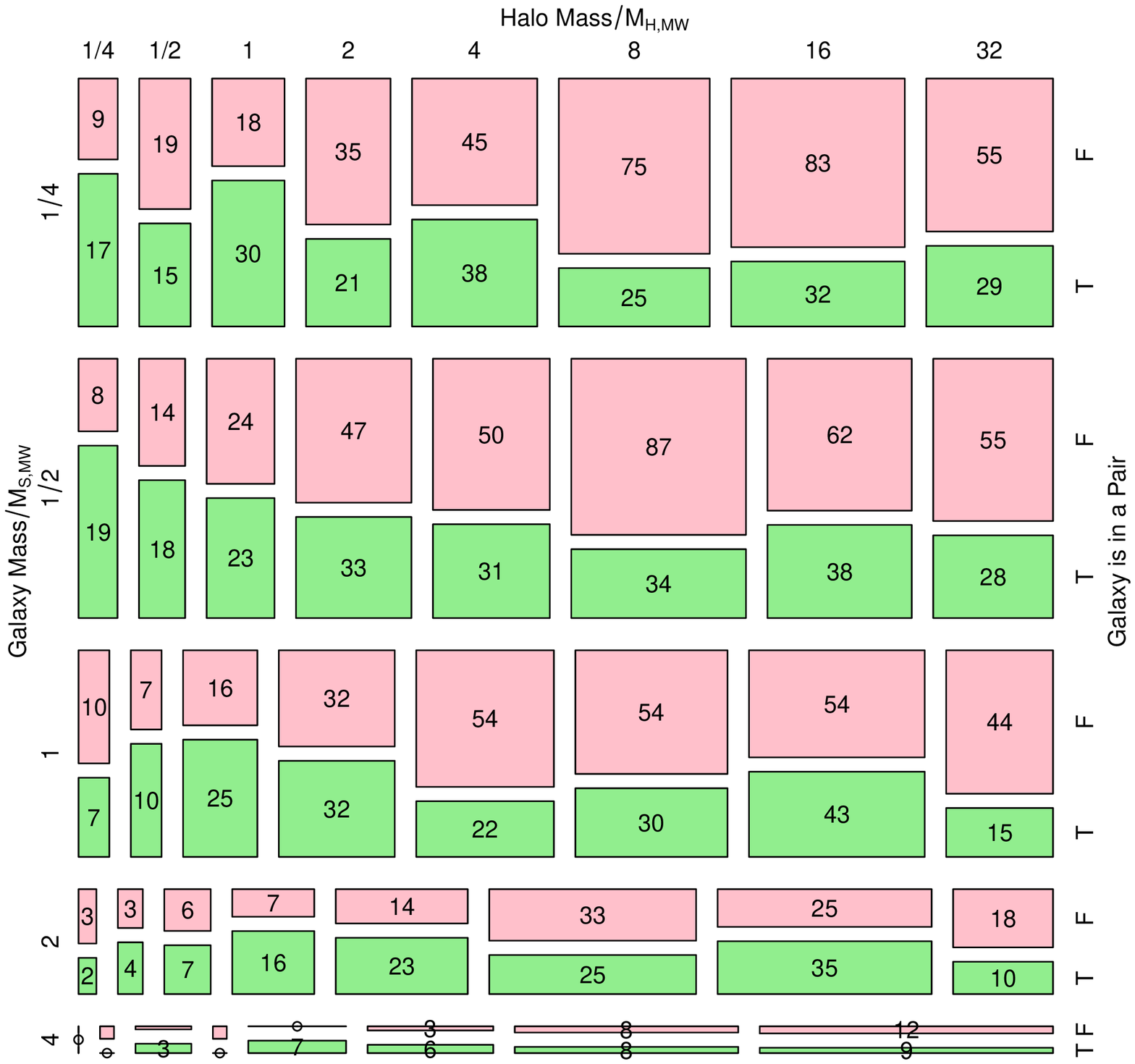}}}
\caption{\small Mosaic plot of Stellar mass of galaxy (Sec \ref{sec:galstelmass})/ Halo mass (Sec \ref{sec:galhalomass})/ Galaxy in pair (Sec \ref{sec:galpaired}). Selection shown is the volume sample (see Sec \ref{sec:selection}).
\newline
{\bf Observations:}
\newline
- Galaxy pairs (by the definition of galaxy pair used in this work) are less likely in the highest mass halos.
\newline
- The peak in MW-LMC-SMC like pair systems for $\mathcal{M}_{S,MW} \ge 1$ galaxies is seen at MW mass halos.
\newline
- The indication is that MW stellar mass galaxies in MW halo mass groups are key sites of dynamical activity.
}
\label{fig:LsHmPair}
\end{figure}

Figure \ref{fig:LsHmPair} shows the prevalence of galaxy pairs as a function of halo mass and stellar mass. A clear characteristic seen for all stellar masses is that the highest fraction of pairs (by our definition) do not occur in the most massive halos. In these environments velocity dispersions are large enough that galaxies are unlikely to be observed in the window of dynamical parameter space that we are sensitive to. Instead, the pair rate is highest in lower mass halos. For galaxies with $\mathcal{M}_{S,MW} \ge 1$ we see that MW mass halos host the highest pair fraction of all, i.e.\ the most common dynamical environment for MW stellar mass galaxies are MW halo mass groups. This lends credence to the idea that the Milky-Way and it's local halo are a cosmologically significant site for Galactic archeology studies--- it is clearly a key environment for understanding the typical experience of stellar mass in large cosmological volumes of the Universe.

\begin{figure}
\centerline{\mbox{\includegraphics[width=3.7in]{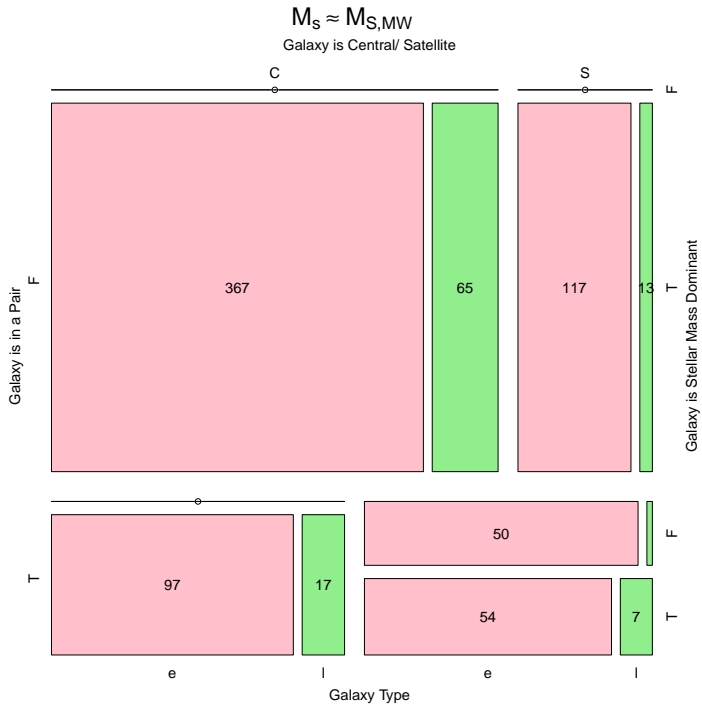}}}
\caption{\small Mosaic plot of Galaxy in pair (Sec \ref{sec:galpaired})/ Galaxy is Central or Satellite (Sec \ref{sec:galbgg})/ Largest galaxy in pair (Sec \ref{sec:galbpg})/ Type (Sec \ref{sec:galtype}). Subset used for this figure is $\mathcal{M}_{s} = \mathcal{M}_{S,MW}$ (Sec \ref{sec:galstelmass}). In this plot, central galaxies include those not in groups (since they are the central galaxy of their own halo).  Selection shown is the mass sample (Sec \ref{sec:selection}).
\newline
{\bf Observations:}
\newline
- The late-type fraction is between 10\%--15\% for all but the minor pair galaxy subset.
\newline
- Central galaxies are always the dominant component of a pair, and the late-type fraction does not vary much. Satellite galaxies show a variety of interaction dominance with an almost 54/46 split between being dominant or minor given our observational limits.
\newline
- For $\mathcal{M}_{s} = \mathcal{M}_{S,MW}$ galaxies, being the minor galaxy in a pair suppresses the star forming fraction more than merely being a satellite galaxy of the same stellar mass in a group.
}
\label{fig:inpairBGGBPGTy}
\end{figure}

Figure \ref{fig:inpairBGGBPGTy} is the final plot tying this work together, showing how combinations of paired and grouped galaxies can predict the late-type fraction for $\mathcal{M}_{S,MW}$ mass galaxies. Since we have found that the morphological type displays the strongest associations with other galaxy parameters, this is used as the final contingency. This plot demonstrates that  a significant fraction of $\mathcal{M}_{S,MW}$ mass galaxies are in closely interacting pairs (down to our detectable limits), and those which are the minor galaxy in a galaxy-galaxy interaction have a hugely suppressed late-type fraction (2\%). This is a smaller fraction than satellite galaxies not in pairs (10\%), or any central galaxy (15\%). Central galaxies are always the dominant galaxy in any interaction, so they do not show much residual effect due to pairwise interactions. Whilst central galaxies might always be dominant in pairwise interactions, satellite galaxies show a mixture of being dominant or minor for the selected stellar mass range. Broadly speaking $\mathcal{M}_{S,MW}$ mass galaxies tend to be the dominant galaxies in all interactions, even when they are in a halo which has a more massive galaxy. In conclusion, the integrated interaction history for satellite galaxies is non-trivial, whilst central galaxies should possess a much simpler evolution since they will have been much less affected by galaxy-galaxy interactions.

{

\section{Discussion and Future Work}
\label{sec:discussion}

This work has opened up a number of brackets on the topic of what influences, and is influenced by, $L^*$ galaxy evolution. Not all of these brackets are closed in this paper, and indeed the expectation is that further analysis and comparisons to detailed simulations will be required to fully untangle the complex causes, effects and mere correlations. There are a few unambiguous remarks that can be made, and these will be discussed first. The more speculative physics hinted at in this work is discussed towards the end of this section.

\subsection{Clear Results}

As discussed at length in the introduction, the fate of any single galaxy will be caused by a complex mixture of inputs. In extreme detail the history of any galaxy is certainly unique, and we can only hope to observe broad systematic trends in statistically large samples. The GAMA survey is suitably deep enough and complete enough to separate out the three major influences on galaxy evolution: stellar mass, galaxy-galaxy interactions and halo mass effects.

There are two particularly clear results presented in this work: galaxy stellar mass is a very good predictor of most galaxy properties, and galaxy properties are generally poor predictors of each other. 

Finding that stellar mass is the best predictor of simple galaxy observables is very much in line with recent extra galactic literature \citep[e.g.][]{bald06,peng10,thom10}, where the burgeoning consensus is that stellar mass has the primary correlation with the properties of all galaxies. Whilst we concur with these findings, it is important to stress that the process of assembling stellar mass to form the galaxies is undoubtably influenced by the environment (e.g.\ halo) that the galaxy resides in. The explanation can be formed in the language of sub-halo abundance matching \citep[SHAM: e.g.][]{simh12}, the halo occupation distribution \citep[HOD: e.g.][]{zeha05,brow08,zeha11} or the conditional luminosity function \citep[CLF: e.g.][]{coor06}, but all share a common theme that the dark matter mass of the parent halo predicts the stellar mass function within the halo. Once the halo has assembled this stellar mass it has a more minor role in dictating the future evolution of the galaxy, and it is certainly true that two galaxies of a given stellar mass are more likely to share observational properties than two galaxies of a given parent halo mass. However, it is less clear that stellar mass {\it drives} the evolution. Turning the argument around, it is just as reasonable to state that stellar mass is the end product (or by product) of a large (but bounded) number of evolutionary paths.

Whilst stellar mass appears to be the best predictor of galaxy properties, other galaxy properties are notably poor predictors. Whilst it has long been observed that galaxies possess strong bimodality in a number of properties \citep[see][and Taylor et~al. in prep]{blan03,bald04,bald06,driv06}, this work has demonstrated quite comprehensively that these bimodalities operate quite independently, with little correlation with each other. The simplest explanation is discussed in the introduction, and posits that differing observational timescales for a common trigger will naturally allow for galaxy bimodality in many simultaneous dimensions, but with few common properties amongst individual galaxies. Accurate SED based `age' estimates are not available for GAMA galaxies, so future work to ascertain the role temporal lag has to play in disconnecting various bimodalities will have to be based on unambiguous transition populations, such as E+A galaxies.

Figure \ref{fig:CoDiTyEmLs0} is particularly important in highlighting the danger of over-simplification of galaxy traits. Here we see that much less than half of all $L^*$ galaxies can be meaningfully described as being `red, disk-less, early-type and passive' and `blue, disky, late-type and star forming'. In fact, between these four characteristics only disk-less late-types are found to never exist, which must be the case for any reasonable definition of late-type. Every other combination of the 4D contingency space is populated at least to some extent. This alerted us to a key realisation: to map out these 4 properties simultaneously with stellar mass, halo mass and pair fraction would be impossible. There are not enough galaxies to properly sample the parameter space, even when using mosaic plots designed for deconstructing multi-dimensional co-dependencies in relatively small datasets. We chose to focus on investigating star formation and morphological type in latter sections, with the caveat that this will not fully capture all secondary galaxy properties.

\subsection{Complex Results}

The analysis presented here is not designed to offer definitive predictions on all mechanisms of galaxy evolution. That said, whilst stellar mass has been highlighted as the main characteristic that predicts the properties of a galaxy, this work also uncovered strong evidence for secondary evolutionary drivers that act independently of stellar mass. We even found compelling evidence for galaxy-galaxy conformity between observational properties in galaxy groups and galaxy pairs. These secondary drivers produce much weaker effects than stellar mass, so are easily washed away without careful multi-dimensional analysis of the data, which in this work was executed with the use of mosaic plots. The mosaic approach has been used because the statistical power of the data (broadly speaking the number of galaxies we have to work with) is at the limit for sub-divisions into multi-dimensional space. It also allows us to explore discrete and/or binary parameter space not accessible with principal component analysis. The full parameter space that captures all the physics of even just an $L^*$ galaxy is extremely large, so here we have chosen the few parameters that we assume to be the most important in influencing galaxy properties.

A common theme of the results in this work is that central and satellite galaxies of a given stellar mass {\it and} halo mass have distinctly different properties--- star formation is relatively suppressed in satellite galaxies, and the late-type fraction is smaller. What is not certain is the mechanism that causes these differences. It is plausible that halo scale effects, for instance the stripping of gas during the accretion of sub-halos that host the future satellite galaxies, could cause such results. However, the work has been extended to investigate the role that pairwise interactions might have on the same galaxy properties, and many of the same trends were observed. The dominant pair galaxy of a given stellar mass will be more likely to be late-type and forming stars than a minor pair galaxy of the same stellar mass. The strength of these differences is of a similar magnitude to the the central/ satellite variation already noted, and a large fraction of grouped galaxies are undergoing these close galaxy-galaxy interactions.

Previous works have found that environment does play a secondary evolutionary role, after the primary role that stellar mass plays in determining galaxy properties. \citet{peng10} note that environment plays an important secondary role in determining the fate of satellite galaxies. This work broadly supports the same view, since we find that ungrouped galaxies display properties most like central galaxies in groups, i.e.\ it is less a case of the halo strongly influencing central galaxies, and a clearer case that it suppresses star formation in satellite galaxies.

What this work demonstrates is that the mechanism of this `environmental' suppression is heavily associated with galaxy-galaxy interaction physics. The population fractions for a given stellar mass galaxy are altered more dramatically by virtue of being the major or minor galaxy in a pair than by whether the galaxy is a central or satellite in a group. Surprisingly, morphological type is more strongly affected by close interactions than star formation, suggesting the latter is not necessarily clear-cut, with both triggering and suppression physics playing a role. The full role, and the full range of mechanisms, that environment has on the evolution of galaxies is not a closed case, but some significant fraction of the role that environment plays (beyond assembling stellar mass into galaxies) is simply assembling different stellar mass galaxies within close proximity of other galaxies with velocities conducive to regular close interactions. This concurs with the final conclusion made in \citet{thom10}, who highlighted the likely role environment plays in stimulating phase transitions via galaxy-galaxy interactions.

Since we have observed the strong instantaneous effect of galaxy-galaxy interactions, the argument can be extended that the integral of many such interactions over the lifetime of a galaxy could well explain its major properties (after the role of intrinsic stellar mass is considered, of course). The importance of galaxy-galaxy interactions is also suggested by the fact that morphological fractions are equally (if not more) affected by close passages. Feedback mechanisms that do not involve gravity are less able to radically redistribute the orbits of stars (a prerequisite for a morphological transformation), and such events are also able to disturb gas and trigger AGN and further feedback \citep[e.g. harassment:][]{moor96}. Future work includes using simulations to investigate in detail the integrated effects of galaxy-galaxy interactions.

We looked at galaxy conformity of morphological type and emission class in groups and pairs, and found evidence for it, to differing degrees, in both cases. In galaxy groups there was strong evidence for morphological type conformity: i.e.\ the morphological type of the central galaxy aided the prediction (and was correlated with) the morphological type of the accompanying satellite galaxies. This finding is very much in line with that of \citet{wein06}, with the caveat that they used colour and sSFR to define `type' whereas we used a purely visual classification. However, we found much less compelling evidence for emission class conformity between central and satellite galaxies. When analysing galaxy pairs in detail we found strong evidence of conformity. In this case, the effect was that the star formation class of one galaxy within a pair could predict its likelihood in the other pair galaxy. The morphological type conformity was considerably weaker, i.e.\ the parameters that display the strongest conformity swapped between considering central and satellite galaxies in groups, and closely interacting galaxies. 

Under the assumption that star formation has a more instantaneous response to stimulation than morphology, this suggests that the morphological type conformity witnessed in galaxy groups is the result of a longer period integrated effect. Such a mechanism could be multiple minor interactions over time. Such dynamical events would slowly rearrange the orbits of the associated galaxies, and the net result of this will be to introduce morphological conformity. On the other hand, the emission class conformity witnessed in galaxy pairs is a transitory response due to a brief close interaction. Only in a minority of cases are the tidal physics of the interaction significant enough to evoke a morphological response in the dynamical window in which we have selected our pairs.

This work was not intended to investigate galaxy occupation statistics in detail. The main analysis considering the typicality of the MW galaxy was presented in \citet{robo12}, where the main conclusion was that only $\sim11.9$\% of MW stellar mass galaxies should host an LMC mass (or more massive) close companion. This bright satellite fraction rate was in close agreement with \citet{boyl11}. More recent work in \citet{boyl11,love12} has suggested that the largest sub-halos in the MW appear to be unoccupied (AKA the `too big too fail' problem). Further work investigating the faint end of the galaxy group LF is underway, with the intention of quantifying how typical the MW satellites are beyond the LMC regime explored in this work.

}

\section{Conclusions}
\label{sec:conclusions}

In this work we have analysed in detail the typical group scale and pair scale environment of galaxies close to the stellar mass of the MW and a factor 4 either side of it. We did this by visually classifying a complete sample of galaxies that had stellar mass $\mathcal{M}_{s} > \mathcal{M}_{S,MW}/4$, and all associated pair galaxies and group galaxies with a $0.01 < z < 0.089$ redshift selection (allowing us to observe down to at least LMC depth in the $r<19.4$ mag GAMA survey). The major findings are summarised below.

\begin{itemize}

\item The correlation between colour, SF, morphological type and presence of a disk is non-trivial, and one parameter cannot be used to accurately predict another. However, some combinations are highly predictive: red, diskless early-type galaxies are almost certain to have no emission line features. General relationships are as expected: bluer galaxies are more likely to be strongly star forming than red galaxies etc. For later mosaic plots the most distinctive of these parameters is chosen to avoid information degeneracy. (See Fig \ref{fig:CoDiTyEmLs0}).

\item Larger stellar mass galaxies are more likely to be grouped (in our volume limited definition of group), and also more likely to be a central galaxy if they are in a group. This result in particular falls in line with halo occupation distribution (HOD) work conducted over the last decade. These have found that the chance of a massive galaxy being a satellite is small, and indirectly proportional to the halo mass \citep[e.g.][]{zeha05,brow08,zeha11}. (See Fig \ref{fig:LsingroupsLineTy}).

\item The emission class fractions vary hugely for galaxies of a given stellar mass and group halo mass depending on whether or not it is the central or satellite galaxy. This suggests that galaxy-galaxy interactions play the most important role in shaping the evolution within a group halo, the main role of halo mass being to gather the galaxies together to allow for the interactions.  (See Fig \ref{fig:LsingroupsLineTy} and \ref{fig:GaHmLsLi}).

\item Galaxy disks are more likely to be present in ungrouped galaxies for galaxies more massive than \MMW, suggesting they can be easily disrupted by dynamical interactions. This has a stronger effect than any enhancement to the disk accretion rate due to there being more material available for minor mergers in galaxy groups. This effect is also enhanced relative to the late-type to early-type fraction (where late-type galaxies always have disks). This relationship is not observed for unpaired galaxies, indicating that the extra degree of isolation offered by being ungrouped, and the longer period to the last major dynamical interaction, is significant in establishing a visible galaxy disk. (See Fig \ref{fig:PopFracs}).

\item Satellite galaxies comprise a larger fraction of stellar mass in larger halo mass groups. This is again in line with the standard HOD analysis seen over the last decade \citep[e.g.][]{zeha05,brow08,zeha11}. (See Fig \ref{fig:GaHmLsLi}).

\item { There is strong evidence for morphological type conformity in groups, where the morphological type of the central galaxy predicts the morphological type of the satellite galaxy. We also find weaker evidence for galaxy emission class conformity (See Fig \ref{fig:SatdepCen}).}

\item The MW is unusually massive, and has an unusually low sSFR, for a late-type galaxy in a MW mass halo. (See Fig \ref{fig:GroupScatter}).

\item The LMC is very close to the MW compared to similar mass satellites relative to central galaxies in MW halo mass groups. (See Fig \ref{fig:SatScatter}).

\item If MW mass galaxies are the major component of a pair experiencing a close passage, the star forming--passive fraction is close to 1--1, but if they are the minor galaxy in of a pair experiencing a close passage then the passive fraction is significantly larger. (See Fig \ref{fig:PairPhaseSF}).

\item Massive galaxies are more likely to be in a pair (given our definition of pair, and selection requirements). (See Fig \ref{fig:LsinpairLine}).

\item The late-type fraction drops more rapidly with stellar mass than the star forming fraction, particularly for minor pair galaxies. (See Fig \ref{fig:LsinpairLine}).

\item The most massive minor pair galaxies show a statistically significant increase in the disk fraction relative to the unpaired case. (See Fig \ref{fig:PairPopFracs}).

\item Dominant pair galaxies only find their emission state greatly altered when the pair ratio is close to 1, otherwise we observe the same galaxies as we would in an unpaired system. (See Figs \ref{fig:G1depG2vcomp} and \ref{fig:PairSFR}).

\item The minor galaxy in a pair is greatly affected by its companion galaxy, and whilst all star formation is suppressed relative to equivalent stellar mass unpaired galaxies, it becomes lower still when the mass ratio increases (i.e.\ the major galaxy becomes more massive and dominant within the pair). (See Fig \ref{fig:G1vG2}).

\item { There is strong evidence for galaxy emission class conformity in galaxy pairs, where star forming emission (or no emission) in one galaxy in a pair strongly predicts the same state in the other galaxy. We find much weaker evidence for morphological type conformity in pairs (See Fig \ref{fig:G1depG2em}).}

\item Given it is a late-type dominant pair galaxy, the MW is unusual in having such a low sSFR. (See Fig \ref{fig:PairScatter}).

\item Group halo mass has a clear effect on the prevalence of galaxy-galaxy interactions. These become rapidly suppressed in the largest halos, and are more common, for a given amount of stellar mass, in lower mass halos. (See Fig \ref{fig:LsHmPair}).

\item Milky-Way stellar mass interactions peak for Milky-Way halo mass groups, suggesting that local Galactic archeology studies will be probing a very important part parameter space for understanding the fate of $L^*$ galaxies in general. (See Fig \ref{fig:LsHmPair}).

\item The observable properties of Milky-Way mass galaxies is more radically altered by being the minor pair galaxy in an interaction than merely being the satellite galaxy in a group halo. (See Fig \ref{fig:inpairBGGBPGTy}).

\end{itemize}

\section*{Acknowledgments}

ASGR acknowledges STFC and SUPA funding that was used to do this work.
GAMA is a joint European-Australasian project based around a spectroscopic campaign using the Anglo-Australian Telescope. The GAMA input catalogue is based on data taken from the Sloan Digital Sky Survey and the UKIRT Infrared Deep Sky Survey. Complementary imaging of the GAMA regions is being obtained by a number of independent survey programs including GALEX MIS, VST KIDS, VISTA VIKING, WISE, Herschel-ATLAS, GMRT and ASKAP providing UV to radio coverage. GAMA is funded by the STFC (UK), the ARC (Australia), the AAO, and the participating institutions. The GAMA website is {\tt http://www.gama-survey.org/}.

\bibliographystyle{mn2e}
\setlength{\bibhang}{2.0em}
\setlength\labelwidth{0.0em}
\bibliography{lgav1}

\begin{appendix}
\section{Contingency Tables and Mosaic Plots}

Data that has a large number of parameters can be confusing to visually represent, and most methods that increase the clarity of the data shown reduce the information presented. In astronomy it is not unusual to have data that has multiple discrete columns. A simple example of a set of discrete criteria for galaxies is:

\begin{itemize}
\item Is the galaxy in a group (True/ False)?
\item Is the galaxy early-type or late-type (Early/ Late)?
\item Does the galaxy have a disk (True/ False)?
\item What kind of H$\alpha$ emission is observed in the spectra (AGN (A)/ star formation (S)/ no emission (X))?
\end{itemize}

An incomplete, but popular, approach to explaining how these different discrete classes depend on each other is to simplify the problem to asking assumed `key' questions such as ``What fraction of galaxies in groups are late-type?''. The reader will run into difficulty if they happen to ask an unexpected question such as: ``given a galaxy does not exhibit star formation and has a disk, what is the likelihood that it is early-type and in a group?'' Such a question is perfectly valid, but not precisely communicated by independent bar charts, tables and a few choice examples.

In statistics, parameters that have multiple discrete outcomes are known as contingencies, and hence the relationships between multiple parameters with discrete outcomes are known as contingency tables. A simple hypothetical 2-way contingency table is given as an example in Table \ref{tab:cont1}, where the contingencies compared are morphological type (early or late) and presence of a disk (True or False). Table \ref{tab:cont1} shows the number of objects that exist in each of the 4 possible logic overlaps. As expected, a large fraction of early-type galaxies do not possess disks, and 100\% of late-type galaxies do.

\begin{table}
\begin{center}
\begin{tabular}{llrrrrrr}
Type 	& \multicolumn{2}{c}{Disk Present}	\\
		& False	& True	\\
  \hline
Early		& 310	& 334  \\
Late 		& 0		& 100  \\
\end{tabular}
\end{center}
\caption{2-way contingency table comparing hypothetical numbers of galaxies that are early/ late-type and do or do not have a disk.}
\label{tab:cont1}
\end{table}

In some scenarios a contingency table is an appropriate form of conveying information. However, to determine what fraction of galaxies are early-type requires the reader to add numbers together that are not given in the Table. In this very simple 2-way contingency example this calculation is trivial ($(310+334)/(310+334+100)=$87\%), however simply adding one more contingency makes this task a serious burden.

Table \ref{tab:cont2} expands Table \ref{tab:cont1} to include an extra contingency (presence of emission line), which has three allowed outcomes: AGN emission (A)/ star formation (S)/ No emission (X). Calculating the early-type fraction is now quite onerous. It is also harder to get a feel for the data--- what is the best predictor of a galaxy possessing strong emission: being early/ late-type or the presence of a disk? Without a large amount of arithmetic the answer is opaque.

\begin{table}
\begin{center}
\begin{tabular}{llllrrrrrr}
Type		& Emission	& \multicolumn{2}{c}{Disk Present}	\\
		& 			&False	& True	\\
  \hline
Early		&  A			& 9		& 44		\\
 		&  S			& 37		& 126	\\
 		&  X			& 264	& 164	\\
\hline
Late		&  A			& 0		& 13		\\
 		&  S			& 0		& 83		\\
 		&  X			& 0		& 4		\\

\end{tabular}
\end{center}
\caption{3-way contingency table comparing hypothetical numbers of galaxies that are early/ late-type, do or do not have a disk and exhibit star formation (AGN emission (A)/ star formation (S)/ No emission (X)).}
\label{tab:cont2}
\end{table}

\begin{figure}
\centerline{\mbox{\includegraphics[width=3.7in]{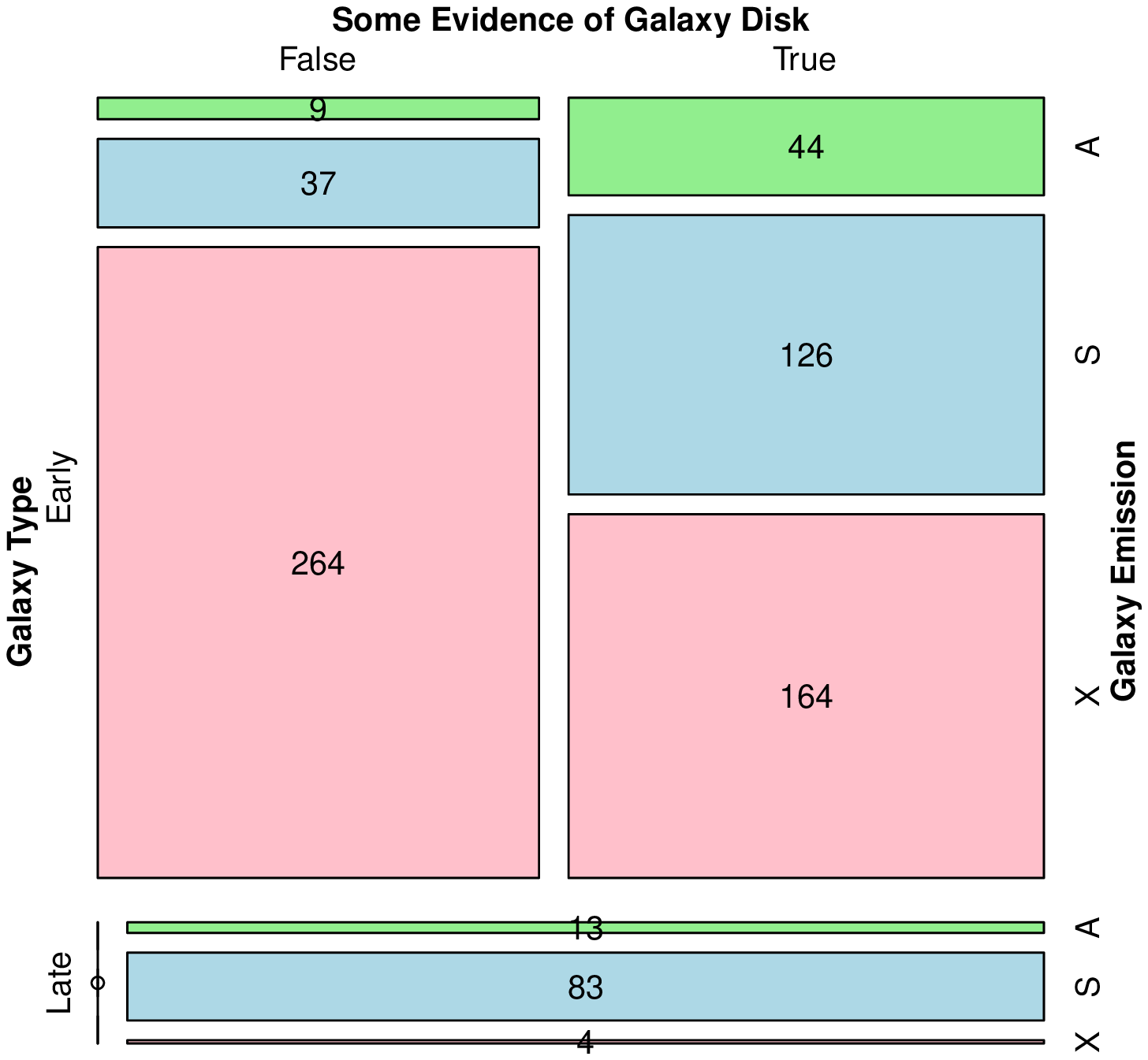}}}
\caption{\small Mosaic plot of the data contained in Table \ref{tab:cont2}. The contingency ordering is: Type--Disk--Emission, where cell colouring is according to the emission contingency.}
\label{fig:cont2}
\end{figure}

For analysing this sort of multi-dimensional data, with numerous interesting dependencies, statisticians have long favoured mosaic plots.  Astronomy has been slow to catch on to this variety of data visualisation, but it provides a very powerful means of presenting the data when datasets are large and the author does not want to over-simplify the information.

A mosaic plot works by iteratively sub-dividing the data in a contingency table into ever small cells, alternating whether the splits are made vertically or horizontally. In this example the first division could be horizontal, showing whether the galaxy is early-type or late-type (87\% to 13\%, as calculated above) creating two cells of the appropriate area. The next could then be a vertical split, showing whether the galaxy has any residual disk, so the two cells become 4. The last split would again be horizontal, showing the emission class (the final contingency). Since this last contingency has 3 possible classes, 12 cells would be plotted in total (assuming all cells are occupied).

Figre \ref{fig:cont2} shows exactly the plot construction described above, where the counts given in Table \ref{tab:cont2} are shown within each cell. The combination of cells containing no data (late-type galaxies with no disk) are shown as a line with a small circle drawn on top. The final contingency used to split the rectangles is also used to colour code the data. All cells of the same colour have the same property for the final contingency--- in this case all red cells show galaxies with no emission. In such a plot a lot of previously buried information becomes immediately, intuitively, available to the reader. In the subset containing all early-type galaxies we can see an almost identical fraction of galaxies do or do not have disks (without needing to add up any numbers). Also, an early-type galaxy with a disk is more likely to be undergoing strong star formation than an early-type galaxy without a disk. The key to being able to make this last claim is that the relative fraction of galaxies with and without disks is no longer relevant, instead we wish to know the fraction of star forming galaxies within each sub-class. The mosaic plot allows us to assess the predictive power of possessing a disk by comparing the relative {\it lengths} of the adjacent rectangles.

\begin{figure}
\centerline{\mbox{\includegraphics[width=3.7in]{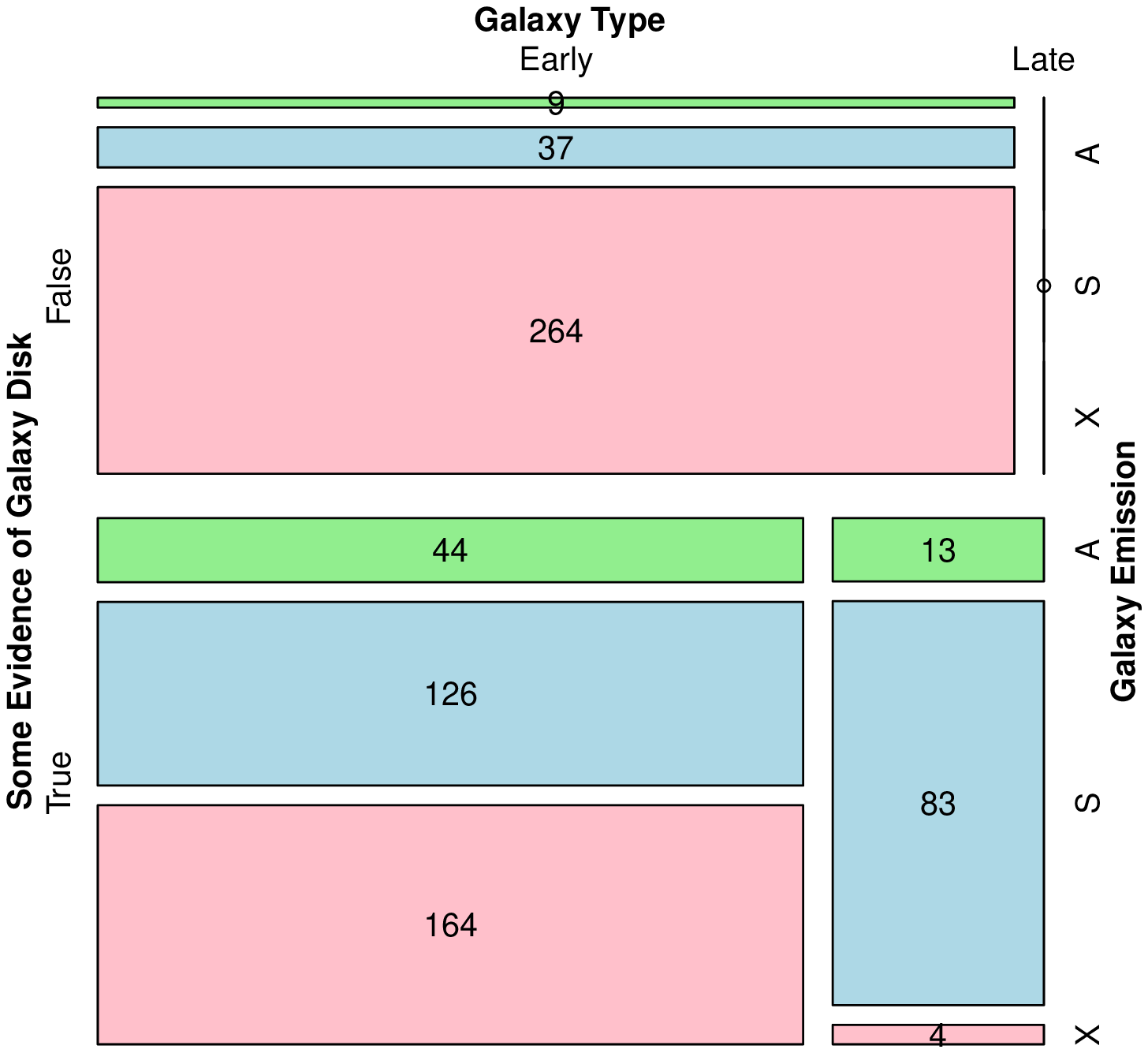}}}
\caption{\small Mosaic plot of the data contained in Table \ref{tab:cont2}. The contingency ordering is: Disk--Type--Emission, where cell colouring is according to the emission contingency. This is a reversal of the first two contingencies shown in Figure \ref{fig:cont2}, allowing some information to be extracted more easily, but others harder.}
\label{fig:cont2v2}
\end{figure}

Care must be taken to construct a mosaic plot in the most useful order, since each subsequent division becomes harder to analyse. Figure \ref{fig:cont2v2} shows exactly the same information as Figure \ref{fig:cont2}, but now the ordering of the first two contingencies is reversed. The questions posed above are no longer as easy to answer with a cursory look at the mosaic since the key cells are no longer side-by-side. However, other observations become easier instead: star forming galaxies are more common in a late-type galaxy with a disk than an early-type galaxy with a disk. Whilst the ordering is important in terms of what can be easily extracted from the data, the mosaic plot gives access to all the contingency information so all the dependencies can be extracted.

Colour coding according to the final contingency is particularly useful for this reason: a visual summation of the three colours clearly shows the red (i.e.\ no emission) to be dominant, followed by star forming and lastly AGN. For gaining rapid access to the most important relationships in multi-dimensional data, mosaic plots are a powerful tool. In terms of information retention, they behave no worse for continuously distributed data than using a running median or other quantiles, this being popular method of displaying relationships in astronomical data with large amounts of intrinsic scatter.

In theory any number of dimensions of contingency data can be plotted on one mosaic plot- the data is just continually subdivided as outlined above. In practice it becomes extremely difficult to extract useful information out of more than 4 dimensions, and the plot labelling becomes confusing (there are four sides to the superset rectangle, so using more than 4 dimensions in a contingency tables means labelling has to be stacked). For this work we will restrict the mosaic plots to showing 4 or fewer dimensions. Also, only one version of each mosaic will be made (i.e.\ one contingency ordering). This will be chosen as the most instructive of the different combinations attempted.

For a detailed description of the generation of mosaic plots the interested reader should refer to \citet{meye06}. This paper discusses the popular `vcd' package for R that was used in this work. This software proved to be the most flexible and extendable of a number tried, and is a good choice for anyone wishing to produce mosaic plots for other datasets. In all the mosaic plots shown in this work the ordering of the mosaic splitting (and thus labelling) is: $1^{st}$ contingency, horizontal split, labelling on the left; $2^{nd}$ contingency, vertical split, labelling on top; $3^{rd}$ contingency, horizontal split, labelling on right; $4^{th}$ contingency, vertical split, labelling on bottom. In cases of crowded labelling, some labels may be omitted/ abbreviated/ truncated. Cells, or combinations of cells, that sum to zero are represented as a line with a circle drawn on top. Cells with only one count do not have a number printed inside.
\end{appendix}

\label{lastpage}

\end{document}